\documentclass[proof]{WileyASNA-v1}
\usepackage{supertabular}
\usepackage{lscape}
\articletype{Article Type}%

\received{09 March 2020}
\revised{11 July 2020}
\accepted{03 September 2020}

\begin{document}

\title{Identification of young nearby runaway stars based on Gaia data and the lithium test\protect\thanks{Based on observations obtained with telescopes of the University Observatory Jena, which is operated by the Astrophysical Institute of the Friedrich-Schiller University, and telescopes of the Fred L.\ Whipple Observatory, which is operated by the Harvard-Smithsonian Center for Astrophysics Cambridge MA.}}

\author[1]{R. Bischoff*}
\author[1]{M. Mugrauer}
\author[2]{G. Torres}
\author[1]{T. Heyne}
\author[1]{O. Lux}
\author[1]{V. Munz}
\author[1]{R. Neuh\"auser}
\author[1]{S. Hoffmann}
\author[1]{A. Trepanovski}

\authormark{Bischoff \textsc{et al.}}

\address[1]{\orgdiv{Astrophysikalisches Institut und Universit\"{a}ts-Sternwarte Jena},  \orgaddress{\state{Schillerg\"{a}{\ss}chen 2, 07745 Jena}, \country{Germany}}}
\address[2]{\orgdiv{Center for Astrophysics $\vert$ Harvard \& Smithsonian}, \orgaddress{\state{60 Garden Street, Cambridge MA 02138}, \country{United States of America}}}

\corres{*R. Bischoff, Astrophysikalisches Institut und Universit\"{a}ts-Sternwarte Jena, Schillerg\"{a}{\ss}chen 2, 07745 Jena, Germany \email{richard.bischoff@uni-jena.de}}

\abstract{Young nearby runaway stars are suitable to search for their place of origin and possibly associated objects, for example neutron stars. \cite{tetzlaff} selected young ($\le 50$ Myr) runaway star candidates from Hipparcos, for which they had estimated the ages from the location in the \text{Hertzsprung-Russell} diagram and evolutionary models. Here, we redetermine or constrain their young ages more precisely not only by using the new \textit{Gaia}\,DR2 data, but also by measuring lithium, which is a youth indicator. For 308 stars, we took spectra to search for the strong resonance doublet of the lithium-7 isotope at 6708\,\AA. The spectra were taken with the \'Echelle spectrograph FLECHAS at the University Observatory Jena between February 2015 and June 2018 and with TRES between April 2011 and June 2017 at the Fred L.\ Whipple Observatory. We found 208 stars with significant occurrence of lithium in their spectra, and five possess a possible age younger or about 50\,Myr. Three of these targets are even closer than GJ\,182, the nearest known runaway star at about 24\,pc. Theses stars are young runaway stars suitable for further investigation of their origin from either a dynamical or supernova ejection.}

\keywords{stars: HRD, fundamental parameters; methods: observational, data analysis; techniques: spectroscopic; astronomical databases: catalogues}

\jnlcitation{\cname{%
\author{R. Bischoff},
\author{M. Mugrauer},
\author{G. Torres},
et al.} (\cyear{2020}),
\ctitle{Identification of young nearby runaway stars based on Gaia data and the lithium test}, \cjournal{Astron.  Nachr.}, 2020.}

\maketitle

\section{Introduction}\label{sec1}

Runaway stars have higher velocities than typical field stars. They can be the result of a supernova explosion in a binary system where the secondary component is ejected as a runaway star \citep{blaauw} or they are ejected due to gravitational interactions between stars in dense stellar systems or clusters \citep{poveda}.

\cite{tetzlaff} did a combined analysis of spatial, tangential and radial velocities for all 118\,218 stars from the \textit{Hipparcos} catalogue \citep{perryman} to find more runaway star candidates. In their study they did not only focus on O- and B-type stars but extended the use of the term \textit{runaway} to every star with an unusual space velocity.
Given the motivation to study runaway stars from core-collapse supernovae in binaries and that such runaway stars cannot be traced back easily (because of the Galatic potential) for more than a few Myr, the age of such runaway stars is below about 50\,Myr (including up to 30\,Myr until the supernova). The age estimation was done by comparing luminosity and effective temperature of the stars to different evolutionary models. The results of their study were published in the catalogue \textit{Young runaway stars within 3\,kpc} which is available at the \texttt{VizieR}\footnote{\url{http://cdsarc.u-strasbg.fr/viz-bin/cat/J/MNRAS/410/190}} database.

For selected stars of this catalogue, our spectroscopic observing program was started in order to determine or confirm their possible young age based on the second data release of the \textit{Gaia} mission (\textit{Gaia}\,DR2) of the European Space Agency \citep{gaiadr2}. Furthermore, we searched for the absorption line of the lithium-7 doublet at 6708\,\AA, which is a youth indicator.

In this paper, we describe in detail the sample selection, as well as the spectroscopic observations and the data reduction in section 2. In section 3, a characterisation of physical properties of the targets, based on their \textit{Gaia}\,DR2 data, is given. The following section 4 yields information about the equivalent width measurements of the Li\,(6708\,\AA) line in the spectra of all observed stars and the lithium abundances of the identified dwarfs. In Section 5 we describe the age estimation of the dwarf stars. The results are discussed in the final section of this paper.

\section{Sample selection, Observations and data reduction}\label{sec2}

For this project, we chose stars from the catalogue by \cite{tetzlaff} that are observable at air masses $X<2.4$ from Jena (Dec $>-14^{\circ}$) and are bright enough ($V\leq8.5$\,mag) to obtain spectra with sufficiently high signal-to-noise-ratio (SNR) $>50$ at short integration times of only a few minutes. In total, 460 stars were selected as targets for spectroscopy. While 308 of them are already observed, analysed and discussed in this arcticle, the remaining 152 will be part of a future observing campaign.

The spectra for this study were taken with the fibre-linked \'{E}chelle spectrograph FLECHAS \citep{mugrauer2014}, mounted at the Nasmyth telescope ($D=90$\,cm, $f/D=15$) of the University Observatory Jena. The observatory is located 10\,km west of the city of Jena \citep{pfau}. During two observing epochs between 20th February 2015 and 3rd February 2016, as well as 21st January 2018 and 4th June 2018, 924 spectra with a total integration time of 123\,h were obtained in the course of our observing campaign at the University Observatory Jena.

The spectra were taken in the 1x1 binning mode of the spectrograph FLECHAS with individual integration times ranging between 10\,s and 1200\,s, according to target brightness. FLECHAS covers the spectral range from  3900\,\AA\,\,to 8100\,\AA\,\,in 29 orders with a resolving power of $R\approx9,300$ \citep{mugrauer2014} and the Li\,(6708\,\AA) line is detected in the middle of the 24th spectral order of the instrument. In order to achieve a sufficiently high SNR and to remove cosmics, three spectra were always taken for each target.
The SNR of all fully reduced spectra was measured at $\lambda=6700$\,\AA.
At this wavelength, most spectra yield at least a $\text{SNR}\geq 50$ and $\text{SNR}=152$ is reached on average. The only exceptions are HIP\,56383 ($\text{SNR}=25$), HIP\,107350 ($\text{SNR}=39$), HIP\,94761 ($\text{SNR}=42$) and HIP\,97198 ($\text{SNR}=42$). The details of all observations are listed in Table\,\ref{tab:obslog}\hspace{-2mm}.

For calibration purposes, three well exposed flat-field frames of a tungsten lamp as well as three spectra of a thorium-argon (ThAr) lamp are recorded immediately before the observation of each target. Each calibration image possess an individual integration time of 5\,s. In the ThAr spectra are about 700 detected emission lines for wavelength calibration. The long-term stability of the wavelength calibration of the instrument was previously shown by \cite{irrgang}, \cite{bischoff} and \cite{heyne}. Furthermore, three dark frames for all used integration times are recorded in every observing night for the dark and bias subtraction.

The FLECHAS software pipeline, which was developed at the Astrophysical Institute Jena, was utilised for dark and bias subtraction, flat-fielding, the extraction and wavelength calibration of the individual spectral orders. Additionally, the pipeline includes averaging and normalisation of the recorded spectra for each target \citep{mugrauer2014}.\\

As part of a separate long-term spectroscopic monitoring program to measure radial velocities and discover binary systems in another sample of runaway stars from \cite{tetzlaff}, overlapping in part with the one in this work, 30 of our objects were also observed between 6th April 2011 and 12th June 2017 with the Tillinghast Reflector Echelle Spectrograph TRES \citep{Szentgyorgyi2007}, \citep{furesz2008}. The spectrograph is attached to the 1.5\,m Tillinghast reflector at the Fred L.\ Whipple Observatory\footnote{\url{http://linmax.sao.arizona.edu/help/FLWO/whipple.html}} on Mount Hopkins (Arizona, USA). This bench-mounted, fiber-fed instrument delivers spectra at a resolving power of $R \approx 44,000$ that cover the wavelength region between 3800\,\AA\,\,and 9100\,\AA\ in 51 orders. Exposure times ranged from 45\,s to about 10 minutes, depending on brightness
and weather conditions, and the SNRs achieved at a mean wavelength of 5200\,\AA\ are between about 20 and 130 per resolution element of 6.8\,km/s. Exposures of a ThAr lamp were taken before and after each science frame, and the observations were reduced with a dedicated pipeline.

\section{Target characterisation with Gaia\,DR2 data}\label{sec3}

\begin{figure}[h!]
	\centering\includegraphics[width=8.25cm,height=8.25cm,keepaspectratio]{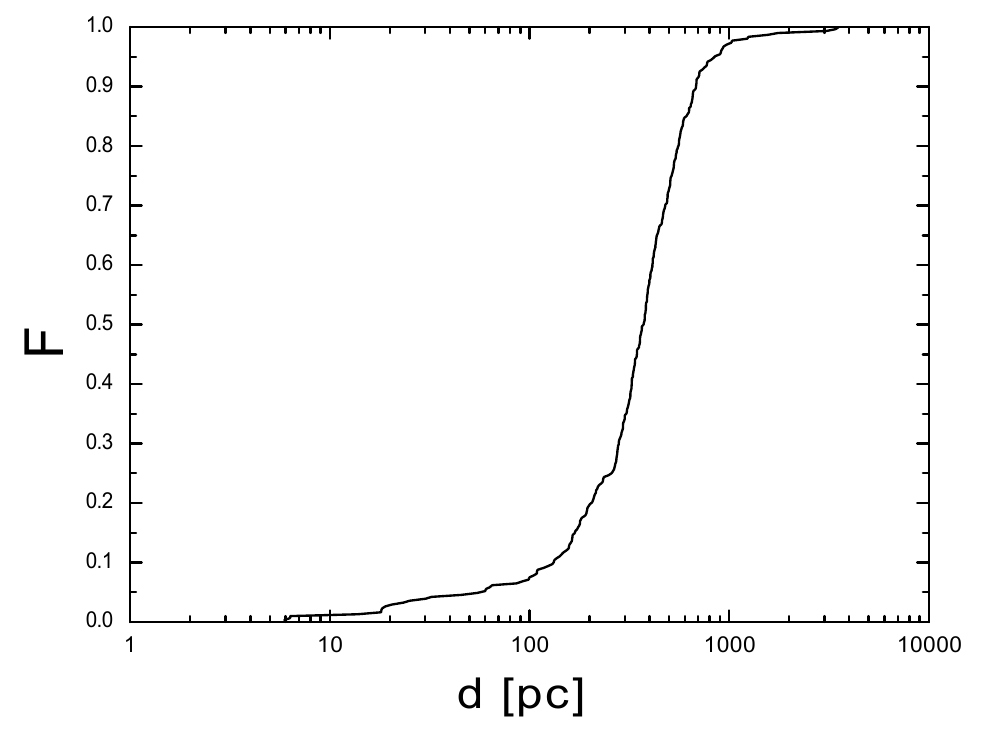}
	\caption{The cumulative distribution function of the distances of our sample according to \cite{bailerjones}.}
	\label{fig:dist}
\end{figure}

We typically used data from \textit{Gaia}\,DR2 for a detailed characterisation of all 308 observed stars. Only entries that have a gold or silver flag (see \cite{andrae2018} for details) were taken into account for the analysis. In Table \ref{tab:gaia_infos} in the appendix, we list the \textit{Gaia}\,DR2 distances $d$ from the catalogue by \cite{bailerjones}, the apparent brightness $G$ and extinction $A_{\text{G}}$ in the $G$-band, effective temperature $T_{\text{eff}}$, stellar radius $R$ and luminosity $L$ of our targets. From theses values, the resulting absolute brightness in $G$-band was calculated. The distance distribution of the whole sample is illustrated in Figure\,\ref{fig:dist} with a median distance of about 370\,pc. The individual distances range between $5.91_{-0.01}^{+0.01}$\,pc in the case of HIP\,94761 and $3536_{-813}^{+1274}$\,pc for HIP\,113561.

However, not all of these parameters are available for every target (marked with different flags) in \textit{Gaia}\,DR2. For six targets without given parallax, for the distance determination, the Hipparcos parallax \citep{vanleeuwen} was taken instead. In the cases of HIP\,18488, HIP\,27989, HIP\,110991 and HIP\,113881 the apparent brightnesses in $V$- and $I$-band from the Hipparcos catalogue \citep{perryman} were converted into $G$-band using the relations by \cite{jordi}.

142 of our targets have no estimation of the interstellar extinction in the $G$-band. For this reason, measurements of their $V$-band extinction from the literature, as listed in \texttt{VizieR} \citep{ochsenbein}, were collected instead to calculate the median and standard deviation of the $V$-band extinction for these targets. This average $V$-band extinction was then converted into $G$-band by using the relations from \cite{jordi} as explained in \cite{mugrauer2019}.

For ten objects no sufficient $T_{\text{eff}}$ was provided by \textit{Gaia}\,DR2. These ten stars were identified as giants according to their bright absolute $G$-band magnitudes. Their temperatures were derived from the spectral type (SpT) listed in the Hipparcos catalogue with the corresponding $T_{\text{eff}}$-SpT relation by \cite{damiani}.\\

\begin{figure*}[h!]
	\centering\includegraphics[width=17.5cm,height=17.5cm,keepaspectratio]{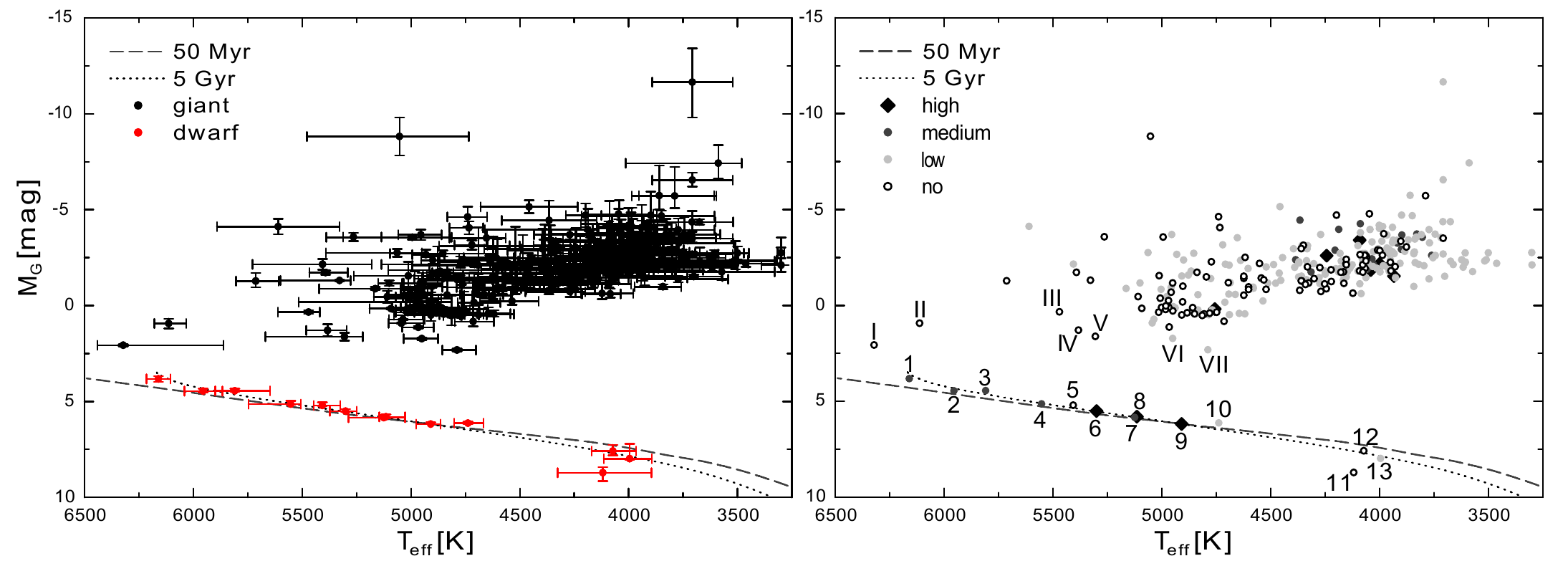}
	\caption{\text{Hertzsprung-Russell} diagram of all observed targets. On the \textbf{left}, we show giant stars marked with a full black dots and dwarf stars with full red dots. On the \textbf{right}, the targets are characterised by the measured equivalent width of the Li\,(6708\,\AA) line: Targets with $EW_{\text{Li}}< 3\cdot \sigma EW_{\text{Li}}$ have \textit{no} significant lithium detection, while objects with $EW_{\text{Li}}\geq 3\cdot\sigma EW_{\text{Li}}$ and $EW_{\text{Li}}<100$\,m\AA\,\,are ranked as \textit{low}, 100\,m\AA\,\,$\leq\,EW_{\text{Li}}<200$\,m\AA\,\,as \textit{medium} and $EW_{\text{Li}}\geq200$\,m\AA\,\,as \textit{high}. Further information about the stars marked with Roman numerals can be found in Table\,\ref{giants} and  for the ones marked numbers with from 1 to 13 in Table\,\ref{clearname}\hspace{-2mm}, respectively. We have also plotted the PARSEC isochrones \citep{bressan} for 50\,Myr and 5\,Gyr with solar metallicity $Z=0.0152$ in both distributions.}
	\label{fig:HRD}
\end{figure*}

We plot all targets in a \text{Hertzsprung-Russell-Diagram} (HRD) which is shown in Figure\,\ref{fig:HRD}\hspace{-2mm}. Most of the investigated stars are clearly on the giant branch and therefore, they can be excluded from the list of possible young runaway stars.
In contrast to this, the targets numbered from 1 to 13 in the right panel of Figure\,\ref{fig:HRD} can be seen as dwarf stars, because they are located close to the isochrones for 50\,Myr and 5\,Gyr.

The remaining objects that are marked with Roman numerals in Figure\,\ref{fig:HRD} cannot be assigned so easily, from their position in the HRD. At comparable temperatures, the radii of those targets, as listed in Table\,\ref{giants}\hspace{-2mm}, are too large compared to those of known dwarfs as given by E. Mamajek\footnote{\url{http://www.pas.rochester.edu/~emamajek/EEM_dwarf_UBVIJHK_colors_Teff.dat}} or \cite{torres2010}. Therefore, they are also giants.

\begin{table}[h!]
\caption{Targets that were identified as giants due to comparison of effective temperature $T_{\text{eff}}$ and radius $R$ from \textit{Gaia}\,DR2 with those of known dwarfs. These objects are marked with Roman numerals in Figure\,\ref{fig:HRD}\hspace{-2mm}.}
\centering
\begin{tabular}{cccc}
\hline
Roman numeral     & Target   & $T_{\text{eff}}$ [K] & $R$ [R$_{\odot}$]   \\
\hline
 I 		& HIP\,118077& $6322_{-460}^{+118}$ & $3.70_{-0.13}^{+0.61}$  \\
 II		& HIP\,96966 & $6113_{-80 }^{+67 }$ & $4.34_{-0.09}^{+0.12}$ \\
 III	& HIP\,21408 & $5470_{-52 }^{+141}$ & $7.60_{-0.38}^{+0.14}$ \\
 IV		& HIP\,64543 & $5383_{-87 }^{+99 }$ & $5.26_{-0.19}^{+0.18}$ \\
 V		& HIP\,46977 & $5305_{-84 }^{+365}$ & $4.72_{-0.59}^{+0.15}$ \\
 VI		& HIP\,63368 & $4950_{-75 }^{+83 }$ & $5.43_{-0.18}^{+0.17}$ \\
 VII	& HIP\,68904 & $4788_{-87 }^{+69 }$ & $4.57_{-0.13}^{+0.17}$ \\
\hline
\end{tabular}                                              		
\label{giants}
\end{table}	
	
However, further analysis is needed to identify young stars among the dwarfs.

\section{Li\,(6708\,\AA) Equivalent width measurements and abundances}\label{sec4}

\begin{figure*}[h!]
	\centering\includegraphics[width=14cm,height=12cm,keepaspectratio]{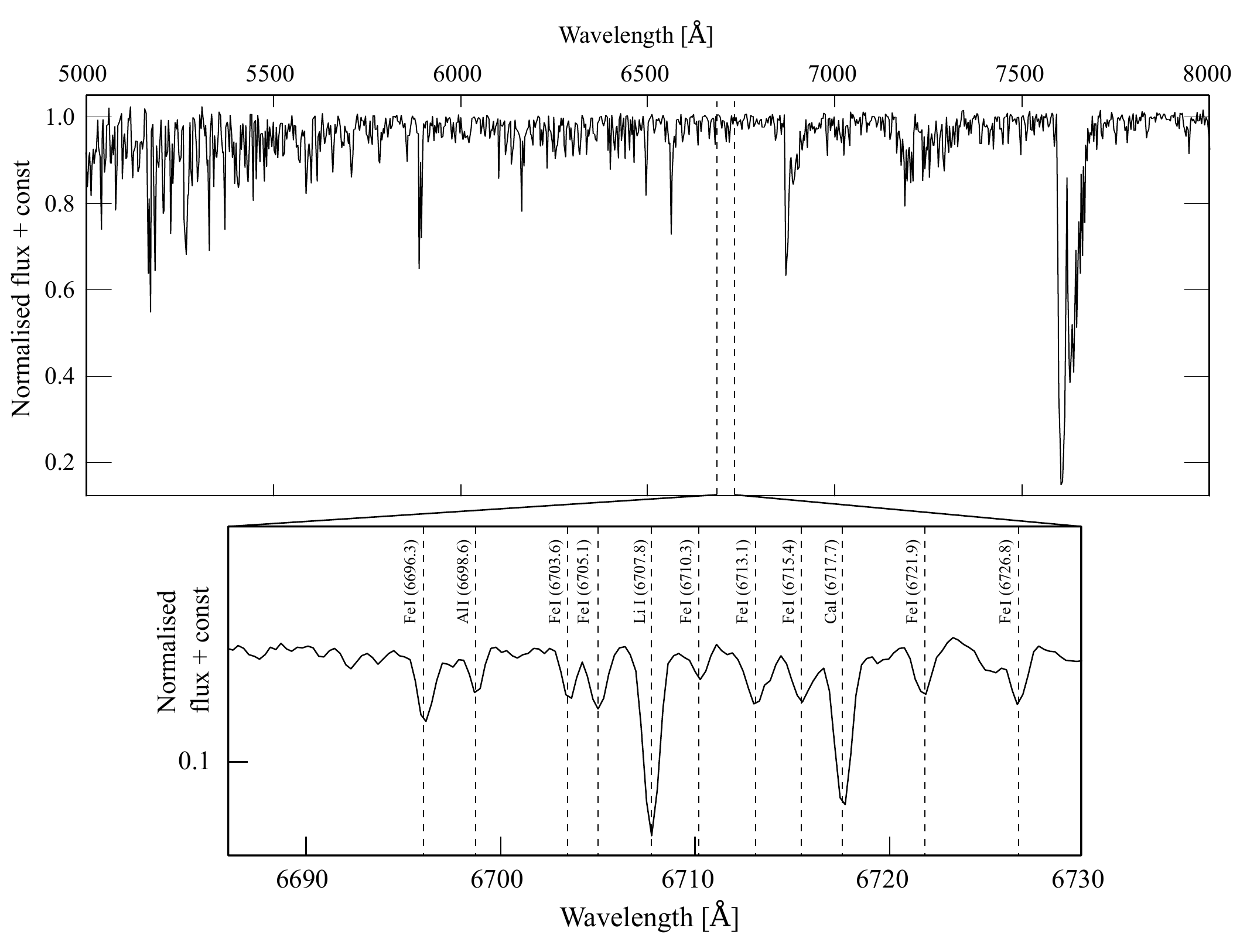}
	\caption{Normalised FLECHAS spectrum of the lithium rich dwarf HIP\,46843. Besides the prominent Ca\,(6718\,\AA) line, the Al\,(6699\,\AA) and many iron lines are identified around the Li\,(6708\,\AA) line.}
	\label{fig:Bsp}
\end{figure*}

The first step to identify the Li\,(6708\,\AA) line was to correct the wavelength shift in every spectrum due to the relative motion between the target and the observing site. Therefore, the position of the Ca\,(6718\,\AA) line was measured, by using Gaussian fitting of the IRAF script \texttt{splot}, and then was shifted to its characteristic wavelength ($\lambda_0 = 6717.685$\,\AA), as given in the ILLSS catalogue \citep{coluzzi}. The Ca\,(6718\,\AA) line was chosen because it is the most prominent spectral line next to Li\,(6708\,\AA), and is detected in the same spectral order as the lithium line. As an example, we show in Figure\,\ref{fig:Bsp} the FLECHAS spectrum of the lithium rich dwarf HIP\,46843.\\

The equivalent width
\begin{equation}
EW =  \sum\limits_{x=1}^{N} \left( -\dfrac{I(x)-C(x)}{C(x)} \right) \text{d}x
\end{equation}
was determined in the reduced spectra by using the IRAF task \texttt{splot} with a direct integration of the line profiles, where $x$ is the pixel center coordinate, $I(x)$ the pixel value, $C(x)$ the pixel value of the continuum and d$x$ is the pixel width in \AA.
\texttt{splot} also includes an error estimation based on a Poisson statistics model, where a constant Gaussian error based on the detected flux in the spectral line is calculated individually for every pixel. This takes into account the typical read-noise (11\,e- at a readout frequency of 700\,kHz) and gain (1.3\,e-/ADU) of the instrument \citep{mugrauer2014}. The final uncertainty is calculated based on Gaussian propagation of the individual pixel uncertainties.

\begin{table}[h!]
	\caption{Targets with the largest measured equivalent widths of the Li\,(6708\,\AA) line $EW_{\text{Li}}$ detected with FLECHAS and their derived spectral types (SpT) using the $T_{\text{eff}}$-SpT-correlations from \cite{damiani}.}
	\centering
	\begin{tabular}{ccc}
		\hline
		Target      & $EW_{\text{Li}}$ [m\AA]   &      SpT       \\
		\hline
		HIP\,41221 & $\,~331\pm10$	& K5\,III\\
		HIP\,43030 & $310\pm9$		& K5\,III\\
		HIP\,27778 & $266\pm7$		& K5\,III\\
		HIP\,46816 & $\,~254\pm11$	& K2\,V\\
		HIP\,80941 & $\,~242\pm12$	& K0\,III\\
		HIP\,115147& $\,~239\pm13$	& K0\,V\\
		HIP\,40628 & $235\pm9$		& K5\,III\\
		HIP\,102377& $\,~235\pm10$	& K4\,III\\
		HIP\,46843 & $\,~200\pm12$	& G8\,V\\
		\hline
	\end{tabular}                                              		
	\label{topnine}
\end{table}

The targets with the largest measured equivalent widths of the Li\,(6708\,\AA) lines are listed in Table\,\ref{topnine}\hspace{-2mm}. The equivalent widths of the Li\,(6708\,\AA) line of the complete sample can be found in Table\,\ref{tab:Li_all}\hspace{-2mm}. We also present in \textit{\nameref{app2}} and \textit{\nameref{app3}}, all spectra of stars which exhibit a significant measurement, that means $EW_{\text{Li}}\geq 3\cdot \sigma EW_{\text{Li}}$. These spectra are sorted by their spectral type derived from the $T_{\text{eff}}$-SpT-relation given by \cite{damiani}. Thereby we distinguish between dwarf and giant stars and the uncertainty of the spectral classification is about two sub-classes.

Furthermore, we compare our equivalent widths from FLECHAS spectra to measurements from TRES. In these additional spectra, we searched for lithium as described above and the results are summarised in Table \ref{tres}\hspace{-2mm}. Due to a mostly lower SNR in the TRES spectra, we list only targets where in every individual measurement the equivalent width is at least three times larger than its uncertainty. The median of each individual measurement per star was then calculated and the standard deviation represents its accuracy. The values are consistent within 1 to 3\,$\sigma$.

\begin{table}[h!]
	\caption{Comparison of the equivalent widths of the Li\,(6708\,\AA) line, measured in the TRES and FLECHAS spectra. We list the median equivalent width $EW_{\text{med}}$ and the number of used TRES measurements $N$ as well as the derived values from the FLECHAS spectra $EW_{\text{FL}}$.}
	\centering
	\begin{tabular}{cccc}
		\hline
		Target   & $EW_{\text{med}}$ [m\AA]   & $N$   &      $EW_{\text{FL}}$ [m\AA]       \\
		\hline
		HIP\,1479  & $~~21\pm7 $   & 3		& $~~46\pm6 $\\
		HIP\,5912  & $126\pm4 $  & 4		& $137\pm6 $\\
		HIP\,16019 & $~~57\pm7 $   & 4		& $~~69\pm7 $\\
		HIP\,17064 & $~~93\pm2 $   & 3		& $~~94\pm7 $\\
		HIP\,25386 & $~~29\pm8 $   & 6		& $~~~~39\pm10 $\\
		HIP\,26743 & $~~17\pm3 $   & 4		& $~~~~61\pm12 $\\
		HIP\,41221 & $~~357\pm12 $  & 3		& $~~331\pm10 $\\
		HIP\,44580 & $~~~~33\pm11 $   & 3		& $~~67\pm9 $\\
		HIP\,50999 & $~~30\pm3 $   & 3		& $~~55\pm9 $\\
		HIP\,74425 & $~~19\pm2$		& 4		& $~~34\pm7$ \\
		HIP\,77178  & $~~22\pm4$	& 5 	& $~~30\pm9$ \\
		HIP\,84038 & $~~12\pm6 $   & 35		& $~~19\pm9 $\\
		HIP\,100180 & $~~26\pm8 $  & 8		& $~~43\pm7 $\\
		HIP\,100534 & $~~31\pm6 $  & 9		& $~~35\pm7 $\\
		HIP\,101219 & $~~~~80\pm11 $ & 21	& $108\pm8 $\\
		HIP\,107325 & $~~38\pm6 $  & 4		& $~~39\pm7 $\\
		\hline
	\end{tabular}                                              		
	\label{tres}
\end{table}

Additionally, the equivalent widths of the 13 dwarf stars were converted into abundances. We used the curves of growth for the Li\,(6708\,\AA) line from \cite{soderblom}, with an abundance scale based on $\log_{10}(N)_{\text{H}}=12$. Therefore, the measured equivalent widths from our observing campaign together with the effective temperatures from \textit{Gaia}\,DR2 were assigned to the corresponding best matching values of $\log_{10}(EW)$ and $T_{\text{eff}}$ in the corresponding Table\,2 in \cite{soderblom}. The results of this conversion are listed in Table\,\ref{abundance}\hspace{-2mm}.

\begin{table}[h!]
\caption{The identified dwarf stars with their numbers as shown in Figure\,\ref{fig:HRD}\hspace{-2mm}, listed with their effective temperatures $T_{\text{eff}}$ from \textit{Gaia}\,DR2 and the measured equivalent widths of the Li\,(6708\,\AA) line $EW_{\text{Li}}$. We list only measurements with $EW_{\text{Li}}\geq 3\cdot \sigma EW_{\text{Li}}$. Based on these values we give the derived abundances $\log_{10}(N_{\text{Li}})$ according to \cite{soderblom}.}
\centering
\resizebox{0.5\textwidth}{!}{
\begin{tabular}{clccc}
\hline
		id. nr. & Target & $T_{\text{eff}}$ [K] & $EW_{\text{Li}}$ [m\AA] & $\log(N_{\text{Li}})$   \\
\hline
1  & HIP\,60831		& $6160_{-55}^{+55}$     & $116\pm11$  & $3.170_{-0.294}^{+0.090}$ \\
2  & HIP\,107350 	& $5955_{-89}^{+87}$     & $138\pm25$  & $3.140_{-0.414}^{+0.106}$ \\
3  & HIP\,44458 	& $5809_{-161}^{+91}$    & $178\pm13$  & $3.130_{-0.125}^{+0.387}$ \\
4  & HIP\,544 		& $5552_{-46}^{+195}$    & $100\pm11$  & $2.393_{-0.070}^{+0.333}$ \\
5  & HIP\,59280     & $5407_{-82}^{+40}$     & -    & - \\
6  & HIP\,46843 	& $5300_{-51}^{+73}$     & $200\pm12$  & $2.715_{-0.135}^{+0.158}$ \\
7  & HIP\,63742 	& $5124_{-97}^{+163}$    & $140\pm16$  & $2.071_{-0.091}^{+0.401}$ \\
8  & HIP\,115147 	& $5115_{-88}^{+30}$     & $239\pm13$  & $2.745_{-0.186}^{+0.320}$ \\
9  & HIP\,46816 	& $4909_{-45}^{+67}$     & $254\pm11$  & $2.745_{-0.341}^{+0.320}$ \\
10  & HIP\,45963 	& $4738_{-71}^{+64}$     & $29\pm8$    & $0.853_{-0.159}^{+0.103}$ \\
11  & HIP\,94761    & $4118_{-224}^{+208}$   & - & - \\
12  & HIP\,45343    & $4072_{-106}^{+98}$    & -    &-  \\
13 & HIP\,120005 	& $3995_{-102}^{+119}$   & $~~59\pm12$ & $0.235_{-0.112}^{+0.125}$ \\
\hline
\end{tabular}
}                                  		
\label{abundance}
\end{table}

\section{Age estimation}\label{sec5}

\begin{figure}[h!]
	\centering\includegraphics[width=8.3cm,height=8.25cm,keepaspectratio]{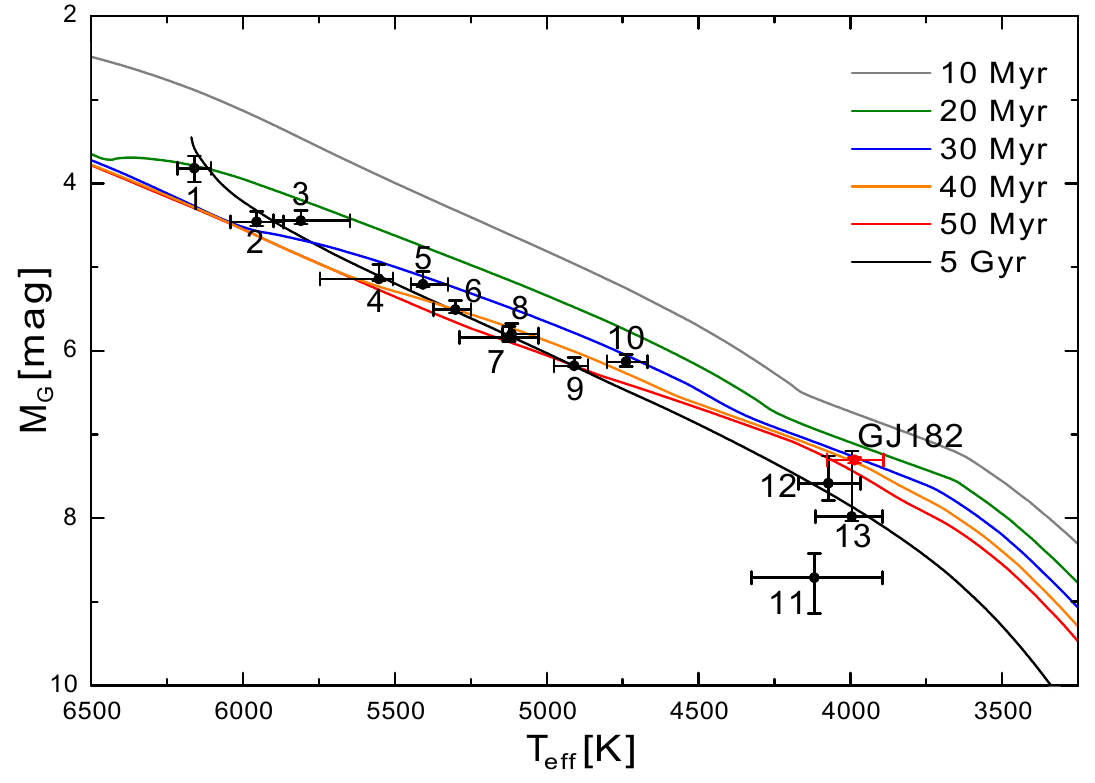}
	\caption{The dwarf stars of our sample and GJ\,182 (in red) are plotted in the \text{Hertzsprung-Russell} diagram with different isochrones using the stellar evolutionary models of \cite{bressan} for solar metallicity $Z=0.0152$.}
	\label{fig:zwerge}
\end{figure}

\begin{figure}[h!]
	\centering\includegraphics[width=8.3cm,height=8.25cm,keepaspectratio]{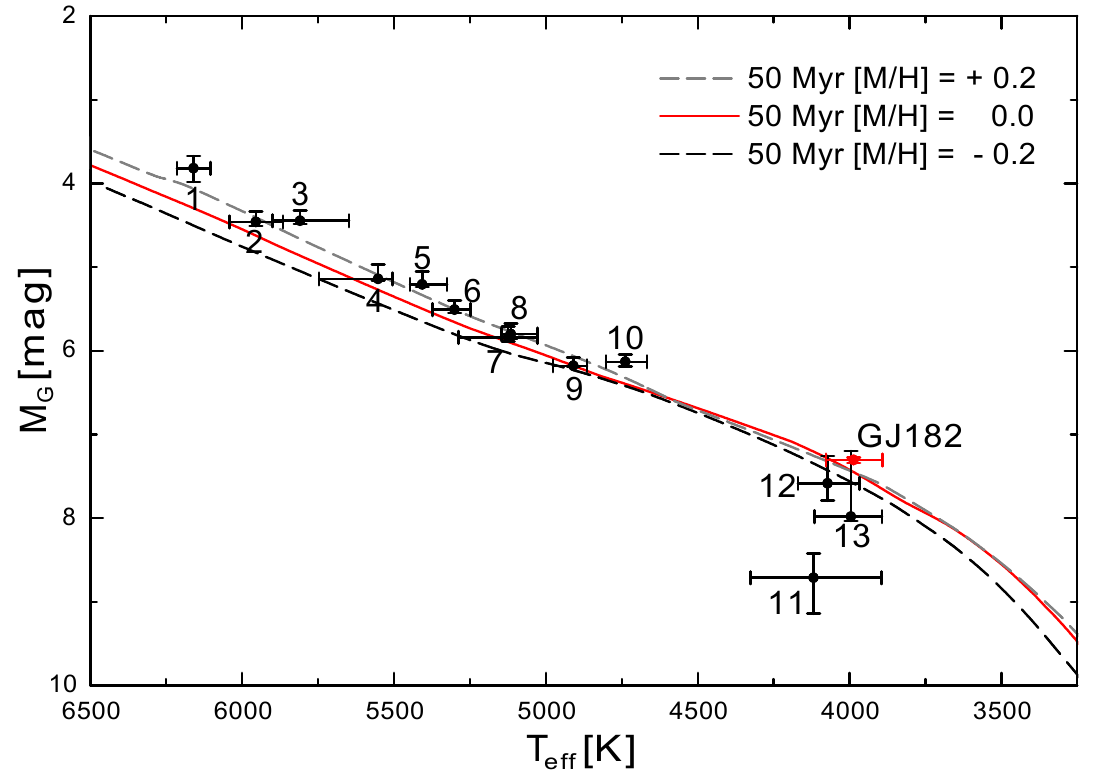}
	\caption{The dwarf stars of our sample and GJ\,182 (in red) are plotted in the \text{Hertzsprung-Russell} diagram with 50\,Myr isochrones for different metallicities $[\text{M}/\text{H}]$ using the stellar evolutionary models of \cite{bressan}.}
	\label{fig:metal}
\end{figure}

The ages of our dwarf stars can be estimated from their location in the HRD  with additional isochrones as illustrated in in Figure\,\ref{fig:zwerge}\hspace{-2mm}. For this we have used again the models of \cite{bressan} with metallicity $Z=0.0152$, because all of these targets have metallicities listed in the \textit{StarHorse} catalogue \citep{anders} which are consistent within 2\,$\sigma$ with solar metallicity. For young targets we would not expect a metallicity $[\text{M}/\text{H}]<0$. Furthermore, we searched for additional information about the metallicity of our dwarf targets in the \texttt{VizieR} database \citep{ochsenbein}. The metallicites for this sample were found in the catalogues of \cite{brewer}, \cite{cassagrande2011}, \cite{franchini}, \cite{gazzano}, \cite{gray2001}, \cite{gray2003}, \cite{gray2006}, \cite{houdebin2016}, \cite{houdebin2017}, \cite{karatas}, \cite{kunder}, \cite{marsden}, \cite{petigura}, \cite{raj}, \cite{rojas}, \cite{stassun}, \cite{valenti} and \cite{worley}, with a median of $[\text{M}/\text{H}]=0.0$ and standard deviation of 0.2\,dex. Taking this scatter of $[\text{M}/\text{H}]$ into account, we analysed how it influences the position of the isochrones in the HRD. The differences between different metallicites are smaller or at least comparable with the uncertainties of effective temperature or absolute brightness and do not effect the age estimation significantly, as illustrated in the example in Figure\,\ref{fig:metal}\hspace{-2mm}.

However, nearly all dwarf stars of the sample are also consistent with the main sequence within their uncertainties, as shown in Figure\,\ref{fig:zwerge}\hspace{-2mm}, and therefore, isochrone fitting can only be used to estimate a lower limit of their age. The results of this are listed in  Table\,\ref{clearname}\hspace{-2mm}.\\

\begin{figure}[h!]
	\centering\includegraphics[width=8.75cm,height=8.75cm,keepaspectratio]{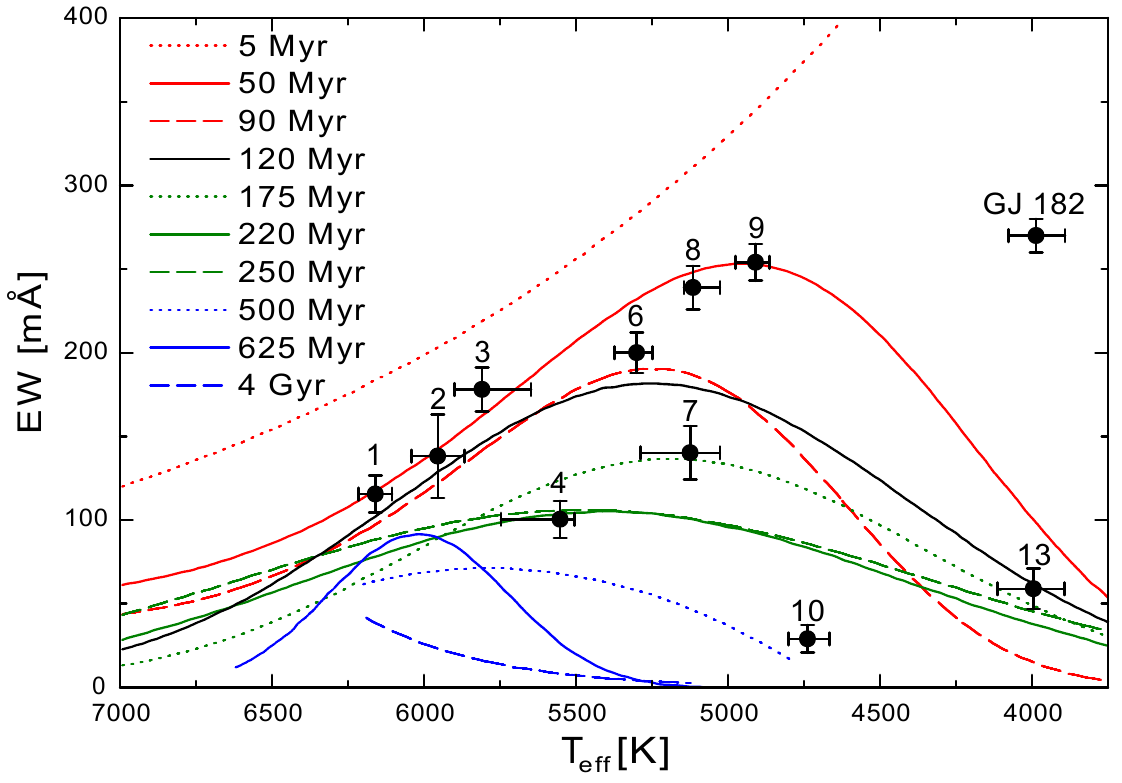}
	\caption{Lithium as age indicator for our ten youngest F\,G\,K\,M stars and GJ\,182. We show the equivalent widths of the Li\,(6708\,\AA) line over the effective temperature for different age curves, that are based on fits done by Eric Mamajek. The derived ages are given in Table\,\ref{clearname}\hspace{-2mm}.}
	\label{fig:Li}
\end{figure}

Spectroscopic investigations are necessary to get more precise age estimates. Therefore, the equivalent widths and $T_{\text{eff}}$ of all identified dwarfs with a significantly detected Li\,(6708\,\AA) line in their spectra were compared to equivalent width distributions of clusters with known ages as seen in Figure\,\ref{fig:Li} (cluster data from E. Mamajek, priv. communication). The curves are polynomial fits to observed $T_{\text{eff}}$ vs. $\text{log}_{10}(EW_{\text{Li}})$ data of each cluster, for example $\alpha$\,Per (90\,Myr) and Hyades (625\,Myr). E. Mamajek`s plot can be found online\footnote{\url{http://www.pas.rochester.edu/~emamajek/images/li.jpg}} and is based on \cite{neuhaeuser}.

In Figure\,\ref{fig:Li}\hspace{-2mm}, HIP\,44458 (\#\,3 in Table\,\ref{clearname}\hspace{-2mm}), HIP\,115147 (\#\,8) and HIP\,46816 (\#\,9) are only consistent with the 50\,Myr curve and therefore their age is estimated to $\sim$\,50\,Myr. For the same reason we specify HIP\,46843 (\#\,6) as $\sim$\,90\,Myr old and HIP\,63742 (\#\,7) as $\sim$\,175\,Myr, respectively. As stated by \cite{soderblom2014} the age errors for this method appear to be $10\%-20\%$ and the detection of lithium in a low-mass star with known effective temperature gives an upper limit to its age.
HIP\,60831 (\#\,1) and HIP\,107350 (\#\,2) cross within their uncertainties, more than one curve and because of that their ages are specified by the range from 50\,Myr to 120\,Myr. In an analogous way, HIP\,544 (\#\,4) is about $175\,...\,250$\,Myr old and HIP\,120005 (\#\,13) about $120\,...\,250$\,Myr. In the case of HIP\,45963 (\#\,10) within its uncertainties it crosses no curve and therefore the next nearby two were set as limits of its age estimation, namely 220\,Myr and 500\,Myr. \\

\begin{table*}[h!]
\caption{Our dwarf stars with their numbers as shown in Figure\,\ref{fig:HRD} and Figure\,\ref{fig:Li} listed with their SpT derived from their $T_{\text{eff}}$ using the SpT-$T_{\text{eff}}$-relation from \cite{damiani}, their distances $d$ according to \cite{bailerjones}, the measured equivalent width of the Hydrogen Balmer line $EW_{\text{H}\alpha}$ (all in absorption) and the measured equivalent width of Li\,(6708\,\AA) $EW_{\text{Li}}$.
Furthermore, we list the estimated age derived from the position in the HRD \textbf{(lower limit)}, as well as the age according to the lithium test \textbf{(upper limit)} from this work and the age from \cite{tetzlaff}.}
\centering
\resizebox{1.0\textwidth}{!}{
\begin{tabular}{clccccccc}
\hline
id. nr. & Target & SpT  & $d$ [pc] &   $EW_{\text{H}\alpha}$ [m\AA]  & $EW_{\text{Li}}$ [m\AA] & age (HRD) [Myr]   & age ($EW_{\text{Li}}$) [Myr] & age [Myr]  \\
	 		&	  	 &  $\ast$    & Bailer-Jones    & $\ast$  &   $\ast$         &     $\ast$   & $\ast$   & Tetzlaff \\
\hline
1  & HIP\,60831		& F6\,V     & $45.5\pm0.1$  & $1898\pm23$ & $116\pm11$ & $>10$   &$50\,...\,120$     & $20.0\pm6.7$  \\
2  & HIP\,107350 	& F8\,V     & $18.1\pm0.1$  & $907\pm31$  & $138\pm25$ & $>20$   &$50\,...\,120$     & $29.1\pm7.9$  \\
3  & HIP\,44458 	& G0\,V     & $30.2\pm0.1$  & $1428\pm27$ &$178\pm13$  & $>10$   &$\sim50$           & $27.0\pm4.7$  \\
4  & HIP\,544 		& G5\,V     & $13.8\pm0.1$  & $1339\pm16$ & $100\pm11$ & $>20$   &$175\,...\,250$    & $20.1\pm6.4$  \\
5  & HIP\,59280     & G8\,V     & $25.1\pm0.1$  & $1442\pm17$ & -          & $>20$   & -                 & $20.1\pm6.4$    \\
6  & HIP\,46843 	& G8\,V     & $18.1\pm0.1$  & $1135\pm20$ & $200\pm12$ & $>30$   &$\sim90$           & $~~51.9\pm23.1$\\
7  & HIP\,63742 	& K0\,V     & $20.5\pm0.4$  & $879\pm24$  & $140\pm16$ & $>30$   &$\sim175$          & $40.9\pm9.9$  \\
8  & HIP\,115147 	& K0\,V     & $19.0\pm0.1$  & $261\pm13$  & $239\pm13$ & $>30$   &$\sim50$           & $~~30.3\pm14.5$\\
9  & HIP\,46816 	& K2\,V     &  $18.3\pm0.1$ & $75\pm11$   & $254\pm11$ & $>30$   &$\sim50$           & $~~51.9\pm17.5$\\
10  & HIP\,45963 	& K2\,V     & $23.4\pm0.1$  & $159\pm11$; $141\pm11^{\dagger}$ & $29\pm8$    &$>20$     & $220\,...\,500$  & $15.0\pm4.8$   \\
11  & HIP\,94761    & K5\,V     &$~~5.91\pm0.01$& $571\pm25$  & -           & $>50$   &-               &$~~~~~~~\leq0.1$   \\
12  & HIP\,45343    & K7\,V     &$~~6.33\pm0.01$& $549\pm10$  & -           & $>20$   &-               &$28.2\pm3.0$    \\
13 & HIP\,120005 	& K7\,V     &$~~6.33\pm0.01$& $555\pm11$  & $~~59\pm12$ & $>10$   &$120\,...\,250$ & $~~51.1\pm32.0$\\
\hline
\end{tabular}
    }
\begin{flushleft}
	$^{\ast}$ this work\\
	$^{\dagger}$ both H$\alpha$-lines of this spectroscopic binary could be measured
\end{flushleft}                              		
	\label{clearname}
\end{table*}

\section{Discussion}\label{sec6}

\begin{figure}[h!]
	\centering\includegraphics[width=8.3cm,height=8.25cm,keepaspectratio]{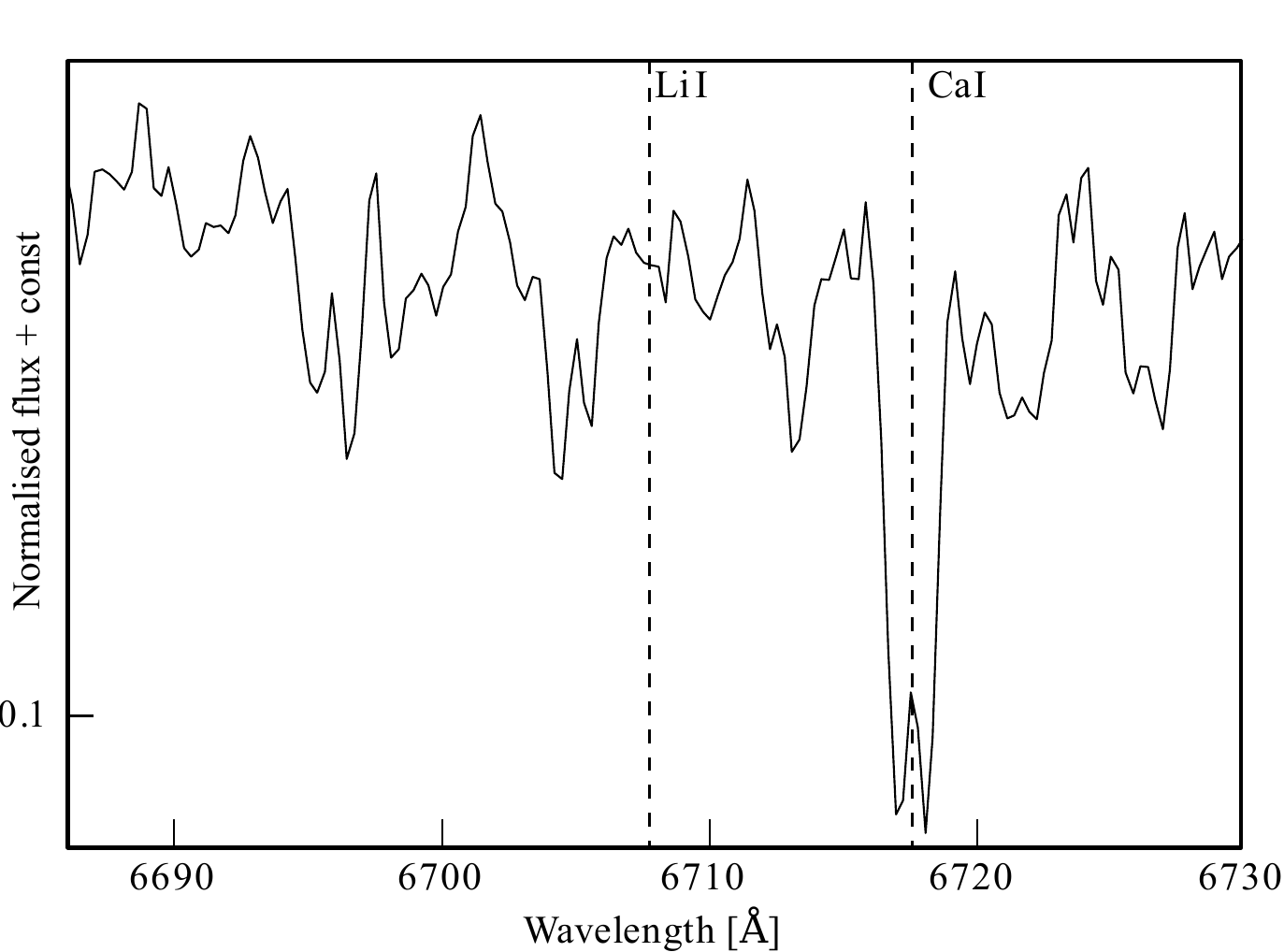}
	\caption{The Spectrum of the double-lined binary system HIP\,118077 was taken at the Barycentric Julian Date $\text{BJD}=2458164.21992$. The barycentric corrected radial velocities, that were derived from the Ca\,(6718\,\AA) line, are $v_{1}=(-12.6\pm5.2)$\,km/s and $v_{2}=(34.2\pm4.5)$\,km/s, respectively.}
	\label{fig:HIP118077}
\end{figure}

We searched for young runaway star candidates within a sample of 308 selected stars from the catalogue \textit{Young runaway stars within 3\,kpc} by \cite{tetzlaff} and estimated the age of the dwarf stars by measuring the equivalent width of the Li\,(6708\,\AA) line in their spectra as well as by comparing their location in the HRD with isochrones of stellar evolutionary models. This comparison relies on data from the \textit{Gaia} DR2 catalogue. In total, the sample consists of 4 F-type, 43 G-type, 208 K-type and 53 M-type stars, as derived from their effective temperatures according to the relations by \cite{damiani}.\\
According to their position in the HRD, 295 targets turned out to be giant stars. If objects were located close to the edge of the giant branch, their effective temperatures and radii were compared to those of known dwarfs for an explicit assignment. For example the classification of HIP\,118077 as an evolved star is consistent with the work of \cite{fekel} where both components of this binary were classified inter alia as F5 subgiant stars. The separated spectral lines of both components of this binary system are also detected in our FLECHAS spectrum in Figure\,\,\ref{fig:HIP118077} and the derived radial velocities agree well with its orbital solution from \cite{imbert}.
	
The identification of the 13 dwarfs as such is consistent with the results of \cite{anders} given in the \textit{StarHorse} catalogue, where all these objects are listed with a surface gravity $\log(g[\text{cm/s}^{2}]) > 3.8$. In reverse, the remaining targets from our sample that are actually included in \textit{StarHorse} catalogue are confirmed as giants due to their $\log(g[\text{cm/s}^{2}]) < 3.8$.

In order to identify possible young runaway star candidates in our sample, we searched for objects that are younger than 50\,Myr via isochrone fitting. We compared their location in the HRD to isochrones based on models of \cite{bressan} assuming solar metallicity. This assumption is justified, because all 13 dwarfs are nearby stars with distances $d<50$\,pc and their metallicities in the \texttt{VizieR} databases scatter around solar one. Possible metallicity effects on the isochrones due to the scatter of $\sigma[\text{M}/\text{H}]=\pm\,0.2$ have no significant impact, as explained in section\,\ref{sec5}\hspace{-2mm}.\hspace{2mm}Due to their location in the HRD, illustrated in Figure\,\ref{fig:HRD} and Figure\,\ref{fig:zwerge}\hspace{-2mm}, they are all clearly older than 10\,Myr.

Hereby, we also consider the multiplicity of  HIP\,45963 (\#\,10) as spectroscopic binary consisting of two stars with equal mass, as described by \cite{halbwachs}. The individual components are about 0.75\,mag fainter than given by \textit{Gaia}\,DR2, because of its equal mass binary nature. Its spectrum was taken at the $\text{BJD}=2458175.48525$ and the barycentric corrected radial velocities, that were again derived from the Ca\,(6718\,\AA) line, are $v_{1}=(-60.1\pm1.2)$\,km/s and $v_{2}=(51.7\pm1.5)$\,km/s, respectively. These velocities agree well with the orbital solution from \cite{halbwachs}.

However, isochrone fitting can only give a lower limit on the age, because all dwarfs in our sample are consistent with the main sequence within their uncertainties of $2\,\sigma$. \\

Another main part of this article was to search for lithium in the spectra of our targets in order to have an additional age indicator besides isochrone fitting. Therefore, we measured the equivalent width of the Li\,(6708\,\AA) lines. The strongest Li\,(6708\,\AA) line was measured in the spectrum of HIP\,41221 with an equivalent width of $(331\pm10)$\,m\AA.\,\,HIP\,27778, HIP\,40628, HIP\,43030, HIP\,80941 and HIP\,102377 are further examples of lithium rich giants that possess an $EW_{\text{Li}}\geq200$\,m\AA. In contrast to this, HIP\,21408, HIP\,46977, HIP\,64543, and HIP\,96966 are examples for giants without any significant detection of the Li\,(6708\,\AA) line.

The measured equivalent widths of the lithium lines in the TRES spectra are in good agreement with the data from FLECHAS and are all consistent within $3\,\sigma$ as seen in Table \ref{tres}\hspace{-2mm}. The additional TRES spectra of the remaining 14 stars were not taken into account because of too low SNR around the Li\,(6708\,\AA) line.\\

HIP\,59280 (\#\,5), HIP\,94761 (\#\,11) and HIP\,45343 (\#\,12) are dwarfs without significant detection of the Li\,(6708\,\AA) line. The other dwarf stars showed significant lithium at 6708\,\AA\,\,in their spectra. Their equivalent widths were converted into abundances by using the curves of growth from \cite{soderblom}. \cite{ramirez} also present lithium abundances for HIP\,544 and HIP\,107350 in their catalogue, namely $\log(N_{\text{Li}})=2.38\pm0.04$ and $\log(N_{\text{Li}})=2.93\pm0.11$, respectively. These abundances agree well with the corresponding values in Table\,\ref{abundance} from this study.

The equivalent width measurements of the Li\,(6708\,\AA) lines and the corresponding effective temperatures of the dwarf stars were then compared to those of clusters with known age as seen in Figure\,\ref{fig:Li}\hspace{-2mm}. These lithium based ages for HIP\,44458 (\#\,3), HIP\,46816 (\#\,9), HIP\,60831 (\#\,1), HIP\,115147 (\#\,8) and HIP\,107350 (\#\,2) in Figure\,\ref{fig:Li} are in good agreement with the derived ages from their location in the HRD in Figure\,\ref{fig:zwerge} as well as with the values from the Tetzlaff catalogue. These five stars can be classified as young according to \cite{tetzlaff} and also from the methods presented in this paper. HIP\,46843 (\#\,6) is located between the 50- and 90\,Myr-curve in Figure\,\ref{fig:Li} and the classification of its age with $\sim90$\,Myr is also consistent with $(51.9\pm23.1)$\,Myr from \cite{tetzlaff}.

In the cases of HIP\,544 (\#\,4) and HIP\,63742 (\#\,7) the estimated values from the lithium method and the HRD, suggest that these stars are significantly older than determined by \cite{tetzlaff}. Both targets were ranked as gold flag by \textit{Gaia} photometry and the differences between the parallaxes of \textit{Hipparcos} and \textit{Gaia} are negligible. Our lithium derived age for HIP\,544 agrees well with the range of $(200\pm100)$\,Myr from \cite{ramirez}.

HIP\,120005 (\#\,13) was considered as 120 to 250\,Myr old from the estimation of the Li\,(6708\,\AA)  equivalent width. However, for cooler temperatures as seen in Figure\,\ref{fig:Li}\hspace{-2mm}, five different curves are close together and therefore HIP\,120005 could, within its error bars, still be in agreement with the derived age of $(51.1\pm32.0)$\,Myr by \cite{tetzlaff}. This is also supported by its position in Figure\,\ref{fig:zwerge} that suggests an age $>10$\,Myr.

As mentioned above, HIP\,45963 (\#\,10) is a spectroscopic binary consisting of two stars with equal mass. This can be confirmed by our equivalent width measurements of the H$\alpha$-line in Table\,\ref{clearname}\hspace{-2mm}. The spectral lines of both components are visible and comparable to each other. This can be also found in its spectrum for the Ca\,(6718\AA) line as illustrated in the appendix. However, the Li\,(6708\AA) line shows no splitting. In that case the derived measurement of $(29\pm8)$\,m\AA\,\,is more likely an upper limit due its equal mass components. Its age range of 220\,Myr to 500\,Myr should be taken with care, because the lithium method does not give very reliable age estimations below 20\,Myr and above 200\,Myr \citep{soderblom2014}. However, our derived ages for HIP\,45963 disagree with the estimate of $(15.0\pm4.8)$\,Myr, from \cite{tetzlaff}.

Another example where we found a significant difference compared to the age from \cite{tetzlaff} is HIP\,94761 (\#\,11). From its position in Figure\,\ref{fig:zwerge} it is expected to be at least older than 50\,Myr, which significantly differs from the age estimate of about 0.1\,Myr by \cite{tetzlaff}. An age of 5\,Gyr for HIP\,94761 was also given by \cite{passeger}.

Further dwarfs without showing lithium in their spectra are HIP\,45343 (\#\,12) and HIP\,59280 (\#\,5). These two objects were estimated to be older than 20\,Myr due to their position in the HRD in Figure\,\ref{fig:zwerge} and are compatible with the ages from \cite{tetzlaff}.

Furthermore, we compare our targets with the young runaway star GJ\,182 (HIP\,23200) as seen in Figure\,\ref{fig:zwerge} and Figure\,\ref{fig:Li}\hspace{-2mm}. For this star, we also used the \textit{Gaia} DR2 data as described above, and adopted the equivalent width of $(270\pm10)$\,m\AA\,\,for the Li\,(6708\,\AA) line from \cite{torres2006}. This object is slightly younger than our targets, as shown in Table\,\ref{gj182}\hspace{-2mm}, and given its distance of $(24.38\pm0.02)$\,pc, there is no other star known, that is younger and nearer than GJ\,182. Our age estimations for GJ\,182 are in good agreement with common values as seen in Table \ref{gj182}\hspace{-2mm}.

\begin{figure}[h!]
	\centering\includegraphics[width=8.3cm,height=8.25cm,keepaspectratio]{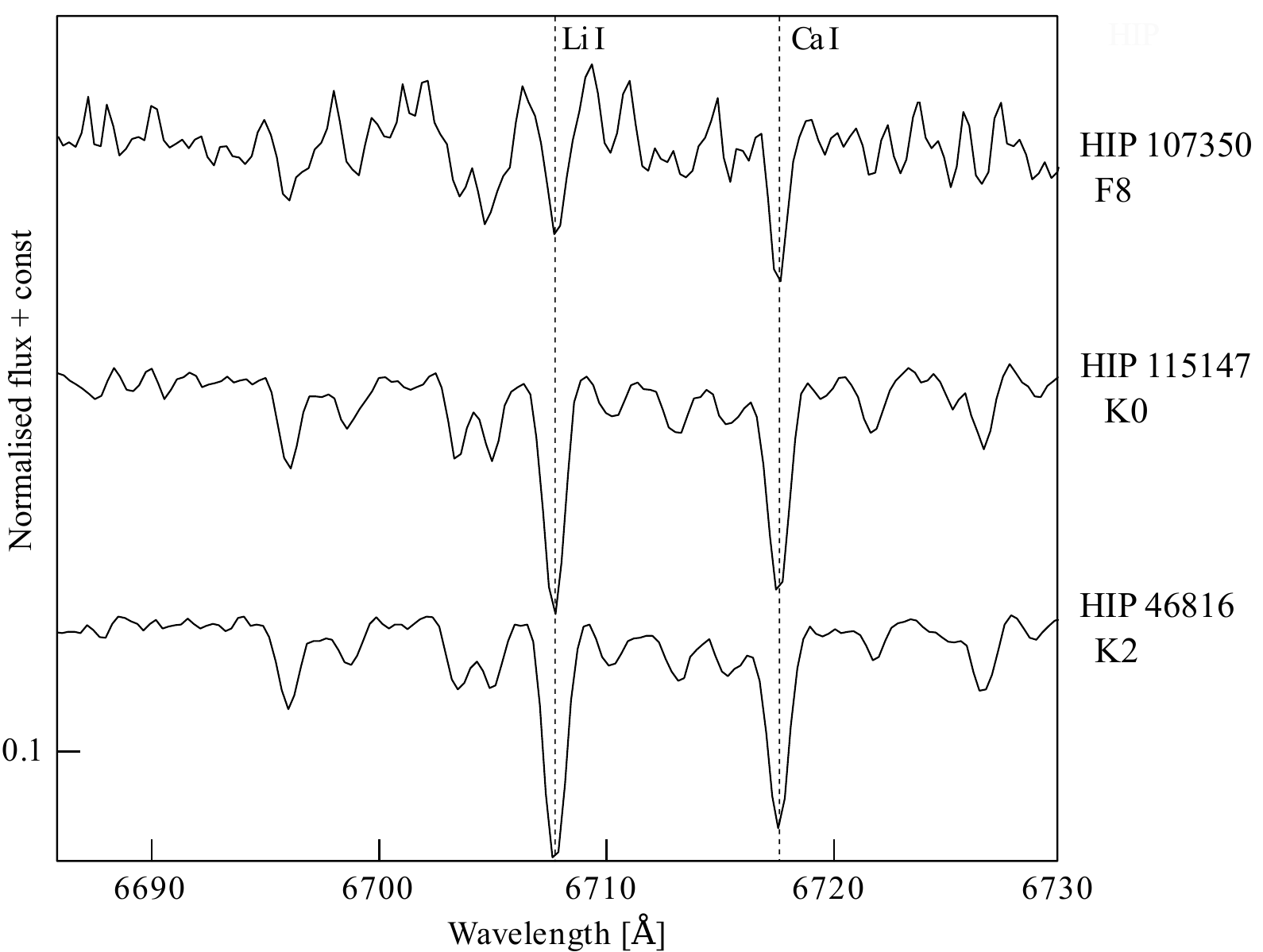}
	\caption{Spectra of the three young and very nearby runaway stars.}
	\label{fig:Li3}
\end{figure}

\begin{table}[h!]
	\caption{Ages and used determination methods for GJ\,182 as well as their references.}
	\centering
	\begin{tabular}{cll}
		\hline
		age [Myr]   & method   & ref.       \\
		\hline
		$~~8\,...\,20$	&  HRD isochrone fitting			& Z\,01	\\	
		$~~0.1\,...\,14.3$	&  HRD isochrone fitting		   	& T\,11 \\
		$17\,...\,26$		&  kinematic &\\
		&\& activity/rotation indicators   	& B\,14\\
		$17\,...\,25$		&  lithium test \\
	                     	& for moving group members	 					& BJ\,14\\
		$21\,...\,27$		&  HRD isochrone fitting	   		& B\,15\\
		$20\,...\,50$		&  HRD isochrone fitting			& \\
			                &  \& lithium test			   		& this work\\
		\hline
	\end{tabular}                                              		
	\label{gj182}
	\begin{flushleft}
		References: Z\,01 \citep{zuckerman}; T\,11 \citep{tetzlaff}; B\,14 \citep{brandt}; BJ\,14 \citep{binks}; B\,15 \citep{bell}
	\end{flushleft}
\end{table}

This study provides a large database of spectra which will be made available in the \texttt{VizieR} database together with the measured equivalent widths of the Li\,(6708\,\AA) line, after publication.	 Furthermore, we identified HIP\,107350 (\#\,2), HIP\,115147 (\#\,8) and HIP\,46816 (\#\,9) as runaway stars with comparable ages to GJ\,182, while they are even closer to Earth. That would mean that these three objects are now the nearest known young runaway stars so far. Their spectra are shown in Figure\,\ref{fig:Li3}\hspace{-2mm}.

HIP\,60831 (\#\,1) and HIP\,44458 (\#\,3) are further examples of classified young runaway stars in this study. These five young targets are suitable for follow-up investigation of their origin from either dynamical or supernova ejection.
Further five objects were identified to range between 50\,Myr and 500\,Myr, while three other are main sequence stars.
The remaining 295 targets are giants according to their position in the HRD as well as their radii, effective temperatures and $\log(g)$.

\section*{Acknowledgments}

 We thank all observers who have been involved in some of the observations of this project, obtained at the University Observatory Jena, in particular H. Gilbert, A. Pannicke, D. Wagner, J. Greif, S. Masda and F. Schiefeneder. We also thank P. Berlind, M. Calkins, and G. Esquerdo for their expert assistance in obtaining the TRES observations, and J. Mink for maintaining the Center for Astrophysics echelle database. \\
This publication makes use of data products of the \texttt{VizieR} databases, operated at CDS, Strasbourg, France. We also thank the \textit{Gaia} Data Processing and Analysis Consortium of the European Space Agency (ESA) for processing and providing the data of the \textit{Gaia} mission.\\
We thank Eric Mamajek for providing the curves of lithium as age indicator for F\,G\,K\,M stars.\\
This work was supported by the Deutsche Forschungsgemeinschaft with financing the projects \fundingNumber{NE 515/58-1} and \fundingNumber{MU 2695/27-1}.

\appendix

\section{Additional target information\label{app1}}

\begin{center}
\vspace{0.8cm}
\tablefirsthead{%
\hline
Target          & \multicolumn{1}{c}{Date [UT]} 	& \multicolumn{1}{c}{T$_{\text{exp}}$ [s]}  & \multicolumn{1}{c}{SNR} \\
\hline}
\tablehead{%
\multicolumn{1}{l}{Continued}\\
\hline
Target          & \multicolumn{1}{c}{Date [UT]} 	& \multicolumn{1}{c}{T$_{\text{exp}}$ [s]}  & \multicolumn{1}{c}{SNR} \\
				\hline}
\tabletail{%
\hline	}
\tablelasttail{\hline}
\tablecaption{Observation log. For each target we list the date and mid-time of the observation,  as well as the individual exposure time (T$_{\text{exp}}$) of each FLECHAS spectrum and the signal-to-noise ratios (SNR) of the three combined spectra, as measured at $\lambda = 6700$\,\AA.}
\begin{supertabular}{cccc}
HIP\,410	& 2018 Feb 23, 18:24 & 300& 112  \\
HIP\,544	& 2018 Feb 23, 18:43 & 300& 133  \\
HIP\,695	& 2018 Feb 05, 20:26 & 300& 123  \\
HIP\,926	& 2018 Mar 02, 18:15 & 600& 50  \\
HIP\,1008	& 2018 Feb 23, 19:05 & 450& 114  \\
HIP\,1118	& 2018 Feb 23, 19:29 & 300& 108  \\
HIP\,1352	& 2018 Feb 23, 19:48 & 300& 146  \\
HIP\,1479	& 2018 Apr 08, 20:36 & 1200& 149  \\
HIP\,1486	& 2018 Feb 24, 20:34 & 300& 130  \\
HIP\,2537	& 2018 Feb 24, 20:53 & 300& 125  \\
HIP\,2583	& 2018 Mar 01, 18:10 & 300& 112  \\
HIP\,2838	& 2018 Apr 08, 21:38 & 1200& 143  \\
HIP\,3083	& 2015 Dec 03, 21:48 & 300& 217  \\
HIP\,3360	& 2018 Mar 01, 18:33 & 450& 78  \\
HIP\,3649	& 2018 Feb 24, 21:12 & 300& 206  \\
HIP\,3693	& 2018 Feb 01, 20:19 & 150& 382  \\
HIP\,3779	& 2018 Mar 01, 19:03 & 600& 107  \\
HIP\,4477	& 2018 Feb 25, 21:23 & 450& 115  \\
HIP\,4961	& 2018 Feb 24, 21:31 & 300& 140  \\
HIP\,5013	& 2018 Feb 25, 21:50 & 450& 98  \\
HIP\,5251	& 2018 Mar 05, 18:42 & 300& 94  \\
HIP\,5372	& 2018 Feb 01, 20:31 & 150& 320  \\
HIP\,5912	& 2015 Mar 19, 19:18 & 600& 155  \\
HIP\,6492	& 2015 Dec 03, 22:10 & 300& 116  \\
HIP\,6811	& 2018 Feb 25, 22:16 & 300& 118  \\
HIP\,7650	& 2018 Feb 01, 20:41 & 150& 172  \\
HIP\,8321	& 2018 Feb 25, 22:38 & 300& 91  \\
HIP\,8979	& 2015 Apr 20, 21:23 & 600& 154  \\
HIP\,9077	& 2018 Mar 05, 19:05 & 450& 86  \\
HIP\,9357	& 2015 Apr 20, 22:37 & 600& 80  \\
HIP\,10855	& 2018 Apr 06, 19:47 & 450& 104  \\
HIP\,11460	& 2018 Apr 06, 21:10 & 900& 102  \\
HIP\,12686	& 2018 Feb 01, 22:37 & 150& 140  \\
HIP\,13098	& 2018 Feb 28, 19:24 & 600& 102  \\
HIP\,13160	& 2018 Feb 22, 18:19 & 300& 110  \\
HIP\,13462	& 2018 Feb 28, 19:00 & 300& 138  \\
HIP\,13645	& 2018 Feb 28, 18:41 & 300& 122  \\
HIP\,13700	& 2018 Feb 24, 02:33 & 300& 196  \\
HIP\,13962	& 2018 Feb 27, 21:19 & 300& 131  \\
HIP\,14350	& 2015 Mar 19, 20:29 & 600& 190  \\
HIP\,14382	& 2018 Feb 01, 22:47 & 150& 229  \\
HIP\,15219	& 2018 Feb 08, 23:15 & 150& 153  \\
HIP\,15795	& 2018 Feb 28, 19:54 & 300& 122  \\
HIP\,15890	& 2018 Feb 02, 00:49 & 150& 250  \\
HIP\,16019	& 2018 Feb 27, 21:43 & 600& 133  \\
HIP\,16165	& 2018 Mar 05, 21:28 & 300& 70  \\
HIP\,16489	& 2018 Feb 01, 20:52 & 150& 148  \\
HIP\,17064	& 2015 Feb 28, 21:06 & 600& 127  \\
HIP\,17342	& 2018 Feb 12, 23:53 & 150& 145  \\
HIP\,18088	& 2018 Feb 13, 00:05 & 300& 105  \\
HIP\,18488	& 2018 Feb 08, 23:31 & 150& 144  \\
HIP\,19057	& 2018 Feb 27, 22:46 & 600& 86  \\
HIP\,19679	& 2018 Feb 27, 17:42 & 300& 135  \\
HIP\,20426	& 2018 Feb 28, 21:37 & 450& 84  \\
HIP\,20513	& 2018 Feb 28, 20:26 & 300& 111  \\
HIP\,20776	& 2015 Apr 12, 22:43 & 600& 275  \\
HIP\,20958	& 2018 Feb 28, 17:53 & 300& 120  \\
HIP\,20974	& 2015 Feb 26, 01:37 & 600& 235  \\
HIP\,21408	& 2018 Feb 21, 22:07 & 300& 115  \\
HIP\,21476	& 2018 Feb 05, 21:32 & 150& 281  \\
HIP\,21601	& 2015 Feb 26, 02:14 & 600& 205  \\
HIP\,22154	& 2015 Apr 19, 20:56 & 600& 134  \\
HIP\,22261	& 2018 Apr 07, 20:50 & 900& 163  \\
HIP\,22453	& 2018 Feb 05, 21:42 & 150& 219  \\
HIP\,22928	& 2018 Feb 24, 22:32 & 600& 158  \\
HIP\,23268	& 2018 Feb 05, 21:56 & 300& 160  \\
HIP\,23359	& 2018 Apr 18, 20:20 & 900& 136  \\
HIP\,23360	& 2018 Feb 24, 23:07 & 600& 120  \\
HIP\,23582	& 2015 Mar 17, 23:55 & 300& 160  \\
HIP\,23766	& 2018 Jan 27, 20:05 & 150& 89  \\
HIP\,24303	& 2015 Mar 19, 22:11 & 600& 114  \\
HIP\,24716	& 2018 Mar 01, 20:47 & 600& 104  \\
HIP\,24914	& 2018 Jan 27, 20:24 & 150& 121  \\
HIP\,25184	& 2015 Feb 28, 20:27 & 600& 238  \\
HIP\,25226	& 2018 Feb 05, 22:15 & 300& 114  \\
HIP\,25386	& 2018 Apr 06, 21:59 & 900& 101  \\
HIP\,25668	& 2018 Feb 23, 21:57 & 450& 106  \\
HIP\,25877	& 2015 Apr 12, 21:32 & 600& 72  \\
HIP\,26070	& 2016 Feb 03, 20:27 & 300& 77  \\
HIP\,26386	& 2018 Jan 27, 20:44 & 150& 59  \\
HIP\,26743	& 2018 Mar 01, 20:01 & 450& 97  \\
HIP\,27227	& 2015 Mar 17, 23:31 & 300& 121  \\
HIP\,27750	& 2018 Jan 27, 21:14 & 150& 69  \\
HIP\,27778	& 2018 Feb 05, 22:34 & 300& 149  \\
HIP\,27989	& 2018 Jan 27, 21:01 & 10& 117  \\
HIP\,28185	& 2018 Feb 23, 23:26 & 600& 113  \\
HIP\,28366	& 2015 Feb 26, 03:33 & 600& 98  \\
HIP\,30341	& 2015 Feb 20, 20:49 & 300& 118  \\
HIP\,32094	& 2018 Feb 18, 20:46 & 300& 139  \\
HIP\,32276	& 2015 Feb 20, 21:09 & 300& 81  \\
HIP\,32631	& 2018 Feb 21, 22:34 & 450& 109  \\
HIP\,33515	& 2018 Apr 08, 19:11 & 300& 96  \\
HIP\,33789	& 2015 Feb 20, 21:27 & 300& 155  \\
HIP\,33927	& 2015 Feb 20, 21:46 & 300& 303  \\
HIP\,33937	& 2015 Feb 20, 22:04 & 300& 263  \\
HIP\,34026	& 2015 Feb 20, 22:23 & 300& 110  \\
HIP\,34055	& 2015 Feb 20, 22:42 & 300& 233  \\
HIP\,34909	& 2015 Feb 20, 23:02 & 300& 402  \\
HIP\,35537	& 2015 Mar 16, 22:56 & 900& 125  \\
HIP\,35551	& 2018 Jan 30, 21:56 & 300& 91  \\
HIP\,35796	& 2015 Feb 20, 23:22 & 300& 133  \\
HIP\,36041	& 2015 Feb 20, 23:41 & 300& 287  \\
HIP\,36629	& 2018 Mar 19, 19:26 & 300& 68  \\
HIP\,37104	& 2018 Mar 19, 19:49 & 450& 103  \\
HIP\,39398	& 2015 Apr 22, 20:59 & 600& 102  \\
HIP\,39958	& 2018 Feb 08, 23:47 & 300& 109  \\
HIP\,40628	& 2015 Apr 22, 20:20 & 600& 112  \\
HIP\,41221	& 2018 Mar 19, 20:18 & 450& 110  \\
HIP\,41283	& 2018 Apr 10, 23:08 & 1200& 99  \\
HIP\,41704	& 2018 Jan 30, 23:00 & 150& 361  \\
HIP\,41896	& 2015 Feb 21, 00:01 & 300& 109  \\
HIP\,42211	& 2018 Apr 08, 22:44 & 1200& 126  \\
HIP\,42331	& 2018 Feb 26, 23:45 & 600& 80  \\
HIP\,42580	& 2018 Mar 24, 20:44 & 900& 122  \\
HIP\,42876	& 2018 Feb 21, 00:32 & 450& 136  \\
HIP\,43030	& 2018 Feb 21, 01:00 & 450& 125  \\
HIP\,44231	& 2018 Jan 30, 23:16 & 300& 125  \\
HIP\,44458	& 2018 Feb 21, 02:55 & 450& 85  \\
HIP\,44580	& 2018 Apr 07, 19:20 & 600& 120  \\
HIP\,44784	& 2018 Apr 04, 23:24 & 900& 111  \\
HIP\,44831	& 2018 Feb 21, 01:57 & 450& 99  \\
HIP\,45104	& 2018 Feb 25, 00:43 & 600& 133  \\
HIP\,45105	& 2018 Apr 05, 23:33 & 900& 77  \\
HIP\,45343	& 2018 Feb 25, 01:17 & 600& 137  \\
HIP\,45963	& 2018 Feb 25, 23:32 & 600& 111  \\
HIP\,46207	& 2018 Mar 19, 21:18 & 900& 126  \\
HIP\,46693	& 2018 Feb 14, 03:51 & 450& 133  \\
HIP\,46816	& 2018 Apr 07, 19:59 & 900& 114  \\
HIP\,46843	& 2018 Feb 09, 00:44 & 450& 105  \\
HIP\,46977	& 2018 Jan 30, 23:35 & 150& 113  \\
HIP\,47193	& 2018 Feb 01, 23:02 & 150& 331  \\
HIP\,48851	& 2018 Feb 21, 03:23 & 450& 110  \\
HIP\,48921	& 2018 Feb 21, 02:26 & 450& 100  \\
HIP\,49688	& 2018 Feb 06, 00:17 & 300& 182  \\
HIP\,49729	& 2018 Feb 05, 23:51 & 300& 109  \\
HIP\,50310	& 2018 Mar 24, 22:32 & 900& 119  \\
HIP\,50999	& 2018 Apr 17, 20:29 & 1200& 118  \\
HIP\,51973	& 2018 Feb 09, 01:37 & 450& 117  \\
HIP\,52032	& 2018 Feb 06, 00:37 & 300& 153  \\
HIP\,52098	& 2018 Feb 02, 01:02 & 150& 253  \\
HIP\,52373	& 2018 Feb 15, 01:14 & 450& 169  \\
HIP\,52556	& 2018 Feb 22, 00:26 & 600& 117  \\
HIP\,52831	& 2018 Mar 19, 22:21 & 900& 130  \\
HIP\,53831	& 2018 Feb 26, 00:01 & 600& 129  \\
HIP\,54024	& 2018 Feb 09, 01:10 & 450& 119  \\
HIP\,55193	& 2018 Feb 05, 01:58 & 300& 91  \\
HIP\,55682	& 2018 Feb 05, 02:19 & 300& 139  \\
HIP\,55945	& 2018 Feb 02, 01:17 & 150& 210  \\
HIP\,56383	& 2018 Feb 06, 00:55 & 300& 25  \\
HIP\,57240	& 2018 Feb 06, 01:14 & 300& 209  \\
HIP\,57261	& 2018 Apr 06, 22:51 & 900& 111  \\
HIP\,58217	& 2018 Feb 09, 02:05 & 450& 112  \\
HIP\,58313	& 2018 Apr 08, 23:48 & 1200& 143  \\
HIP\,58661	& 2018 Feb 06, 01:34 & 300& 120  \\
HIP\,59280	& 2018 Feb 22, 01:41 & 600& 121  \\
HIP\,59501	& 2018 Feb 02, 01:28 & 150& 153  \\
HIP\,59760	& 2018 Feb 09, 02:31 & 450& 111  \\
HIP\,60731	& 2018 Feb 14, 02:02 & 450& 141  \\
HIP\,60831	& 2018 Feb 14, 04:16 & 450& 104  \\
HIP\,61290	& 2018 Feb 09, 03:01 & 450& 117  \\
HIP\,61617	& 2018 Apr 07, 22:41 & 1200& 132  \\
HIP\,61799	& 2018 Apr 09, 23:43 & 1200& 122  \\
HIP\,62455	& 2018 Feb 15, 01:42 & 450& 123  \\
HIP\,62595	& 2018 Apr 07, 23:48 & 1200& 109  \\
HIP\,63356	& 2018 Feb 06, 01:53 & 300& 117  \\
HIP\,63368	& 2018 Mar 29, 23:18 & 900& 94  \\
HIP\,63742	& 2018 Feb 27, 00:21 & 600& 61  \\
HIP\,63803	& 2018 Feb 14, 02:29 & 450& 139  \\
HIP\,64149	& 2018 Mar 30, 00:07 & 900& 155  \\
HIP\,64523	& 2018 Apr 06, 00:24 & 900& 107  \\
HIP\,64543	& 2018 Feb 06, 02:13 & 300& 120  \\
HIP\,65192	& 2018 Mar 19, 23:45 & 900& 136  \\
HIP\,65435	& 2018 Feb 27, 01:01 & 600& 96  \\
HIP\,65915	& 2018 Apr 11, 00:19 & 1200& 108  \\
HIP\,66467	& 2018 Feb 09, 04:23 & 450& 133  \\
HIP\,66690	& 2018 Feb 02, 03:58 & 300& 174  \\
HIP\,67300	& 2018 Feb 02, 04:23 & 300& 127  \\
HIP\,67385	& 2018 Feb 09, 04:01 & 300& 113  \\
HIP\,68879	& 2018 Feb 27, 01:38 & 600& 113  \\
HIP\,68904	& 2018 Feb 27, 23:27 & 600& 107  \\
HIP\,69624	& 2018 Apr 09, 00:53 & 1200& 160  \\
HIP\,70000	& 2018 Apr 08, 00:51 & 1200& 110  \\
HIP\,70108	& 2018 Feb 02, 04:45 & 300& 127  \\
HIP\,70145	& 2018 Feb 02, 05:05 & 300& 145  \\
HIP\,70349	& 2018 Feb 09, 04:50 & 450& 148  \\
HIP\,71436	& 2018 Feb 28, 00:36 & 600& 129  \\
HIP\,72499	& 2018 Feb 06, 03:23 & 300& 127  \\
HIP\,72578	& 2018 Feb 06, 02:32 & 300& 119  \\
HIP\,73977	& 2018 Feb 14, 04:40 & 300& 164  \\
HIP\,74070	& 2018 Apr 06, 01:14 & 900& 96  \\
HIP\,74425	& 2018 Feb 14, 04:59 & 450& 144  \\
HIP\,74938	& 2018 Apr 12, 00:53 & 1200& 155  \\
HIP\,75257	& 2018 Feb 02, 01:41 & 150& 181  \\
HIP\,76733	& 2018 Feb 06, 03:55 & 300& 146  \\
HIP\,76947	& 2018 Apr 06, 23:43 & 900& 143  \\
HIP\,77092	& 2018 Mar 01, 01:43 & 600& 120  \\
HIP\,77178	& 2018 Feb 28, 22:34 & 600& 111  \\
HIP\,78802	& 2018 Apr 18, 22:41 & 1200& 142  \\
HIP\,79187	& 2018 Feb 21, 03:52 & 300& 105  \\
HIP\,79357	& 2018 Feb 02, 01:53 & 150& 132  \\
HIP\,80021	& 2018 Feb 06, 04:15 & 300& 173  \\
HIP\,80941	& 2018 Apr 19, 22:03 & 1200& 116  \\
HIP\,81104	& 2018 Feb 14, 02:56 & 450& 157  \\
HIP\,81289	& 2018 Feb 02, 03:09 & 150& 145  \\
HIP\,81922	& 2018 Apr 19, 23:58 & 1200& 136  \\
HIP\,82324	& 2018 Feb 06, 04:36 & 300& 141  \\
HIP\,82385	& 2018 Feb 14, 03:22 & 450& 118  \\
HIP\,82504	& 2018 Feb 02, 03:20 & 150& 213  \\
HIP\,82604	& 2018 Feb 17, 02:33 & 300& 62  \\
HIP\,83254	& 2018 Feb 02, 03:30 & 150& 147  \\
HIP\,84038	& 2018 Feb 28, 23:28 & 600& 116  \\
HIP\,84239	& 2018 Apr 20, 22:30 & 1200& 122  \\
HIP\,84380	& 2018 Feb 06, 03:43 & 30& 238  \\
HIP\,84671	& 2018 Feb 02, 05:42 & 150& 237  \\
HIP\,85200	& 2018 Apr 20, 23:50 & 1200& 143  \\
HIP\,85560	& 2018 Feb 28, 01:13 & 600& 120  \\
HIP\,86153	& 2018 Feb 06, 04:55 & 300& 173  \\
HIP\,86476	& 2018 Mar 20, 03:42 & 300& 185  \\
HIP\,86483	& 2018 Apr 21, 23:30 & 450& 89  \\
HIP\,86625	& 2018 Mar 25, 02:19 & 600& 127  \\
HIP\,86709	& 2018 Feb 22, 02:43 & 450& 119  \\
HIP\,87107	& 2018 Apr 22, 00:01 & 600& 92  \\
HIP\,87244	& 2018 Apr 22, 02:44 & 600& 141  \\
HIP\,87251	& 2018 Mar 20, 01:31 & 600& 134  \\
HIP\,88411	& 2018 May 01, 22:20 & 1200& 97  \\
HIP\,88518	& 2018 Feb 15, 02:36 & 450& 88  \\
HIP\,88984	& 2018 May 26, 00:55 & 450& 105  \\
HIP\,91200	& 2018 Feb 21, 04:12 & 300& 110  \\
HIP\,92249	& 2018 May 04, 21:42 & 1200& 117  \\
HIP\,92404	& 2018 Mar 01, 02:19 & 600& 89  \\
HIP\,92651	& 2018 Mar 25, 01:16 & 600& 112  \\
HIP\,92713	& 2018 Feb 21, 04:33 & 300& 134  \\
HIP\,93801	& 2018 May 26, 01:22 & 300& 90  \\
HIP\,93913	& 2018 Mar 20, 00:54 & 600& 73  \\
HIP\,94220	& 2018 May 05, 01:25 & 450& 82  \\
HIP\,94229	& 2018 Apr 06, 02:51 & 600& 116  \\
HIP\,94518	& 2018 Apr 23, 22:39 & 1200& 104  \\
HIP\,94761	& 2015 Sep 24, 18:58 & 600& 43  \\
HIP\,95297	& 2018 Apr 29, 21:57 & 1200& 97  \\
HIP\,95744	& 2018 Feb 22, 03:31 & 450& 100  \\
HIP\,96203	& 2018 May 06, 21:58 & 1200& 129  \\
HIP\,96966	& 2018 May 07, 01:29 & 1200& 105  \\
HIP\,97135	& 2015 Apr 23, 01:39 & 600& 186  \\
HIP\,97198	& 2015 Sep 24, 19:32 & 600& 42  \\
HIP\,97359	& 2018 Mar 02, 04:06 & 450& 107  \\
HIP\,97402	& 2015 Sep 24, 20:00 & 300& 161  \\
HIP\,98073	& 2018 Feb 02, 03:42 & 150& 243  \\
HIP\,98443	& 2015 Sep 24, 20:57 & 300& 82  \\
HIP\,98610	& 2015 Apr 23, 23:26 & 600& 161  \\
HIP\,98762	& 2015 Apr 23, 22:07 & 600& 126  \\
HIP\,99853	& 2018 Feb 23, 05:18 & 150& 139  \\
HIP\,100172	& 2018 Mar 02, 04:36 & 600& 75  \\
HIP\,100180	& 2018 May 05, 21:53 & 1200& 130  \\
HIP\,100390	& 2018 Mar 25, 03:45 & 300& 136  \\
HIP\,100534	& 2018 May 07, 22:15 & 900& 127  \\
HIP\,100684	& 2018 Mar 01, 04:14 & 600& 86  \\
HIP\,101219	& 2018 May 05, 02:19 & 600& 116  \\
HIP\,101412	& 2018 Feb 23, 01:36 & 300& 124  \\
HIP\,101692	& 2018 May 29, 01:41 & 150& 182  \\
HIP\,101841	& 2018 May 06, 00:15 & 600& 141  \\
HIP\,101953	& 2018 Apr 30, 01:31 & 450& 82  \\
HIP\,102377	& 2018 Feb 01, 18:43 & 300& 115  \\
HIP\,102440	& 2018 May 26, 01:43 & 300& 167  \\
HIP\,102912	& 2018 Feb 23, 03:05 & 300& 104  \\
HIP\,103035	& 2018 Feb 23, 01:59 & 450& 88  \\
HIP\,103242	& 2018 Apr 30, 01:52 & 300& 124  \\
HIP\,103263	& 2018 Feb 01, 19:01 & 300& 80  \\
HIP\,103637	& 2018 Feb 24, 03:51 & 300& 101  \\
HIP\,103868	& 2018 Jun 04, 01:28 & 300& 121  \\
HIP\,104172	& 2018 Mar 30, 02:37 & 300& 119  \\
HIP\,105182	& 2018 Feb 23, 03:29 & 450& 99  \\
HIP\,105205	& 2018 May 29, 00:18 & 300& 117  \\
HIP\,105669	& 2018 Feb 27, 03:46 & 600& 98  \\
HIP\,105949	& 2018 Feb 01, 19:19 & 300& 249  \\
HIP\,106306	& 2018 May 29, 00:37 & 300& 123  \\
HIP\,106848	& 2018 Feb 24, 04:43 & 450& 94  \\
HIP\,106973	& 2018 Feb 24, 05:06 & 300& 105  \\
HIP\,106974	& 2018 Jun 04, 00:24 & 450& 126  \\
HIP\,107205	& 2018 Feb 27, 04:21 & 600& 109  \\
HIP\,107259	& 2018 Feb 07, 17:08 & 150& 464  \\
HIP\,107315	& 2018 May 31, 00:55 & 30& 113  \\
HIP\,107325	& 2018 May 08, 21:41 & 1200& 117  \\
HIP\,107350	& 2018 May 31, 01:03 & 150& 39  \\
HIP\,107723	& 2018 Feb 05, 18:15 & 300& 152  \\
HIP\,107923	& 2018 May 06, 01:25 & 300& 163  \\
HIP\,108030	& 2018 Mar 25, 00:18 & 450& 72  \\
HIP\,108202	& 2018 May 08, 01:27 & 900& 123  \\
HIP\,108233	& 2018 May 08, 00:59 & 300& 125  \\
HIP\,108296	& 2018 May 29, 00:55 & 300& 159  \\
HIP\,108378	& 2018 Feb 05, 19:54 & 300& 124  \\
HIP\,109247	& 2018 May 06, 02:10 & 450& 128  \\
HIP\,109492	& 2018 Feb 07, 17:19 & 150& 454  \\
HIP\,109602	& 2018 May 29, 01:11 & 150& 119  \\
HIP\,110504	& 2018 Mar 25, 03:08 & 300& 174  \\
HIP\,110991	& 2018 Feb 13, 17:17 & 150& 380  \\
HIP\,110992	& 2018 Jun 03, 01:14 & 150& 170  \\
HIP\,111810	& 2018 Jun 04, 00:45 & 150& 92  \\
HIP\,112098	& 2018 Feb 24, 18:38 & 300& 205  \\
HIP\,112248	& 2018 Feb 24, 19:05 & 600& 160  \\
HIP\,112987	& 2018 Jun 04, 00:58 & 300& 120  \\
HIP\,113561	& 2018 Feb 05, 20:11 & 150& 120  \\
HIP\,113881	& 2018 Jun 03, 01:02 & 30& 253  \\
HIP\,114155	& 2018 Feb 13, 17:29 & 150& 235  \\
HIP\,115147	& 2018 Feb 27, 20:48 & 600& 93  \\
HIP\,117299	& 2018 Feb 13, 17:39 & 150& 208  \\
HIP\,117956	& 2018 Feb 14, 17:29 & 150& 139  \\
HIP\,118077	& 2018 Feb 14, 17:19 & 150& 147  \\
HIP\,120005	& 2018 Feb 25, 01:51 & 600& 125  \\
\end{supertabular}
\label{tab:obslog}
\end{center}

\newpage

\begin{center}
\vspace{0.8cm}
\tablefirsthead{%
\hline
Target          & \multicolumn{1}{c}{$EW_{\text{Li}}$ [m\AA]} &  \multicolumn{1}{c}{Target} &	 \multicolumn{1}{c}{ $EW_{\text{Li}}$ [m\AA]} \\
\hline}
\tablehead{%
\multicolumn{1}{l}{Continued}\\
\hline
Target          & \multicolumn{1}{c}{$EW_{\text{Li}}$ [m\AA]} &  \multicolumn{1}{c}{Target} &	 \multicolumn{1}{c}{ $EW_{\text{Li}}$ [m\AA]} \\ 	
				\hline}
\tabletail{%
\hline	}
\tablelasttail{\hline}
\tablecaption{Measured equivalent widths ($EW_{\text{Li}}$) of the Li\,(6708\,\AA) line. Targets without significant detection ($EW_{\text{Li}}< 3\cdot\sigma EW_{\text{Li}}$) are listed with "-"}
\begin{supertabular}{cccc}
HIP\,410	& $~~~-  $ 				 & HIP\,59280	& $~~~-  $    				   \\
HIP\,544	& $~100  \pm 11  $       & HIP\,59501	& $~~26   \pm 7   $       \\
HIP\,695	& $~~~-  $   			 & HIP\,59760	& $~~~~53   \pm 11  $     \\
HIP\,926	& $~~~-  $   			 & HIP\,60731	& $~~21   \pm 7   $       \\
HIP\,1008	& $\,\,~~45   \pm 10  $  & HIP\,60831	& $~~116  \pm 11  $       \\
HIP\,1118	& $~~~-  $    			 & HIP\,61290	& $~~38   \pm 8   $       \\
HIP\,1352	& $~83   \pm 8   $       & HIP\,61617	& $~~~-  $    				    \\
HIP\,1479	& $~46   \pm 6   $       & HIP\,61799	& $~~~-  $    				      \\
HIP\,1486	& $~~~-  $    			 & HIP\,62455	& $~~~~61   \pm 10  $     \\
HIP\,2537	& $~~~-  $   			 & HIP\,62595	& $~~~-  $    				      \\
HIP\,2583	& $~~~-  $   			 & HIP\,63356	& $~~~-  $   				    \\
HIP\,2838	& $~29   \pm 7   $       & HIP\,63368	& $~~~~98   \pm 11  $     \\
HIP\,3083	& $~47   \pm 5   $       & HIP\,63742	& $~~140  \pm 16  $       \\
HIP\,3360	& $~~~-  $   			 & HIP\,63803	& $~~27   \pm 8   $       \\
HIP\,3649	& $~~~-  $   			 & HIP\,64149	& $~~38   \pm 7   $       \\
HIP\,3693	& $~\,71   \pm 3   $     & HIP\,64523	& $~~~~42   \pm 10  $     \\
HIP\,3779	& $~\,\,\,~38   \pm 12  $& HIP\,64543	& $~~~-  $    				     \\
HIP\,4477	& $~~151  \pm 11  $      & HIP\,65192	& $~~27   \pm 7   $       \\
HIP\,4961	& $~~~-  $   			 & HIP\,65435	& $~~~-  $    				    \\
HIP\,5013	& $~~~~57   \pm 13  $    & HIP\,65915	& $~~~-  $   				     \\
HIP\,5251	& $~~157  \pm 12  $      & HIP\,66467	& $~~27   \pm 8   $       \\
HIP\,5372	& $~~46   \pm 4   $      & HIP\,66690	& $~~25   \pm 7   $       \\
HIP\,5912	& $137  \pm 6   $        & HIP\,67300	& $~~43   \pm 9   $       \\
HIP\,6492	& $~~~~50   \pm 10  $    & HIP\,67385	& $~~~-  $   				       \\
HIP\,6811	& $~~161  \pm 11  $      & HIP\,68879	& $~~~-  $   				     \\
HIP\,7650	& $~~43   \pm 7   $      & HIP\,68904	& $~~79   \pm 9   $       \\
HIP\,8321	& $~~~-  $   			 & HIP\,69624	& $~~~-  $    				       \\
HIP\,8979	& $~~39   \pm 6   $      & HIP\,70000	& $~~~-  $    				       \\
HIP\,9077	& $~~~~64   \pm 12  $    & HIP\,70108	& $~~~-  $   				    \\
HIP\,9357	& $~~~~65   \pm 12  $    & HIP\,70145	& $~~21   \pm 7   $       \\
HIP\,10855	& $~~~-  $    			 & HIP\,70349	& $~~48   \pm 8   $       \\
HIP\,11460	& $~~~-  $    			 & HIP\,71436	& $~~~-  $    				       \\
HIP\,12686	& $~~56   \pm 7   $      & HIP\,72499	& $~~59   \pm 9   $       \\
HIP\,13098	& $~~27   \pm 9   $      & HIP\,72578	& $~~30   \pm 8   $       \\
HIP\,13160	& $~~~-  $    			 & HIP\,73977	& $~~22   \pm 6   $       \\
HIP\,13462	& $~~33   \pm 9   $      & HIP\,74070	& $~~~~35   \pm 11  $     \\
HIP\,13645	& $~~42   \pm 8   $      & HIP\,74425	& $~~21   \pm 7   $       \\
HIP\,13700	& $~~45   \pm 6   $      & HIP\,74938	& $~~48   \pm 7   $       \\
HIP\,13962	& $~~~-  $    			 & HIP\,75257	& $~~43   \pm 6   $       \\
HIP\,14350	& $~~32   \pm 5   $      & HIP\,76733	& $~~23   \pm 7   $       \\
HIP\,14382	& $~~25   \pm 5   $      & HIP\,76947	& $~~54   \pm 7   $       \\
HIP\,15219	& $~~27   \pm 7   $      & HIP\,77092	& $~~~-  $    				     \\
HIP\,15795	& $~~55   \pm 8   $      & HIP\,77178	& $~~30   \pm 9   $       \\
HIP\,15890	& $~~87   \pm 4   $      & HIP\,78802	& $~~~-  $    				     \\
HIP\,16019	& $~~69   \pm 7   $      & HIP\,79187	& $~~~~39   \pm 10  $     \\
HIP\,16165	& $~~~-  $    			 & HIP\,79357	& $~~24   \pm 8   $       \\
HIP\,16489	& $~~~-  $    			 & HIP\,80021	& $~~20   \pm 6   $       \\
HIP\,17064	& $~~94  \pm 7   $       & HIP\,80941	& $~~242  \pm 12  $       \\
HIP\,17342	& $~~33   \pm 8   $      & HIP\,81104	& $~~26   \pm 7   $       \\
HIP\,18088	& $~~~-  $   			 & HIP\,81289	& $~~31   \pm 7   $       \\
HIP\,18488	& $184  \pm 9   $        & HIP\,81922	& $~~32   \pm 8   $       \\
HIP\,19057	& $~~~-  $   			 & HIP\,82324	& $~~~-  $    				    \\
HIP\,19679	& $~~~-  $   			 & HIP\,82385	& $~~~-  $   				      \\
HIP\,20426	& $~~~~44   \pm 12  $    & HIP\,82504	& $~~45   \pm 6   $       \\
HIP\,20513	& $~~109  \pm 10  $      & HIP\,82604	& $~~~~63   \pm 16  $     \\
HIP\,20776	& $~~48   \pm 4   $      & HIP\,83254	& $~~60   \pm 8   $       \\
HIP\,20958	& $~~45   \pm 9   $      & HIP\,84038	& $~~~-  $   				      \\
HIP\,20974	& $~~19   \pm 5   $      & HIP\,84239	& $~~~-  $  				       \\
HIP\,21408	& $~~~-  $    			 & HIP\,84380	& $~~34   \pm 5   $       \\
HIP\,21476	& $~~44   \pm 4   $      & HIP\,84671	& $~~26   \pm 4   $       \\
HIP\,21601	& $~~42   \pm 6   $      & HIP\,85200	& $~~30   \pm 7   $       \\
HIP\,22154	& $~~31   \pm 8   $      & HIP\,85560	& $~~~-  $   				      \\
HIP\,22261	& $~~46   \pm 6   $      & HIP\,86153	& $~~35   \pm 6   $       \\
HIP\,22453	& $108  \pm 5   $        & HIP\,86476	& $~~45   \pm 6   $       \\
HIP\,22928	& $~~39   \pm 7   $      & HIP\,86483	& $~~~~60   \pm 10  $     \\
HIP\,23268	& $159  \pm 7   $        & HIP\,86625	& $~~29   \pm 7   $       \\
HIP\,23359	& $~~24   \pm 7   $      & HIP\,86709	& $~~53   \pm 9   $       \\
HIP\,23360	& $~~~-  $   			 & HIP\,87107	& $~~~~39   \pm 11  $     \\
HIP\,23582	& $~~87   \pm 7   $      & HIP\,87244	& $~~54   \pm 7   $       \\
HIP\,23766	& $~~~-  $    			 & HIP\,87251	& $~~~-  $    				      \\
HIP\,24303	& $~~~~76   \pm 10  $    & HIP\,88411	& $~~~-  $   				     \\
HIP\,24716	& $~~~~44   \pm 11  $    & HIP\,88518	& $~~~~30   \pm 10  $     \\
HIP\,24914	& $~~~~37   \pm 10  $    & HIP\,88984	& $~~49   \pm 9   $       \\
HIP\,25184	& $135  \pm 5   $        & HIP\,91200	& $~~~-  $   				       \\
HIP\,25226	& $~~40   \pm 9   $      & HIP\,92249	& $~~31   \pm 9   $       \\
HIP\,25386	& $~~~~39   \pm 10  $    & HIP\,92404	& $~~~~44   \pm 13  $     \\
HIP\,25668	& $~~~-  $   			 & HIP\,92651	& $~~68   \pm 9   $       \\
HIP\,25877	& $~~~-  $    			 & HIP\,92713	& $~~29   \pm 8   $       \\
HIP\,26070	& $~~~-  $    			 & HIP\,93801	& $~~~-  $    				    \\
HIP\,26386	& $~~~-  $   			 & HIP\,93913	& $~~~-  $   				     \\
HIP\,26743	& $~~~~61  \pm 12  $     & HIP\,94220	& $~~~~68   \pm 11  $     \\
HIP\,27227	& $-  $    				 & HIP\,94229	& $~~~-  $    				   \\
HIP\,27750	& $~~~~58   \pm 14  $    & HIP\,94518	& $~~~-  $    				      \\
HIP\,27778	& $266  \pm 7   $        & HIP\,94761	& $~~~-  $    				     \\
HIP\,27989	& $~~32   \pm 9   $      & HIP\,95297	& $~~~-  $    				    \\
HIP\,28185	& $~~~~42   \pm 10  $    & HIP\,95744	& $~~~~31   \pm 10  $     \\
HIP\,28366	& $~~~-  $    			 & HIP\,96203	& $~~85   \pm 8   $       \\
HIP\,30341	& $~~~~57   \pm 11  $    & HIP\,96966	& $~~~-  $  				     \\
HIP\,32094	& $~~91   \pm 7   $      & HIP\,97135	& $~~69   \pm 6   $       \\
HIP\,32276	& $~~179  \pm 14  $      & HIP\,97198	& $~~~-  $   				     \\
HIP\,32631	& $~~32   \pm 9   $      & HIP\,97359	& $~~27   \pm 9   $       \\
HIP\,33515	& $~~~~46   \pm 11  $    & HIP\,97402	& $~~~-  $   				        \\
HIP\,33789	& $~~38   \pm 7   $      & HIP\,98073	& $~~42   \pm 4   $       \\
HIP\,33927	& $~~16   \pm 4   $      & HIP\,98443	& $~~~-  $    				   \\
HIP\,33937	& $~~82   \pm 4   $      & HIP\,98610	& $~~40   \pm 6   $       \\
HIP\,34026	& $~~~-  $    			 & HIP\,98762	& $~~35   \pm 8   $       \\
HIP\,34055	& $~~23   \pm 5   $      & HIP\,99853	& $~~46   \pm 9   $       \\
HIP\,34909	& $~~34   \pm 3   $      & HIP\,100172	& $~~~-  $    				     \\
HIP\,35537	& $~~~-  $    			 & HIP\,100180	& $~~43   \pm 7   $       \\
HIP\,35551	& $~~~-  $    			 & HIP\,100390	& $111  \pm 8   $         \\
HIP\,35796	& $~~36   \pm 9   $      & HIP\,100534	& $~~35   \pm 7   $       \\
HIP\,36041	& $~~24   \pm 4   $      & HIP\,100684	& $~~~-  $    				     \\
HIP\,36629	& $~~~-  $    			 & HIP\,101219	& $108  \pm 8   $         \\
HIP\,37104	& $~~~-  $    			 & HIP\,101412	& $~~50   \pm 8   $       \\
HIP\,39398	& $~~152  \pm 10  $      & HIP\,101692	& $~~33   \pm 6   $       \\
HIP\,39958	& $~~~-  $    			 & HIP\,101841	& $~~76   \pm 7   $       \\
HIP\,40628	& $235  \pm 9   $        & HIP\,101953	& $~~~~34   \pm 11  $     \\
HIP\,41221	& $~~331  \pm 10  $      & HIP\,102377	& $~~235  \pm 10  $       \\
HIP\,41283	& $~~~~36   \pm 10  $    & HIP\,102440	& $~~74   \pm 7   $       \\
HIP\,41704	& $~~15   \pm 3   $      & HIP\,102912	& $~~30   \pm 9   $       \\
HIP\,41896	& $~~~-  $    			 & HIP\,103035	& $~~~-  $   				     \\
HIP\,42211	& $~~~-  $    			 & HIP\,103242	& $~~40   \pm 8   $       \\
HIP\,42331	& $~~~-  $    			 & HIP\,103263	& $~~~-  $  				      \\
HIP\,42580	& $~~~~45   \pm 10  $    & HIP\,103637	& $~~~-  $    				     \\
HIP\,42876	& $~~42   \pm 8   $      & HIP\,103868	& $150  \pm 8   $         \\
HIP\,43030	& $310  \pm 9   $        & HIP\,104172	& $~~~-  $   				    \\
HIP\,44231	& $140  \pm 8   $        & HIP\,105182	& $~~~~67   \pm 12  $     \\
HIP\,44458	& $~~178  \pm 13  $      & HIP\,105205	& $~~~-  $   				     \\
HIP\,44580	& $~~67   \pm 9   $      & HIP\,105669	& $~~~-  $   				     \\
HIP\,44784	& $~~~-  $    			 & HIP\,105949	& $~~83   \pm 4   $       \\
HIP\,44831	& $~~~-  $    			 & HIP\,106306	& $~~34   \pm 8   $       \\
HIP\,45104	& $176  \pm 9   $        & HIP\,106848	& $~~~-  $    				     \\
HIP\,45105	& $~~~-  $    			 & HIP\,106973	& $~~~~60   \pm 10  $     \\
HIP\,45343	& $~~~-  $    			 & HIP\,106974	& $~~62   \pm 9   $       \\
HIP\,45963	& $~~29   \pm 8   $      & HIP\,107205	& $~~27   \pm 9   $       \\
HIP\,46207	& $~~45   \pm 8   $      & HIP\,107259	& $~~25   \pm 2   $       \\
HIP\,46693	& $116  \pm 9   $        & HIP\,107315	& $~~~~37   \pm 10  $     \\
HIP\,46816	& $~~254  \pm 11  $      & HIP\,107325	& $~~39   \pm 7  $     \\
HIP\,46843	& $~~200  \pm 12  $      & HIP\,107350	& $~~138  \pm 25  $       \\
HIP\,46977	& $~~~-  $    			 & HIP\,107723	& $114  \pm 7   $         \\
HIP\,47193	& $~~46   \pm 3   $      & HIP\,107923	& $114  \pm 7   $         \\
HIP\,48851	& $~~32   \pm 9   $      & HIP\,108030	& $~~~~43   \pm 14  $     \\
HIP\,48921	& $~~~-  $    			 & HIP\,108202	& $~~93   \pm 8   $       \\
HIP\,49688	& $~~65   \pm 6   $      & HIP\,108233	& $~~78   \pm 9   $       \\
HIP\,49729	& $~~~-  $    			 & HIP\,108296	& $161  \pm 7   $         \\
HIP\,50310	& $148  \pm 9   $        & HIP\,108378	& $~~36   \pm 9   $       \\
HIP\,50999	& $~~55   \pm 9   $      & HIP\,109247	& $~~47   \pm 8   $       \\
HIP\,51973	& $~~~-  $    			 & HIP\,109492	& $~~61   \pm 3   $       \\
HIP\,52032	& $~~25   \pm 7   $      & HIP\,109602	& $171  \pm 8   $         \\
HIP\,52098	& $~~13   \pm 4   $      & HIP\,110504	& $~~57   \pm 7   $       \\
HIP\,52373	& $~~90   \pm 6   $      & HIP\,110991	& $~~~-  $    				     \\
HIP\,52556	& $~~~-  $    			 & HIP\,110992	& $~~41   \pm 6   $       \\
HIP\,52831	& $~~60   \pm 9   $      & HIP\,111810	& $~~~-  $    				     \\
HIP\,53831	& $~~25   \pm 8   $      & HIP\,112098	& $~~27   \pm 5   $       \\
HIP\,54024	& $~~~-  $    			 & HIP\,112248	& $~~58   \pm 8   $       \\
HIP\,55193	& $~~~~45   \pm 12  $    & HIP\,112987	& $~~37   \pm 9   $       \\
HIP\,55682	& $~~89   \pm 8   $      & HIP\,113561	& $~~~-  $    				    \\
HIP\,55945	& $~~25   \pm 5   $      & HIP\,113881	& $~~53   \pm 4   $       \\
HIP\,56383	& $~~101  \pm 35  $      & HIP\,114155	& $~~30   \pm 5   $       \\
HIP\,57240	& $~~32   \pm 5   $      & HIP\,115147	& $~~239  \pm 13  $       \\
HIP\,57261	& $~~29   \pm 8   $      & HIP\,117299	& $~~44   \pm 6   $       \\
HIP\,58217	& $~~~~50   \pm 10  $    & HIP\,117956	& $~~62   \pm 8   $       \\
HIP\,58313	& $~~27   \pm 7   $      & HIP\,118077	& $~~~-  $    				     \\
HIP\,58661	& $~~~-  $    			 & HIP\,120005	& $~~~~59   \pm 12  $     \\
\end{supertabular}
\label{tab:Li_all}
\end{center}

\newpage
\onecolumn

\begin{center}
\vspace{0.8cm}
\tablefirsthead{%
	\hline
	Target          & \multicolumn{1}{c}{$d$ [pc]} &  \multicolumn{1}{c}{$G$ [mag]} &	 \multicolumn{1}{c}{$A_{\text{G}}$ [mag]}  & \multicolumn{1}{c}{\textit{$M_{\text{G}}$} [mag]}  & \multicolumn{1}{c}{$T_{\text{eff}}$ [K] }  & \multicolumn{1}{c}{$R$ [R$_{\odot}$]}  & \multicolumn{1}{c}{$L$ [L$_{\odot}$] } &\multicolumn{1}{c}{rem.}\\
	\hline}
\tablehead{%
\multicolumn{1}{l}{Continued}\\
\hline
	Target          & \multicolumn{1}{c}{$d$ [pc]} &  \multicolumn{1}{c}{$G$ [mag]} &	 \multicolumn{1}{c}{$A_{\text{G}}$ [mag]}  & \multicolumn{1}{c}{\textit{$M_{\text{G}}$} [mag]}  & \multicolumn{1}{c}{$T_{\text{eff}}$ [K] }  & \multicolumn{1}{c}{$R$ [R$_{\odot}$]}  & \multicolumn{1}{c}{$L$ [L$_{\odot}$] } &\multicolumn{1}{c}{rem.}\\
	\hline}
\tabletail{%
	\hline	}
\tablelasttail{
	\hline}
\tablecaption{Physical properties of the targets, as provided by \textit{Gaia}\,DR2. We list the apparent brightness $G$, the extinction in the $G$-band $A_{\text{G}}$, the effective temperature $T_{\text{eff}}$, stellar radius $R$, and luminosity $L$ of the target stars (if available). From these parameters together with the distances $d$ by \cite{bailerjones}, the absolute $G$-band brightness $M_{\text{G}}$ of the targets was calculated. The last column yields further remarks, with details listed in the footnote of this table.}
\begin{supertabular}{ccccccccc}
HIP\,410	&  $157 _{-3  }^{+3}   $ & $6.199\pm0.001$ & $0.116_{-0.048}^{+0.048}$ & $ 0.101_{-0.086}^{+0.086}$& $4968_{-40}^{+55}  $ & $11.2_{-0.3 }^{+0.2} $   & $68.4_{-1.4}^{+1.4}$ & c\\
HIP\,544	&  $13.8_{-0.1}^{+0.1} $ & $5.861\pm0.001$ & $0.025_{-0.014}^{+0.172}$ & $ 5.141_{-0.030}^{+0.016}$& $5552_{-46 }^{+195}$ & $0.87_{-0.06}^{+0.02}$   & $0.651_{-0.001}^{+0.001}$& \\
HIP\,695	&  $373 _{-16 }^{+17}  $ & $6.087\pm0.001$ & $0.501_{-0.196}^{+0.196}$ & $-2.271_{-0.292}^{+0.290}$& $4772_{-51 }^{+97 }$ & $31.1_{-1.2 }^{+0.7 }$   & $451  _{-23   }^{+23   }$& c \\
HIP\,926	&  $530 _{-12 }^{+12}  $ & $6.834\pm0.001$ & $0.823_{-0.435}^{+0.767}$ & $-2.610_{-0.485}^{+0.485}$& $4068_{-114}^{+67 }$ & $50.8_{-1.6 }^{+3.0 }$   & $636  _{-21   }^{+21   }$& \\
HIP\,1008	&  $515 _{-15 }^{+16}  $ & $6.869\pm0.001$ & $0.616_{-0.255}^{+1.176}$ & $-2.307_{-0.321}^{+0.320}$& $4175_{-200}^{+149}$ & $44.7_{-3.0 }^{+4.6 }$   & $546  _{-22   }^{+22   }$& \\
HIP\,1118	&  $317 _{-7  }^{+8 }  $ & $6.565\pm0.001$ & $0.467_{-0.403}^{+1.599}$ & $-1.405_{-0.455}^{+0.455}$& $4623_{-65 }^{+167}$ & $23.1_{-1.6 }^{+0.7 }$   & $220  _{-7    }^{+7    }$& \\
HIP\,1352	&  $560 _{-30 }^{+33}  $ & $5.516\pm0.002$ & $0.308_{-0.124}^{+0.124}$ & $-3.533_{-0.245}^{+0.244}$& $3717_{-118}^{+174}$ & $132 _{-12  }^{+9   }$   & $3009 _{-200  }^{+200  }$& c\\
HIP\,1479	&  $657 _{-13 }^{+13}  $ & $7.410\pm0.001$ & $0.511_{-0.206}^{+0.206}$ & $-2.187_{-0.250}^{+0.249}$& $3905_{-185}^{+192}$ & $55.4_{-5.1 }^{+5.6 }$   & $644  _{-21   }^{+21   }$& c\\
\end{supertabular}
\label{tab:gaia_infos}
\end{center}
\hspace{10pt}a - Hipparcos parallax \citep{vanleeuwen} was used for distance determination\\
b - $V$-, and $I$-band magnitude from Hipparcos \citep{perryman} were transformed into $G$-band magnitude \\
\hspace{12pt}according to \cite{jordi} \\
c - median $A_{\text{V}}$ value taken from literature and converted to $A_{\text{G}}$ \\
d - $T_{\text{eff}}$ \citep{damiani} according to the Hipparcos spectral type\\
e - $R$ and $L$ from \textit{Gaia} rejected because no gold/silver flag target\\
f - spectroscopic binary of two equal mass stars (therefore about 0.75\,mag fainter compared to \textit{Gaia}\,DR2 entry)   \\
\newpage
\twocolumn

\section{Target spectra - dwarfs}\label{app2}

\includegraphics[width=11cm,height=20.5cm,keepaspectratio]{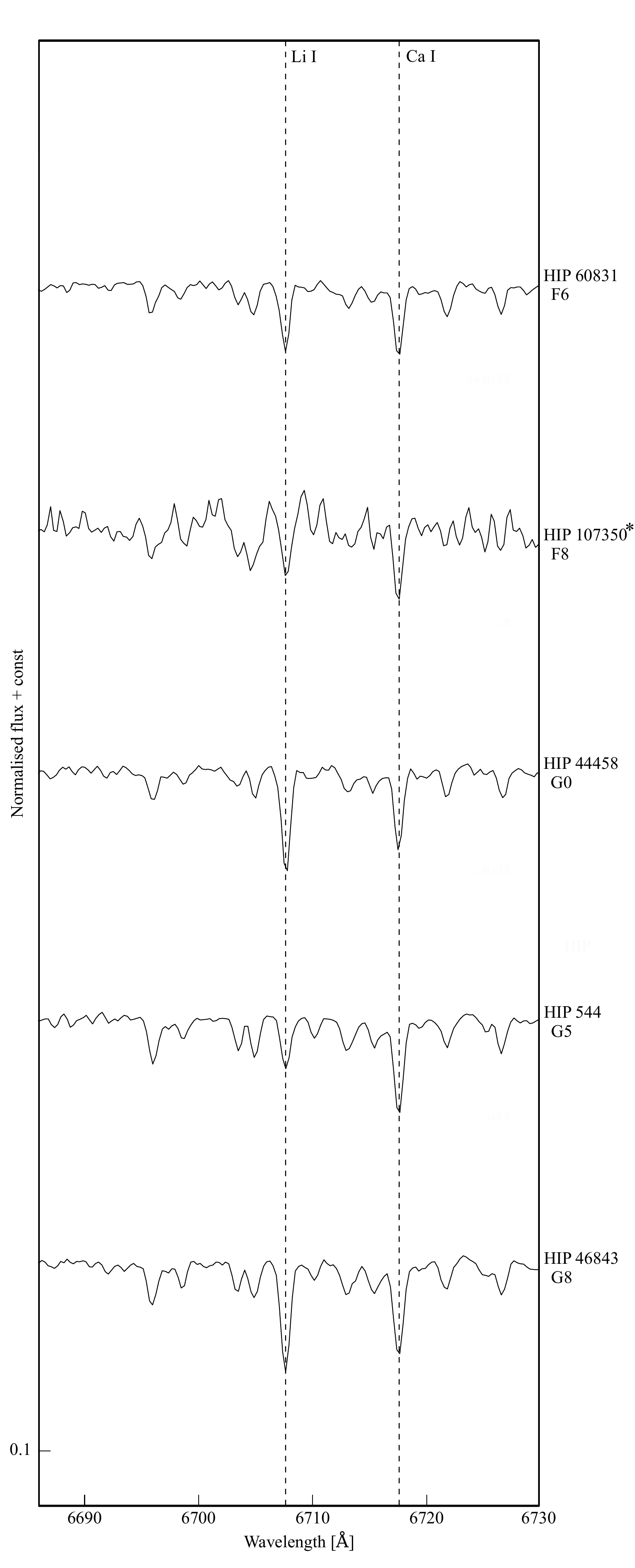}
\\$^{*}$ This spectrum has only a signal-to-noise-ratio of 39.

\newpage
\vspace*{15pt}
\includegraphics[width=11cm,height=20.5cm,keepaspectratio]{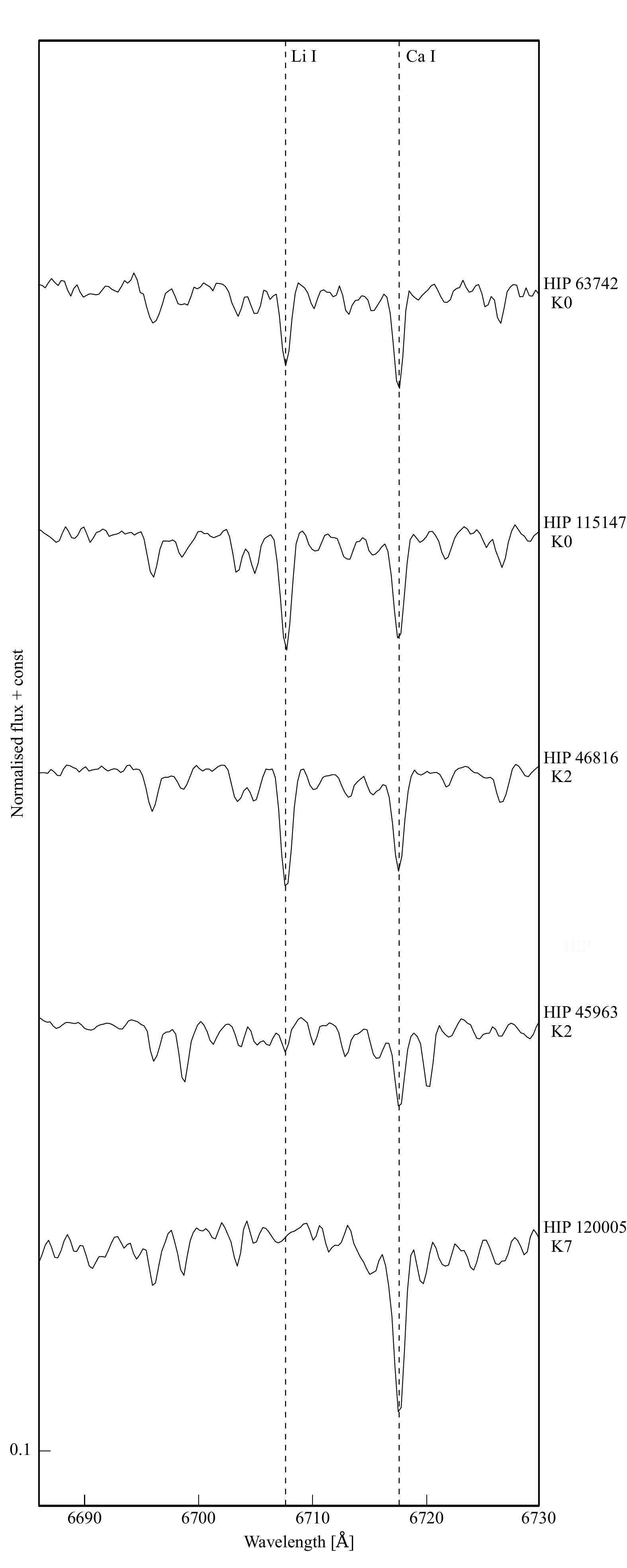}

\newpage

\section{Target spectra - giants}\label{app3}

\includegraphics[width=11cm,height=20.5cm,keepaspectratio]{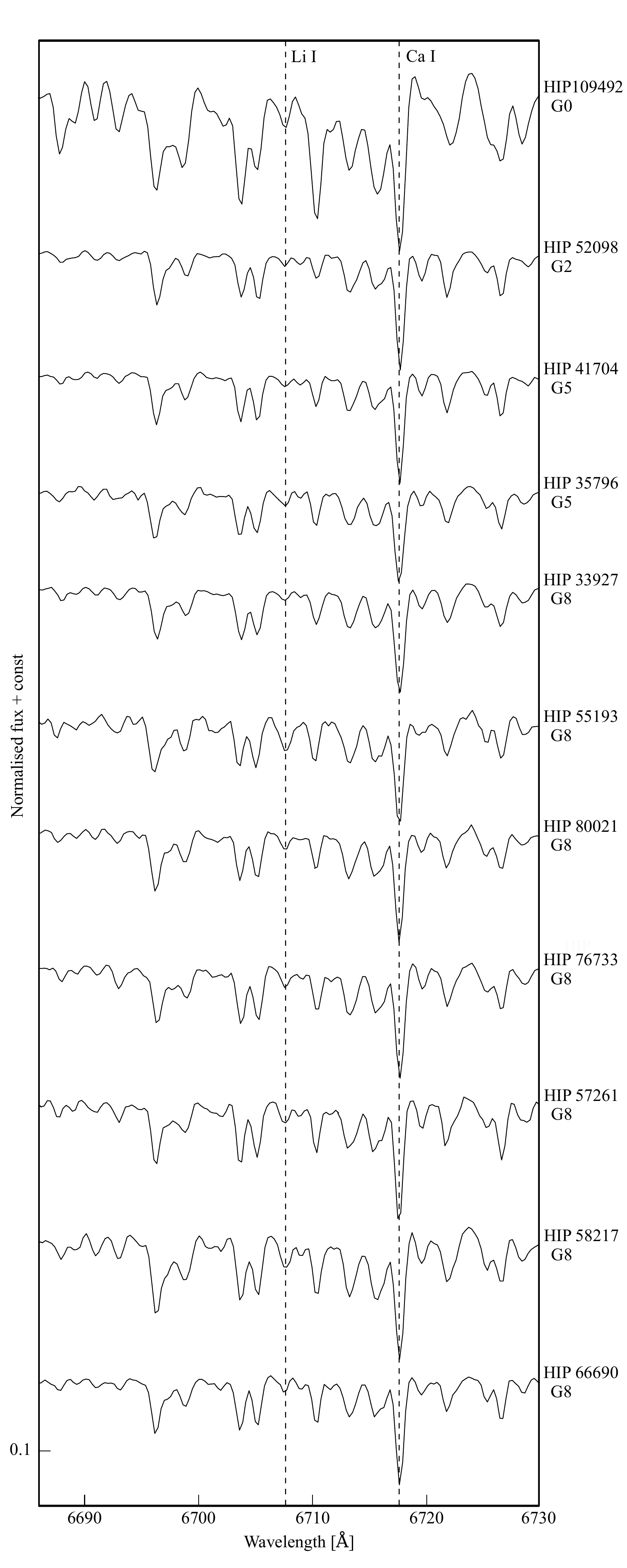}
\newpage
\vspace*{15pt}
\includegraphics[width=11cm,height=20.5cm,keepaspectratio]{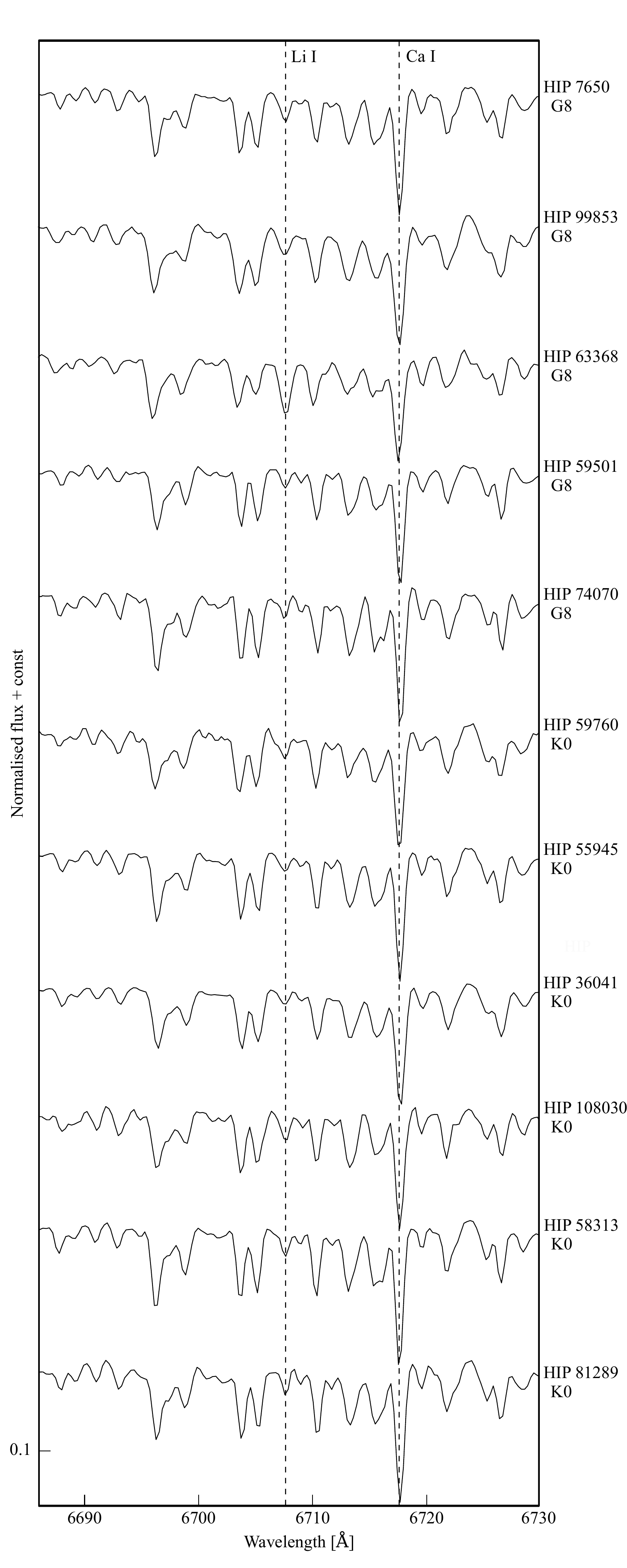}
\newpage

\includegraphics[width=11cm,height=20.5cm,keepaspectratio]{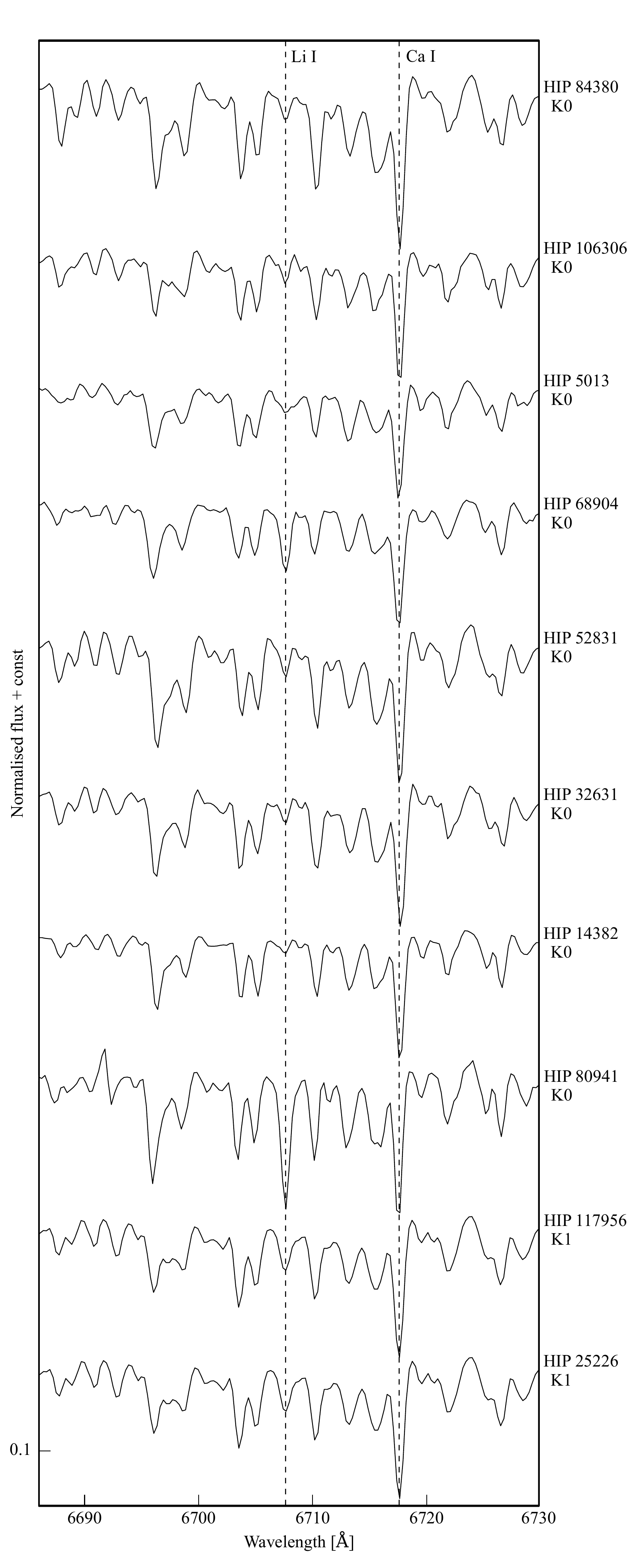}
\newpage
\includegraphics[width=11cm,height=20.5cm,keepaspectratio]{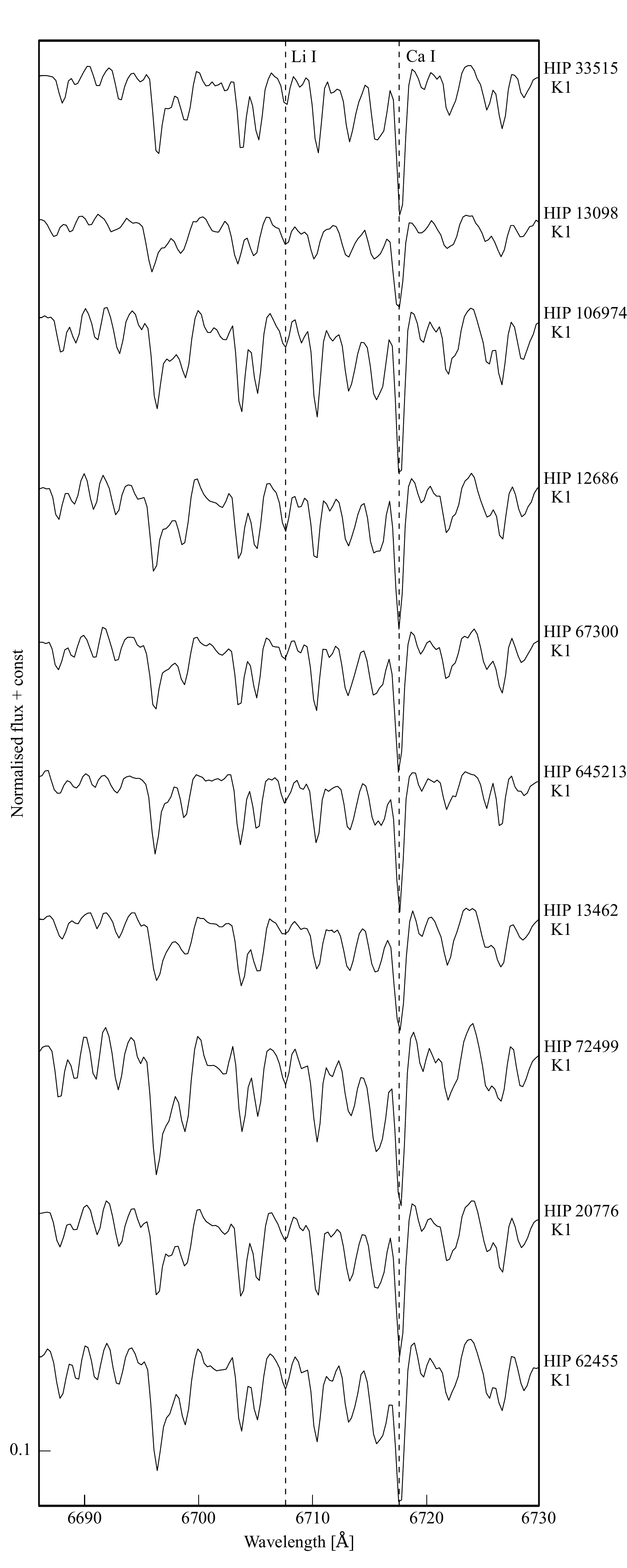}
\newpage

\includegraphics[width=11cm,height=20.5cm,keepaspectratio]{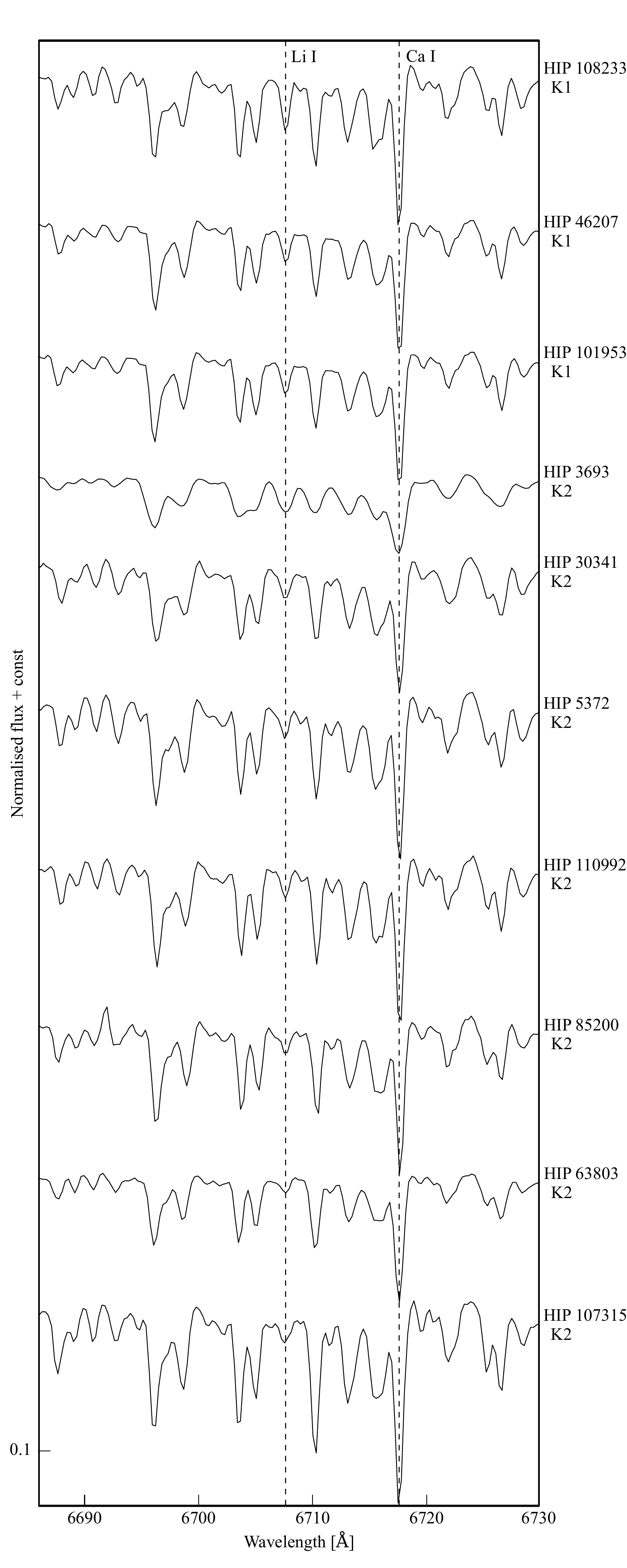}
\newpage
\includegraphics[width=11cm,height=20.5cm,keepaspectratio]{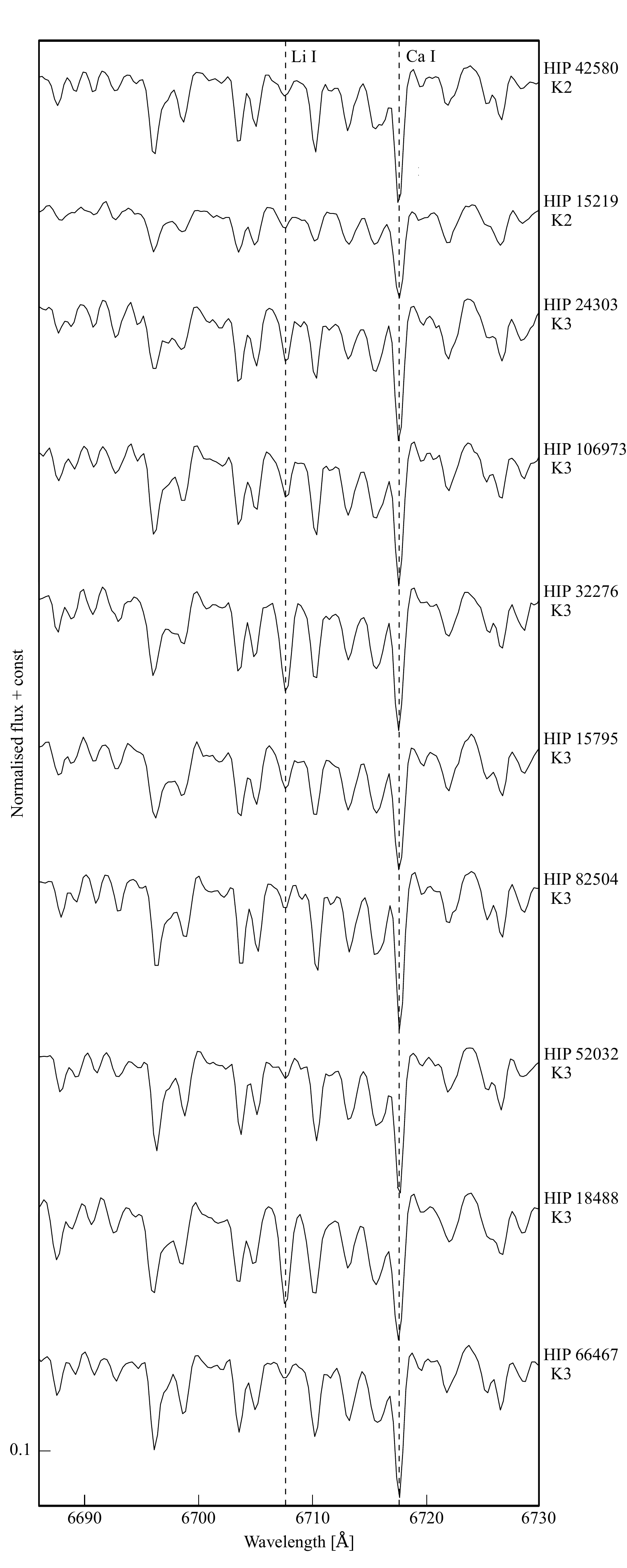}
\newpage

\includegraphics[width=11cm,height=20.5cm,keepaspectratio]{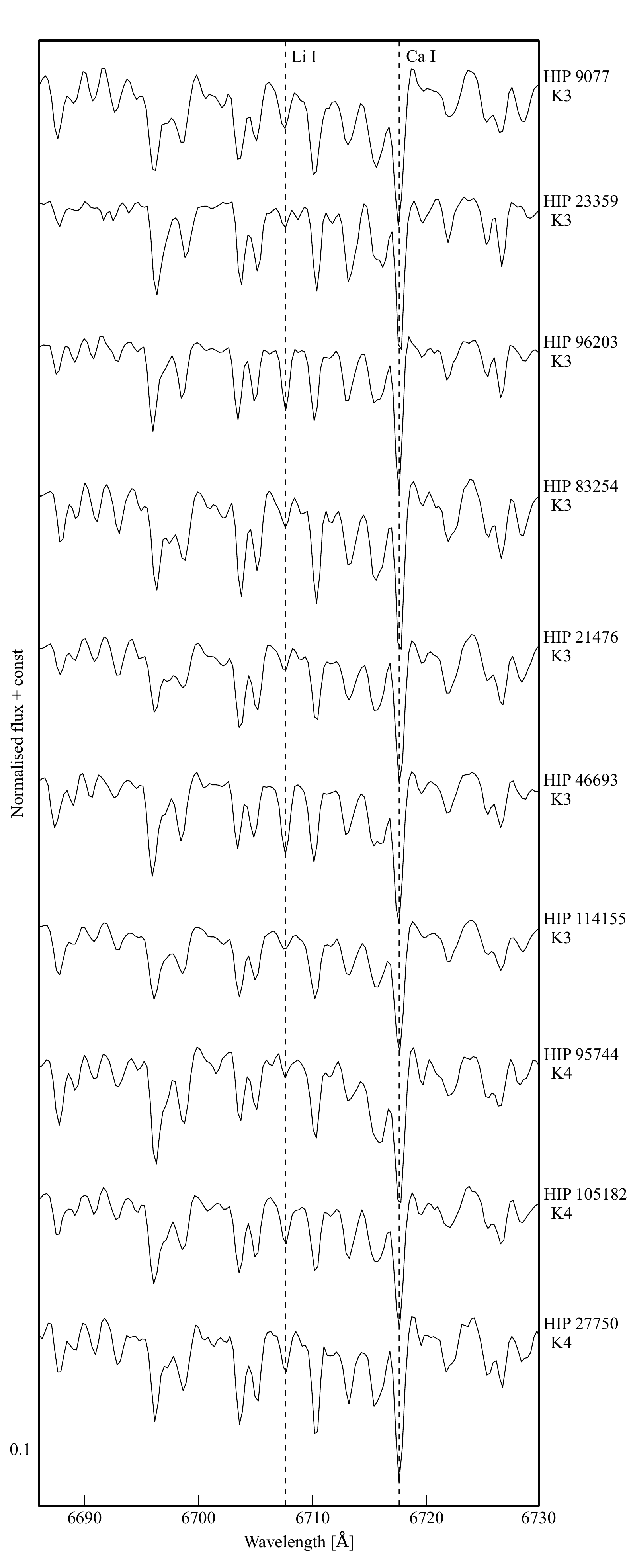}
\newpage
\includegraphics[width=11cm,height=20.5cm,keepaspectratio]{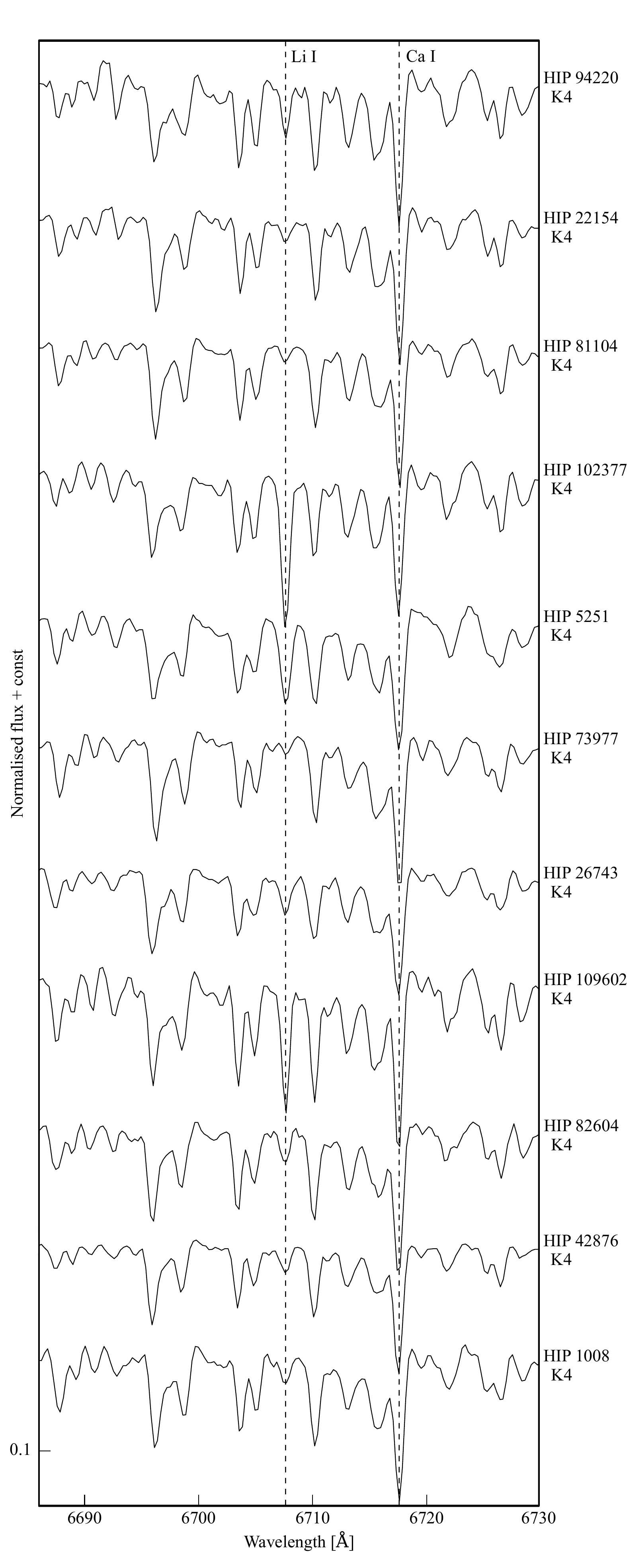}
\newpage

\includegraphics[width=11cm,height=20.5cm,keepaspectratio]{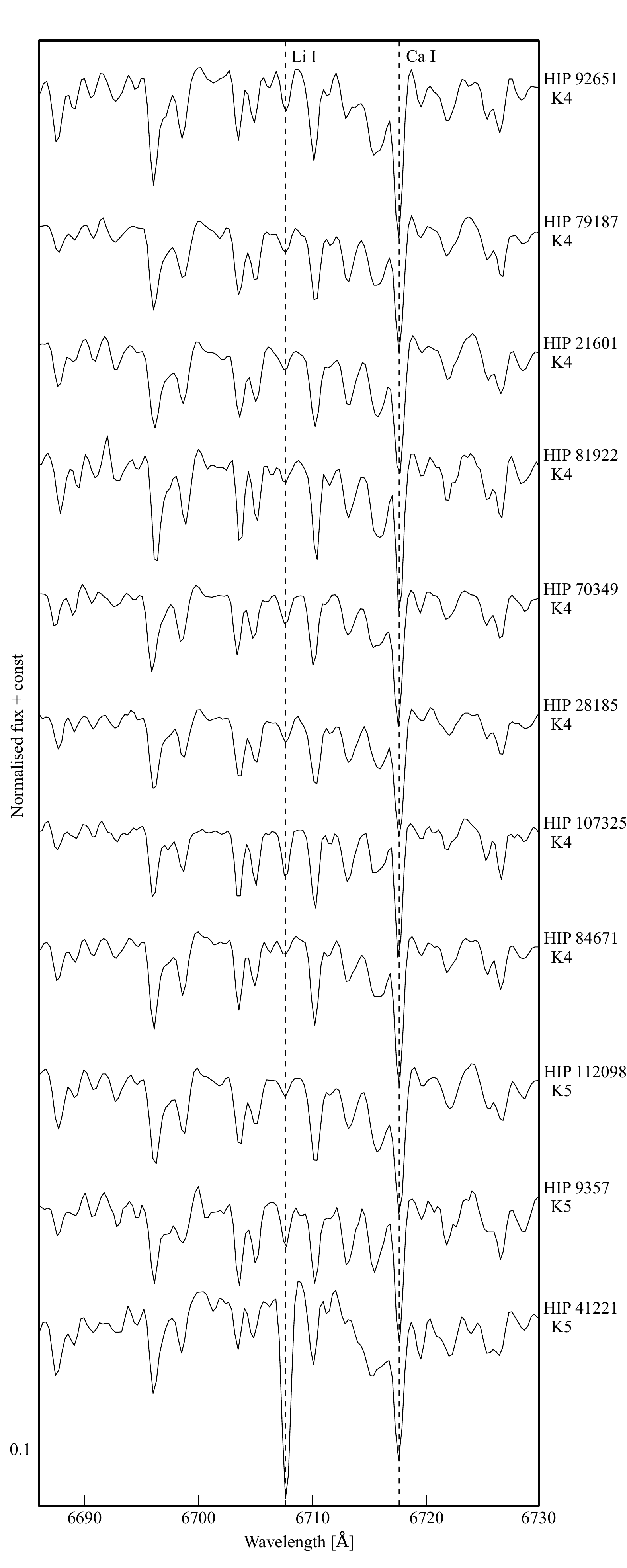}
\newpage
\includegraphics[width=11cm,height=20.5cm,keepaspectratio]{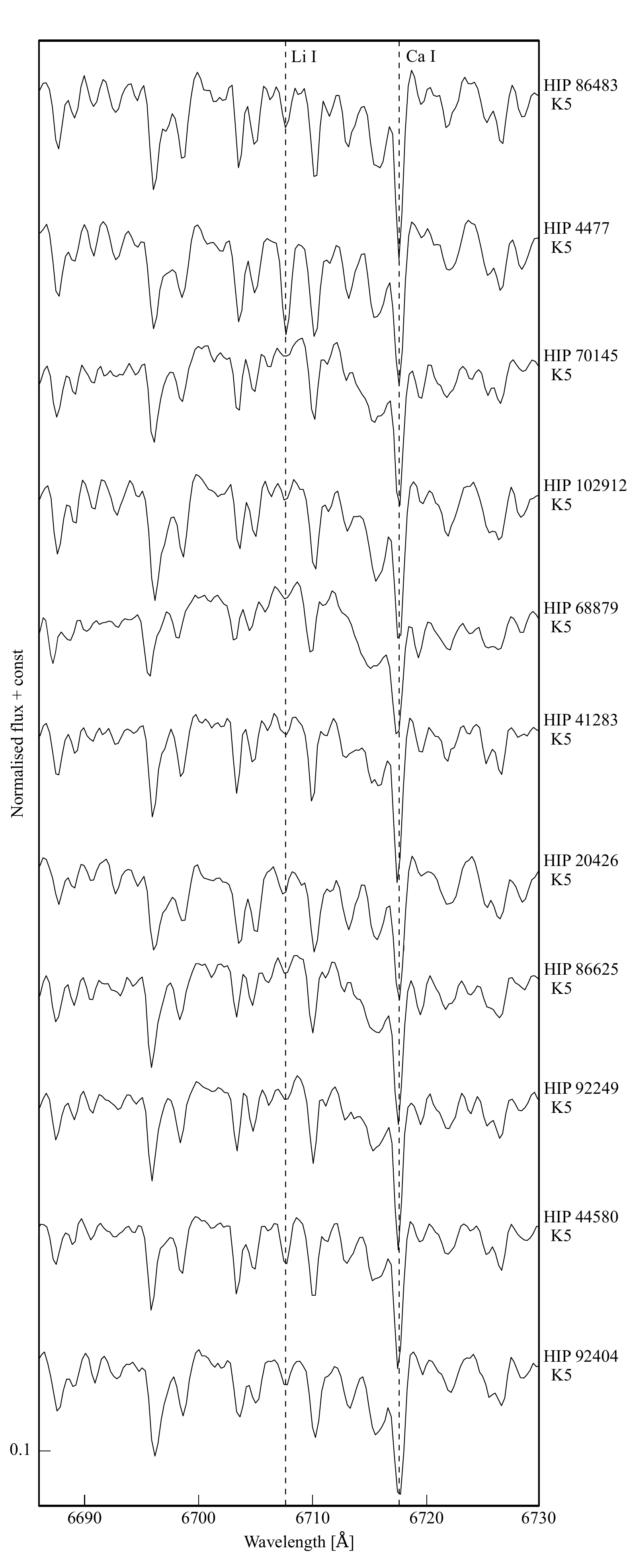}
\newpage

\includegraphics[width=11cm,height=20.5cm,keepaspectratio]{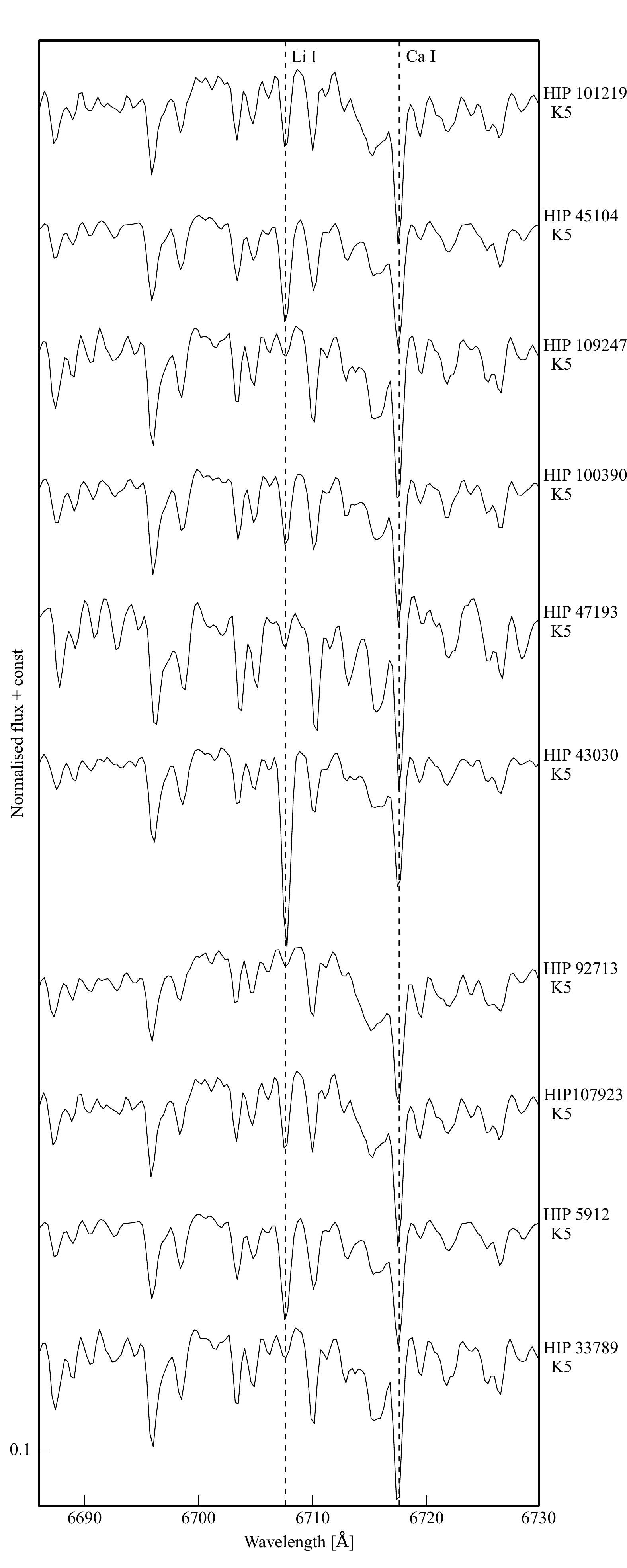}
\newpage
\includegraphics[width=11cm,height=20.5cm,keepaspectratio]{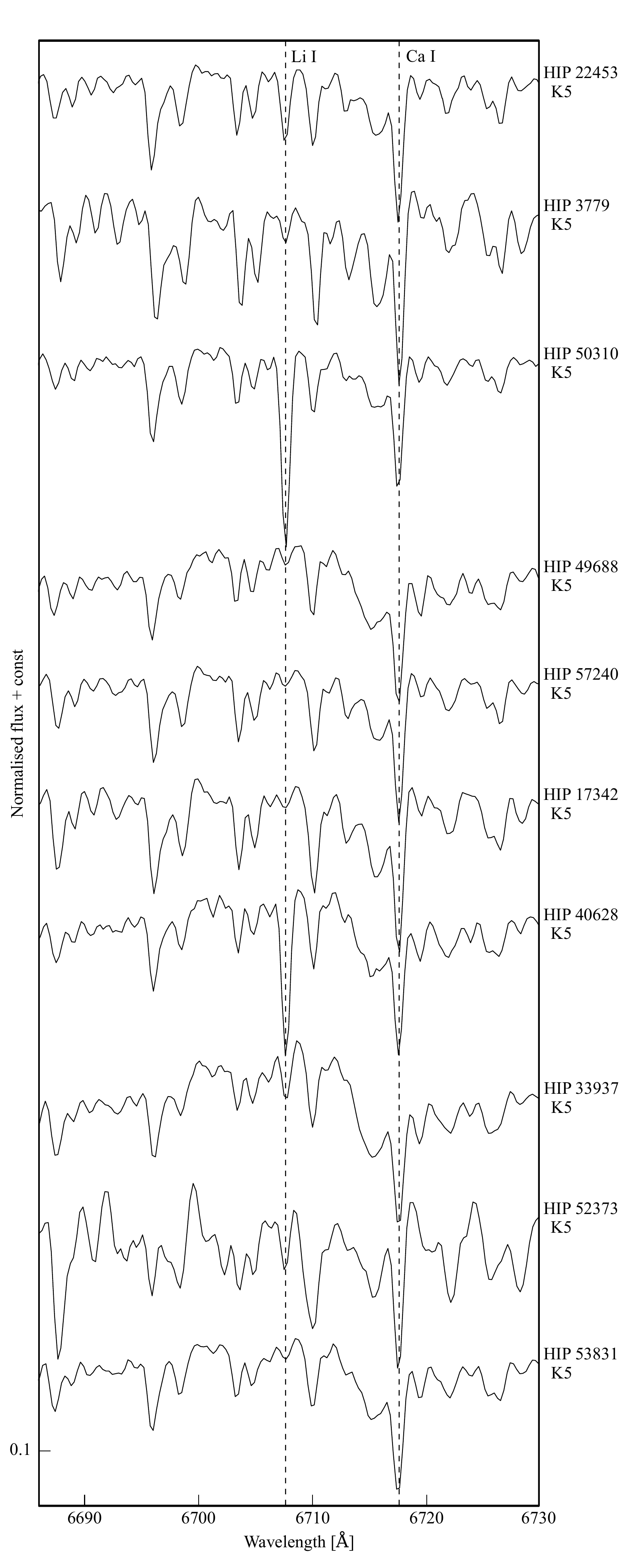}
\newpage

\includegraphics[width=11cm,height=20.5cm,keepaspectratio]{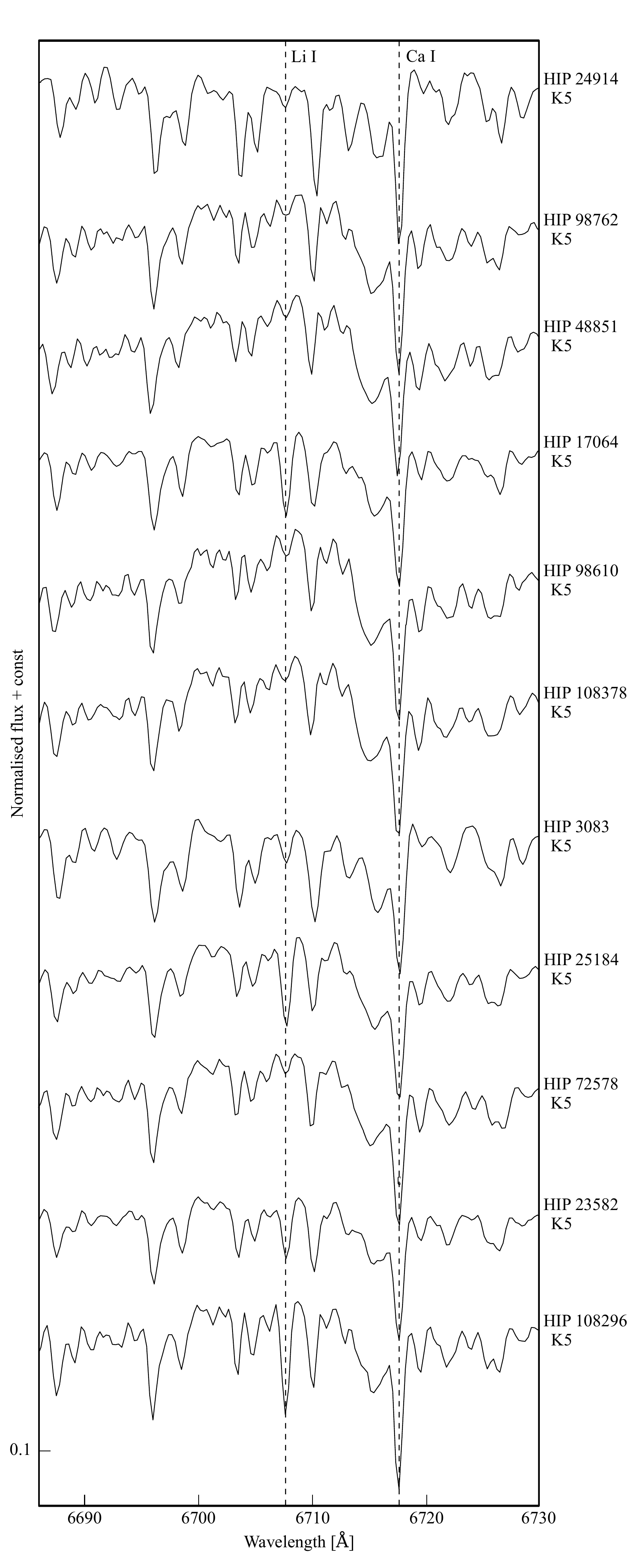}
\newpage
\includegraphics[width=11cm,height=20.5cm,keepaspectratio]{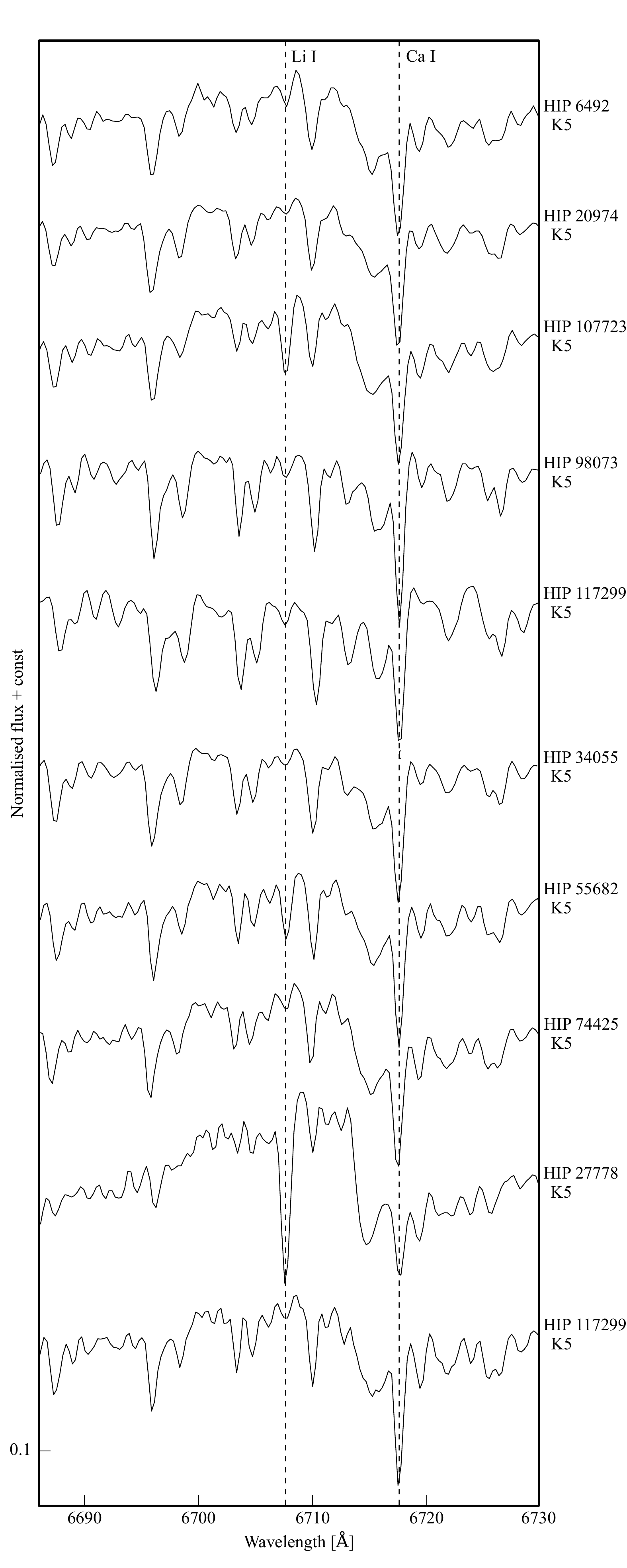}
\newpage

\includegraphics[width=11cm,height=20.5cm,keepaspectratio]{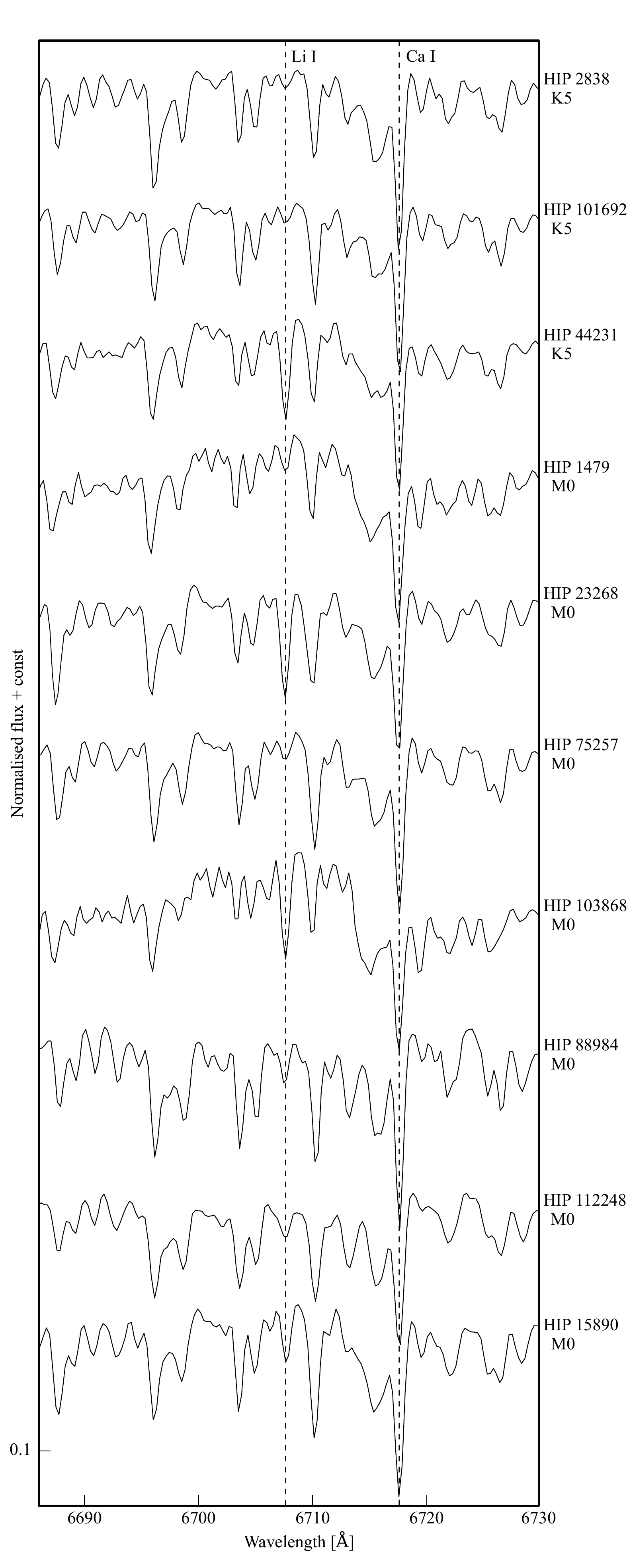}
\newpage
\includegraphics[width=11cm,height=20.5cm,keepaspectratio]{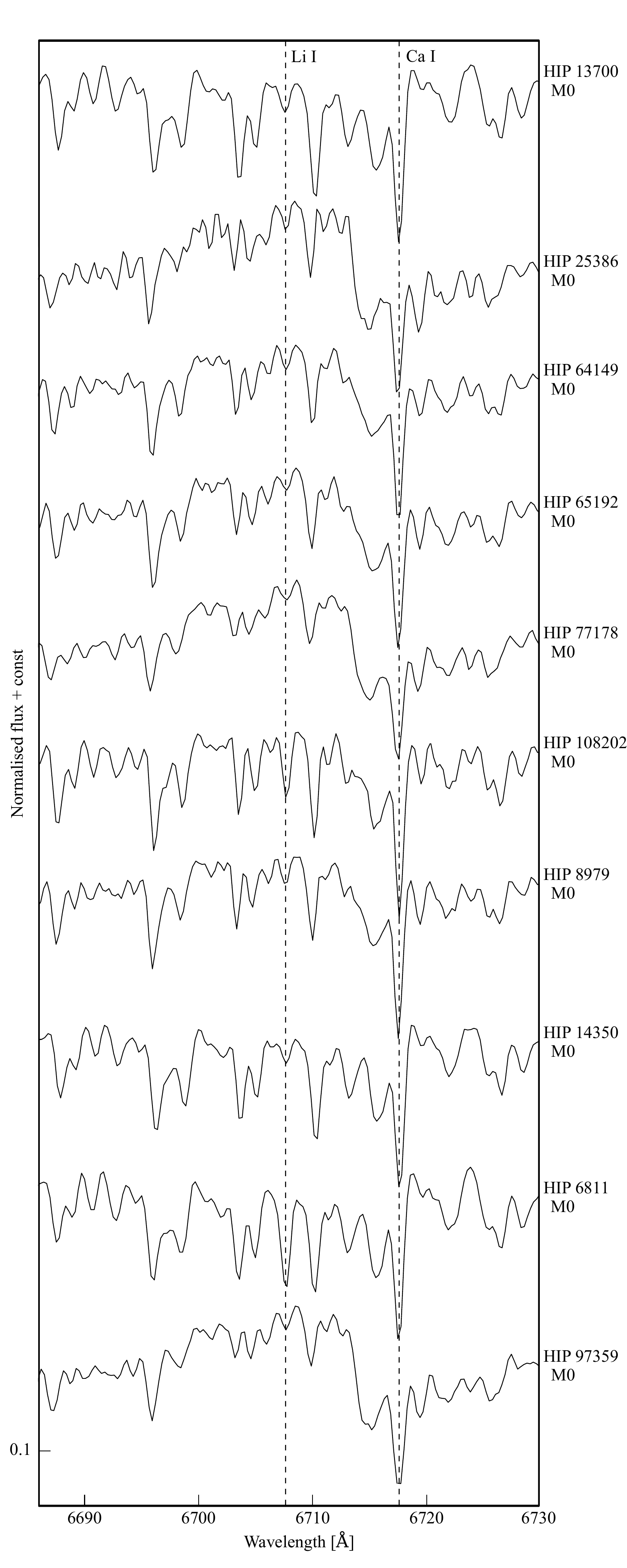}
\newpage

\includegraphics[width=11cm,height=20.5cm,keepaspectratio]{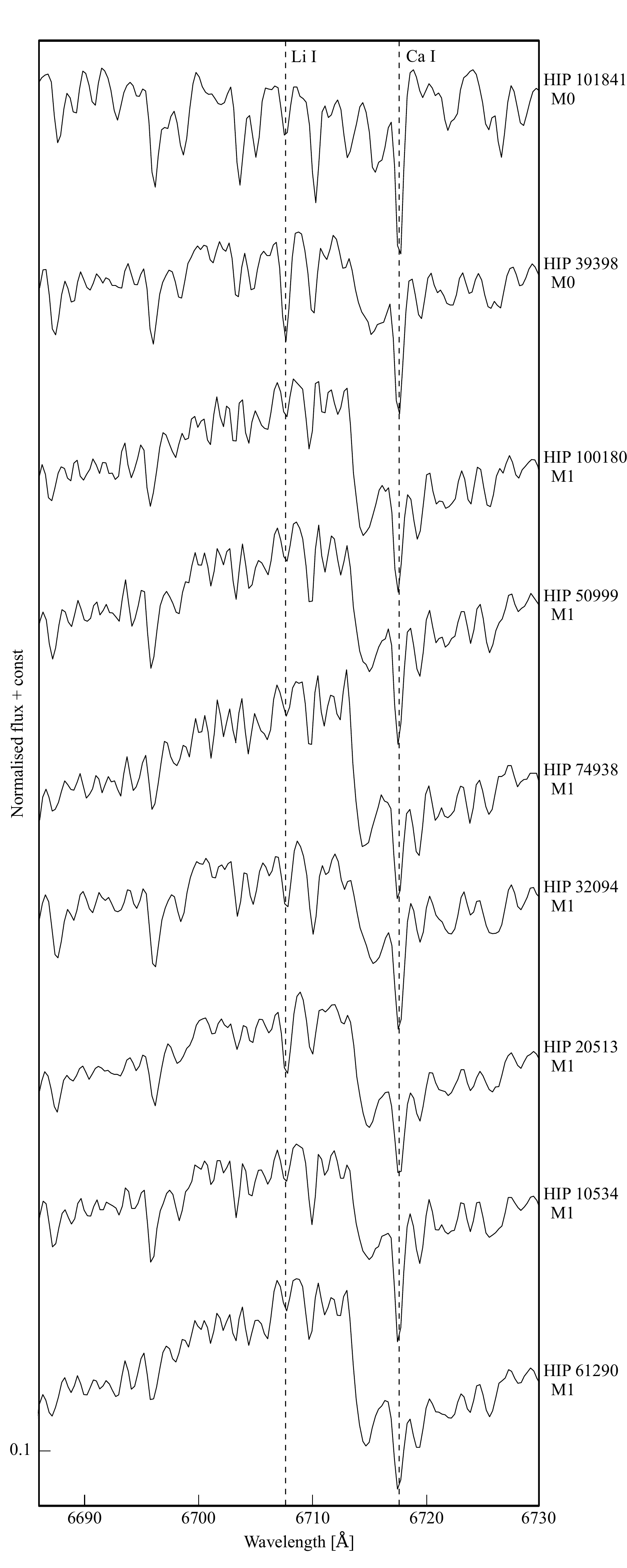}
\newpage
\includegraphics[width=11cm,height=20.5cm,keepaspectratio]{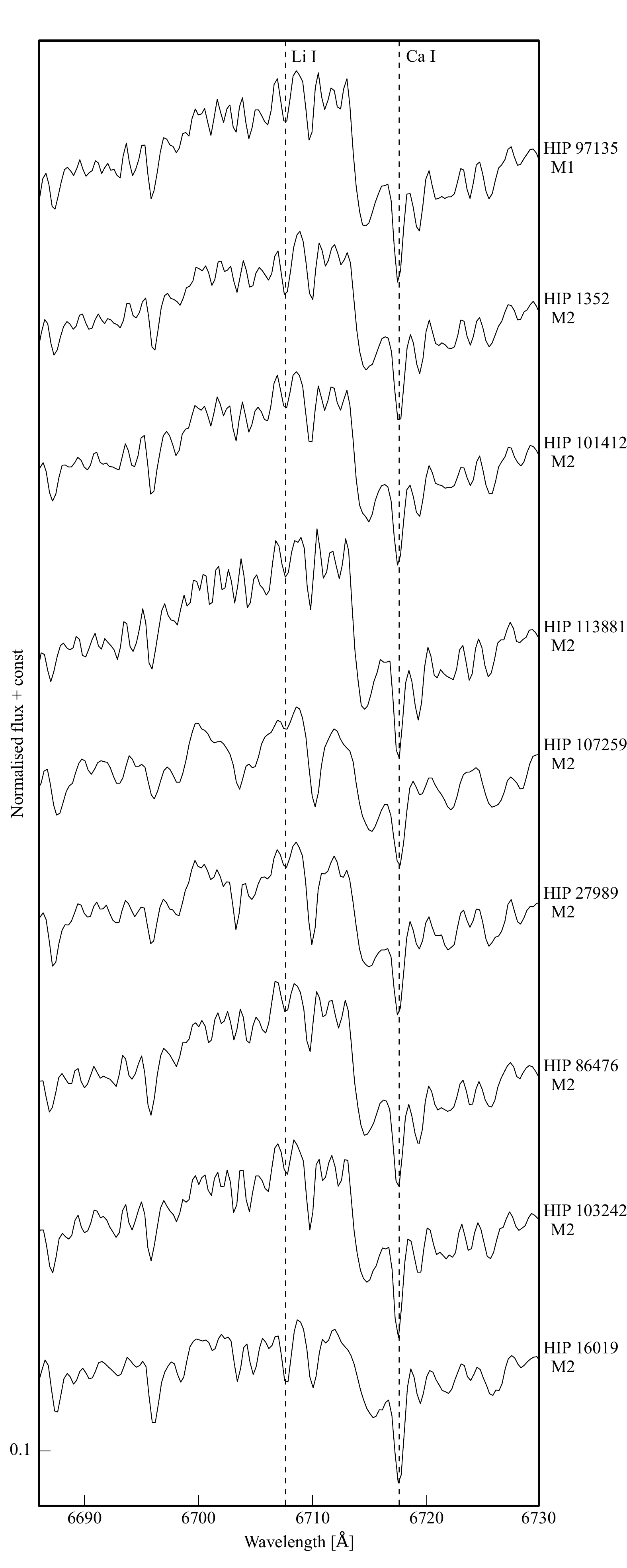}
\newpage

\includegraphics[width=11cm,height=20.5cm,keepaspectratio]{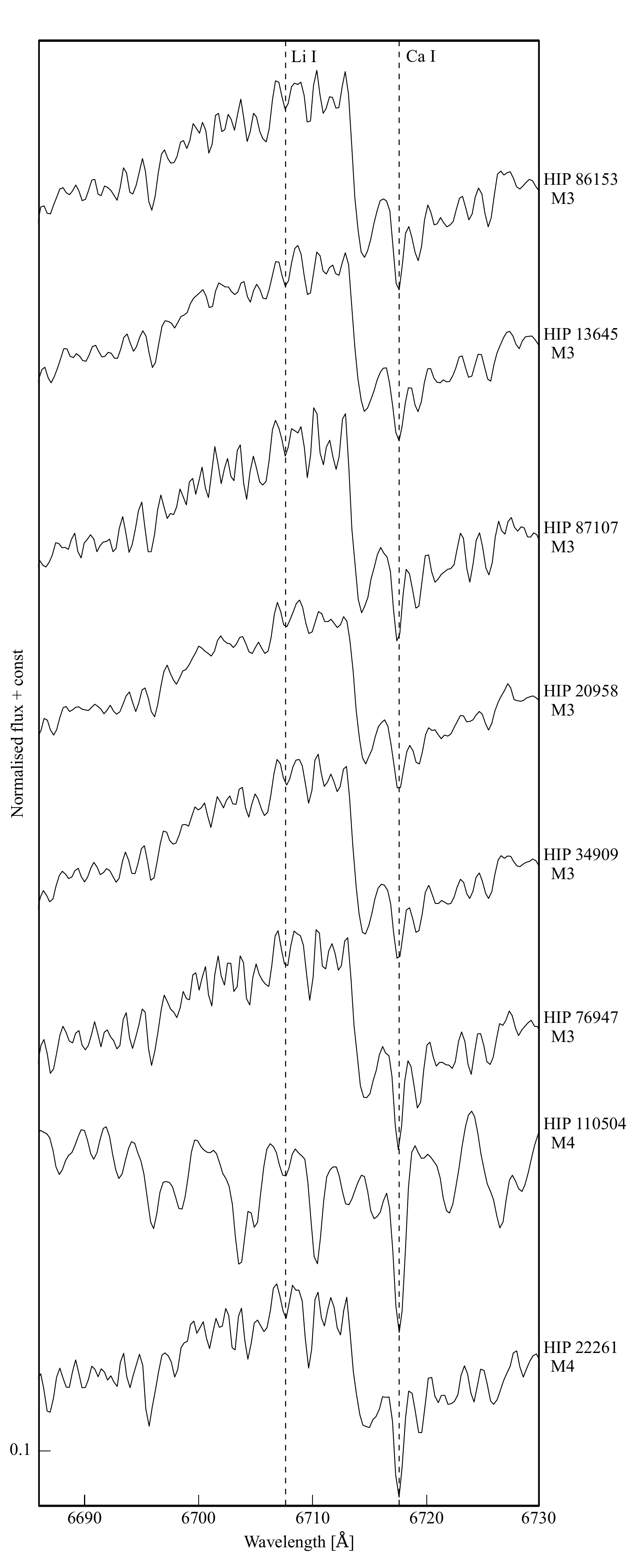}
\newpage
\includegraphics[width=11cm,height=20.5cm,keepaspectratio]{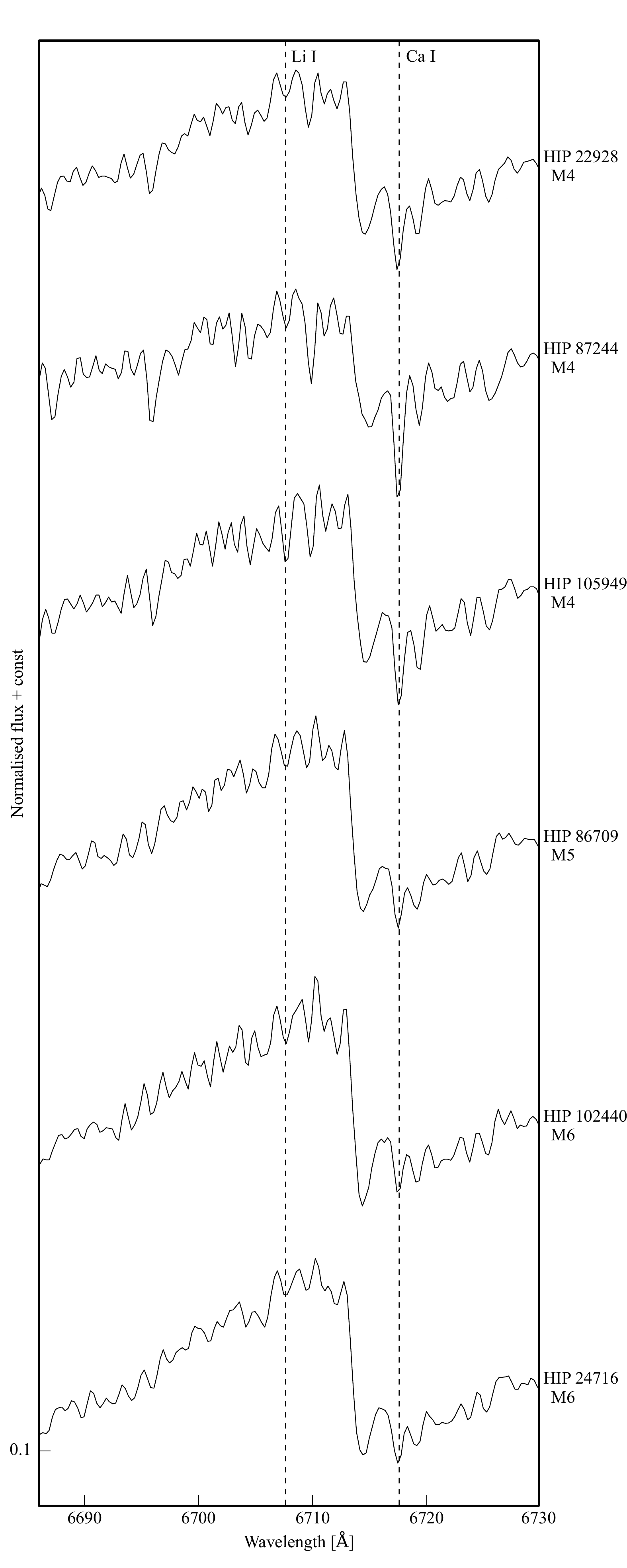}
\newpage

\nocite{*}

\bibliography{bischoff}

\begin{thebibliography}{}

\bibitem [\protect \citeauthoryear {%
{Anders}%
\ \protect \BOthers {.}}{%
{Anders}%
\ \protect \BOthers {.}}{%
{\protect \APACyear {2019}}%
}]{%
anders}
\APACinsertmetastar {%
anders}%
\begin{APACrefauthors}%
{Anders}, F.%
, {Khalatyan}, A.%
, {Chiappini}, C.%
\ et al.\end{APACrefauthors}%
\unskip\
\newblock
\APACrefYearMonthDay{2019}{{\APACmonth{08}}}{},
\newblock
\unskip
\newblock
\APACjournalVolNumPages{\aap}{628}{}{A94}.
\PrintBackRefs{\CurrentBib}

\bibitem [\protect \citeauthoryear {%
{Andrae}%
\ \protect \BOthers {.}}{%
{Andrae}%
\ \protect \BOthers {.}}{%
{\protect \APACyear {2018}}%
}]{%
andrae2018}
\APACinsertmetastar {%
andrae2018}%
\begin{APACrefauthors}%
{Andrae}, R.%
, {Fouesneau}, M.%
, {Creevey}, O.%
\ et al.\end{APACrefauthors}%
\unskip\
\newblock
\APACrefYearMonthDay{2018}{Aug}{},
\newblock
\unskip
\newblock
\APACjournalVolNumPages{\aap}{616}{}{A8}.
\PrintBackRefs{\CurrentBib}

\bibitem [\protect \citeauthoryear {%
{Bailer-Jones}%
, {Rybizki}%
, {Fouesneau}%
, {Mantelet}%
\BCBL {}\ \BBA {} {Andrae}%
}{%
{Bailer-Jones}%
\ \protect \BOthers {.}}{%
{\protect \APACyear {2018}}%
}]{%
bailerjones}
\APACinsertmetastar {%
bailerjones}%
\begin{APACrefauthors}%
{Bailer-Jones}, C\BPBI A\BPBI L.%
, {Rybizki}, J.%
, {Fouesneau}, M.%
, {Mantelet}, G.%
\BCBL {}\ \BBA {} {Andrae}, R.%
\end{APACrefauthors}%
\unskip\
\newblock
\APACrefYearMonthDay{2018}{Aug}{},
\newblock
\unskip
\newblock
\APACjournalVolNumPages{\aj}{156}{2}{58}.
\PrintBackRefs{\CurrentBib}

\bibitem [\protect \citeauthoryear {%
{Bell}%
, {Mamajek}%
\BCBL {}\ \BBA {} {Naylor}%
}{%
{Bell}%
\ \protect \BOthers {.}}{%
{\protect \APACyear {2015}}%
}]{%
bell}
\APACinsertmetastar {%
bell}%
\begin{APACrefauthors}%
{Bell}, C\BPBI P\BPBI M.%
, {Mamajek}, E\BPBI E.%
\BCBL {}\ \BBA {} {Naylor}, T.%
\end{APACrefauthors}%
\unskip\
\newblock
\APACrefYearMonthDay{2015}{Nov}{},
\newblock
\unskip
\newblock
\APACjournalVolNumPages{\mnras}{454}{1}{593-614}.
\PrintBackRefs{\CurrentBib}

\bibitem [\protect \citeauthoryear {%
{Binks}%
\ \BBA {} {Jeffries}%
}{%
{Binks}%
\ \BBA {} {Jeffries}%
}{%
{\protect \APACyear {2014}}%
}]{%
binks}
\APACinsertmetastar {%
binks}%
\begin{APACrefauthors}%
{Binks}, A\BPBI S.%
\BCBT {}\ \BBA {} {Jeffries}, R\BPBI D.%
\end{APACrefauthors}%
\unskip\
\newblock
\APACrefYearMonthDay{2014}{Feb}{},
\newblock
\unskip
\newblock
\APACjournalVolNumPages{\mnras}{438}{1}{L11-L15}.
\PrintBackRefs{\CurrentBib}

\bibitem [\protect \citeauthoryear {%
{Bischoff}%
\ \protect \BOthers {.}}{%
{Bischoff}%
\ \protect \BOthers {.}}{%
{\protect \APACyear {2017}}%
}]{%
bischoff}
\APACinsertmetastar {%
bischoff}%
\begin{APACrefauthors}%
{Bischoff}, R.%
, {Mugrauer}, M.%
, {Zehe}, T.%
\ et al.\end{APACrefauthors}%
\unskip\
\newblock
\APACrefYearMonthDay{2017}{Jul}{},
\newblock
\unskip
\newblock
\APACjournalVolNumPages{Astronomische Nachrichten}{338}{6}{671-679}.
\PrintBackRefs{\CurrentBib}

\bibitem [\protect \citeauthoryear {%
{Blaauw}%
}{%
{Blaauw}%
}{%
{\protect \APACyear {1961}}%
}]{%
blaauw}
\APACinsertmetastar {%
blaauw}%
\begin{APACrefauthors}%
{Blaauw}, A.%
\end{APACrefauthors}%
\unskip\
\newblock
\APACrefYearMonthDay{1961}{May}{},
\newblock
\unskip
\newblock
\APACjournalVolNumPages{\bain}{15}{}{265}.
\PrintBackRefs{\CurrentBib}

\bibitem [\protect \citeauthoryear {%
{Brandt}%
\ \protect \BOthers {.}}{%
{Brandt}%
\ \protect \BOthers {.}}{%
{\protect \APACyear {2017}}%
}]{%
brandt}
\APACinsertmetastar {%
brandt}%
\begin{APACrefauthors}%
{Brandt}, T\BPBI D.%
, {McElwain}, M\BPBI W.%
, {Turner}, E\BPBI L.%
\ et al.\end{APACrefauthors}%
\unskip\
\newblock
\APACrefYearMonthDay{2017}{May}{},
\newblock
\unskip
\newblock
\APACjournalVolNumPages{VizieR Online Data Catalog}{}{}{J/ApJ/794/159}.
\PrintBackRefs{\CurrentBib}

\bibitem [\protect \citeauthoryear {%
{Bressan}%
\ \protect \BOthers {.}}{%
{Bressan}%
\ \protect \BOthers {.}}{%
{\protect \APACyear {2012}}%
}]{%
bressan}
\APACinsertmetastar {%
bressan}%
\begin{APACrefauthors}%
{Bressan}, A.%
, {Marigo}, P.%
, {Girardi}, L.%
, {Salasnich}, B.%
, {Dal Cero}, C.%
, {Rubele}, S.%
\BCBL {}\ \BBA {} {Nanni}, A.%
\end{APACrefauthors}%
\unskip\
\newblock
\APACrefYearMonthDay{2012}{{\APACmonth{11}}}{},
\newblock
\unskip
\newblock
\APACjournalVolNumPages{\mnras}{427}{1}{127-145}.
\PrintBackRefs{\CurrentBib}

\bibitem [\protect \citeauthoryear {%
{Brewer}%
, {Fischer}%
, {Valenti}%
\BCBL {}\ \BBA {} {Piskunov}%
}{%
{Brewer}%
\ \protect \BOthers {.}}{%
{\protect \APACyear {2016}}%
}]{%
brewer}
\APACinsertmetastar {%
brewer}%
\begin{APACrefauthors}%
{Brewer}, J\BPBI M.%
, {Fischer}, D\BPBI A.%
, {Valenti}, J\BPBI A.%
\BCBL {}\ \BBA {} {Piskunov}, N.%
\end{APACrefauthors}%
\unskip\
\newblock
\APACrefYearMonthDay{2016}{{\APACmonth{08}}}{},
\newblock
\unskip
\newblock
\APACjournalVolNumPages{\apjs}{225}{2}{32}.
\PrintBackRefs{\CurrentBib}

\bibitem [\protect \citeauthoryear {%
{Casagrande}%
\ \protect \BOthers {.}}{%
{Casagrande}%
\ \protect \BOthers {.}}{%
{\protect \APACyear {2011}}%
}]{%
cassagrande2011}
\APACinsertmetastar {%
cassagrande2011}%
\begin{APACrefauthors}%
{Casagrande}, L.%
, {Sch{\"o}nrich}, R.%
, {Asplund}, M.%
\ et al.\end{APACrefauthors}%
\unskip\
\newblock
\APACrefYearMonthDay{2011}{{\APACmonth{06}}}{},
\newblock
\unskip
\newblock
\APACjournalVolNumPages{\aap}{530}{}{A138}.
\PrintBackRefs{\CurrentBib}

\bibitem [\protect \citeauthoryear {%
{Coluzzi}%
}{%
{Coluzzi}%
}{%
{\protect \APACyear {1993}}%
}]{%
coluzzi}
\APACinsertmetastar {%
coluzzi}%
\begin{APACrefauthors}%
{Coluzzi}, R.%
\end{APACrefauthors}%
\unskip\
\newblock
\APACrefYearMonthDay{1993}{Jul}{},
\newblock
\unskip
\newblock
\APACjournalVolNumPages{Bulletin d'Information du Centre de Donnees
  Stellaires}{43}{}{7}.
\PrintBackRefs{\CurrentBib}

\bibitem [\protect \citeauthoryear {%
{Damiani}%
\ \protect \BOthers {.}}{%
{Damiani}%
\ \protect \BOthers {.}}{%
{\protect \APACyear {2016}}%
}]{%
damiani}
\APACinsertmetastar {%
damiani}%
\begin{APACrefauthors}%
{Damiani}, C.%
, {Meunier}, J\BPBI C.%
, {Moutou}, C.%
, {Deleuil}, M.%
, {Ysard}, N.%
, {Baudin}, F.%
\BCBL {}\ \BBA {} {Deeg}, H.%
\end{APACrefauthors}%
\unskip\
\newblock
\APACrefYearMonthDay{2016}{Nov}{},
\newblock
\unskip
\newblock
\APACjournalVolNumPages{\aap}{595}{}{A95}.
\PrintBackRefs{\CurrentBib}

\bibitem [\protect \citeauthoryear {%
{Fekel}%
, {Tomkin}%
\BCBL {}\ \BBA {} {Williamson}%
}{%
{Fekel}%
\ \protect \BOthers {.}}{%
{\protect \APACyear {2010}}%
}]{%
fekel}
\APACinsertmetastar {%
fekel}%
\begin{APACrefauthors}%
{Fekel}, F\BPBI C.%
, {Tomkin}, J.%
\BCBL {}\ \BBA {} {Williamson}, M\BPBI H.%
\end{APACrefauthors}%
\unskip\
\newblock
\APACrefYearMonthDay{2010}{Apr}{},
\newblock
\unskip
\newblock
\APACjournalVolNumPages{\aj}{139}{4}{1579-1591}.
\PrintBackRefs{\CurrentBib}

\bibitem [\protect \citeauthoryear {%
F\H{u}r\'esz%
}{%
F\H{u}r\'esz%
}{%
{\protect \APACyear {2008}}%
}]{%
furesz2008}
\APACinsertmetastar {%
furesz2008}%
\begin{APACrefauthors}%
F\H{u}r\'esz, G.%
\end{APACrefauthors}%
\unskip\
\newblock
\APACrefYear{2008}.
\unskip\
\newblock
\APACtypeAddressSchool {{PhD} dissertation}{}{}.
\unskip\
\newblock
\APACaddressSchool {}{University of Szeged, Hungary}.
\PrintBackRefs{\CurrentBib}

\bibitem [\protect \citeauthoryear {%
{Franchini}%
, {Morossi}%
, {di Marcantonio}%
, {Malagnini}%
\BCBL {}\ \BBA {} {Chavez}%
}{%
{Franchini}%
\ \protect \BOthers {.}}{%
{\protect \APACyear {2014}}%
}]{%
franchini}
\APACinsertmetastar {%
franchini}%
\begin{APACrefauthors}%
{Franchini}, M.%
, {Morossi}, C.%
, {di Marcantonio}, P.%
, {Malagnini}, M\BPBI L.%
\BCBL {}\ \BBA {} {Chavez}, M.%
\end{APACrefauthors}%
\unskip\
\newblock
\APACrefYearMonthDay{2014}{{\APACmonth{07}}}{},
\newblock
\unskip
\newblock
\APACjournalVolNumPages{\mnras}{442}{1}{220-228}.
\PrintBackRefs{\CurrentBib}

\bibitem [\protect \citeauthoryear {%
{Gaia Collaboration}%
\ \protect \BOthers {.}}{%
{Gaia Collaboration}%
\ \protect \BOthers {.}}{%
{\protect \APACyear {2018}}%
}]{%
gaiadr2}
\APACinsertmetastar {%
gaiadr2}%
\begin{APACrefauthors}%
{Gaia Collaboration}%
, {Brown}, A\BPBI G\BPBI A.%
, {Vallenari}, A.%
\ et al.\end{APACrefauthors}%
\unskip\
\newblock
\APACrefYearMonthDay{2018}{Aug}{},
\newblock
\unskip
\newblock
\APACjournalVolNumPages{\aap}{616}{}{A1}.
\PrintBackRefs{\CurrentBib}

\bibitem [\protect \citeauthoryear {%
{Gazzano}%
\ \protect \BOthers {.}}{%
{Gazzano}%
\ \protect \BOthers {.}}{%
{\protect \APACyear {2010}}%
}]{%
gazzano}
\APACinsertmetastar {%
gazzano}%
\begin{APACrefauthors}%
{Gazzano}, J\BPBI C.%
, {de Laverny}, P.%
, {Deleuil}, M.%
\ et al.\end{APACrefauthors}%
\unskip\
\newblock
\APACrefYearMonthDay{2010}{{\APACmonth{11}}}{},
\newblock
\unskip
\newblock
\APACjournalVolNumPages{\aap}{523}{}{A91}.
\PrintBackRefs{\CurrentBib}

\bibitem [\protect \citeauthoryear {%
{Gray}%
\ \protect \BOthers {.}}{%
{Gray}%
\ \protect \BOthers {.}}{%
{\protect \APACyear {2006}}%
}]{%
gray2006}
\APACinsertmetastar {%
gray2006}%
\begin{APACrefauthors}%
{Gray}, R\BPBI O.%
, {Corbally}, C\BPBI J.%
, {Garrison}, R\BPBI F.%
\ et al.\end{APACrefauthors}%
\unskip\
\newblock
\APACrefYearMonthDay{2006}{{\APACmonth{07}}}{},
\newblock
\unskip
\newblock
\APACjournalVolNumPages{\aj}{132}{1}{161-170}.
\PrintBackRefs{\CurrentBib}

\bibitem [\protect \citeauthoryear {%
{Gray}%
, {Corbally}%
, {Garrison}%
, {McFadden}%
\BCBL {}\ \BBA {} {Robinson}%
}{%
{Gray}%
\ \protect \BOthers {.}}{%
{\protect \APACyear {2003}}%
}]{%
gray2003}
\APACinsertmetastar {%
gray2003}%
\begin{APACrefauthors}%
{Gray}, R\BPBI O.%
, {Corbally}, C\BPBI J.%
, {Garrison}, R\BPBI F.%
, {McFadden}, M\BPBI T.%
\BCBL {}\ \BBA {} {Robinson}, P\BPBI E.%
\end{APACrefauthors}%
\unskip\
\newblock
\APACrefYearMonthDay{2003}{{\APACmonth{10}}}{},
\newblock
\unskip
\newblock
\APACjournalVolNumPages{\aj}{126}{4}{2048-2059}.
\PrintBackRefs{\CurrentBib}

\bibitem [\protect \citeauthoryear {%
{Gray}%
, {Graham}%
\BCBL {}\ \BBA {} {Hoyt}%
}{%
{Gray}%
\ \protect \BOthers {.}}{%
{\protect \APACyear {2001}}%
}]{%
gray2001}
\APACinsertmetastar {%
gray2001}%
\begin{APACrefauthors}%
{Gray}, R\BPBI O.%
, {Graham}, P\BPBI W.%
\BCBL {}\ \BBA {} {Hoyt}, S\BPBI R.%
\end{APACrefauthors}%
\unskip\
\newblock
\APACrefYearMonthDay{2001}{{\APACmonth{04}}}{},
\newblock
\unskip
\newblock
\APACjournalVolNumPages{\aj}{121}{4}{2159-2172}.
\PrintBackRefs{\CurrentBib}

\bibitem [\protect \citeauthoryear {%
{Halbwachs}%
, {Mayor}%
\BCBL {}\ \BBA {} {Udry}%
}{%
{Halbwachs}%
\ \protect \BOthers {.}}{%
{\protect \APACyear {2018}}%
}]{%
halbwachs}
\APACinsertmetastar {%
halbwachs}%
\begin{APACrefauthors}%
{Halbwachs}, J\BPBI L.%
, {Mayor}, M.%
\BCBL {}\ \BBA {} {Udry}, S.%
\end{APACrefauthors}%
\unskip\
\newblock
\APACrefYearMonthDay{2018}{Nov}{},
\newblock
\unskip
\newblock
\APACjournalVolNumPages{\aap}{619}{}{A81}.
\PrintBackRefs{\CurrentBib}

\bibitem [\protect \citeauthoryear {%
{Heyne}%
\ \protect \BOthers {.}}{%
{Heyne}%
\ \protect \BOthers {.}}{%
{\protect \APACyear {2020}}%
}]{%
heyne}
\APACinsertmetastar {%
heyne}%
\begin{APACrefauthors}%
{Heyne}, T.%
, {Mugrauer}, M.%
, {Bischoff}, R.%
\ et al.\end{APACrefauthors}%
\unskip\
\newblock
\APACrefYearMonthDay{2020}{{\APACmonth{01}}}{},
\newblock
\unskip
\newblock
\APACjournalVolNumPages{Astronomische Nachrichten}{341}{1}{99-117}.
\PrintBackRefs{\CurrentBib}

\bibitem [\protect \citeauthoryear {%
{Houdebine}%
, {Mullan}%
, {Bercu}%
, {Paletou}%
\BCBL {}\ \BBA {} {Gebran}%
}{%
{Houdebine}%
\ \protect \BOthers {.}}{%
{\protect \APACyear {2017}}%
}]{%
houdebin2017}
\APACinsertmetastar {%
houdebin2017}%
\begin{APACrefauthors}%
{Houdebine}, E\BPBI R.%
, {Mullan}, D\BPBI J.%
, {Bercu}, B.%
, {Paletou}, F.%
\BCBL {}\ \BBA {} {Gebran}, M.%
\end{APACrefauthors}%
\unskip\
\newblock
\APACrefYearMonthDay{2017}{{\APACmonth{03}}}{},
\newblock
\unskip
\newblock
\APACjournalVolNumPages{\apj}{837}{1}{96}.
\PrintBackRefs{\CurrentBib}

\bibitem [\protect \citeauthoryear {%
{Houdebine}%
, {Mullan}%
, {Paletou}%
\BCBL {}\ \BBA {} {Gebran}%
}{%
{Houdebine}%
\ \protect \BOthers {.}}{%
{\protect \APACyear {2016}}%
}]{%
houdebin2016}
\APACinsertmetastar {%
houdebin2016}%
\begin{APACrefauthors}%
{Houdebine}, E\BPBI R.%
, {Mullan}, D\BPBI J.%
, {Paletou}, F.%
\BCBL {}\ \BBA {} {Gebran}, M.%
\end{APACrefauthors}%
\unskip\
\newblock
\APACrefYearMonthDay{2016}{{\APACmonth{05}}}{},
\newblock
\unskip
\newblock
\APACjournalVolNumPages{\apj}{822}{2}{97}.
\PrintBackRefs{\CurrentBib}

\bibitem [\protect \citeauthoryear {%
{Imbert}%
}{%
{Imbert}%
}{%
{\protect \APACyear {1977}}%
}]{%
imbert}
\APACinsertmetastar {%
imbert}%
\begin{APACrefauthors}%
{Imbert}, M.%
\end{APACrefauthors}%
\unskip\
\newblock
\APACrefYearMonthDay{1977}{{\APACmonth{09}}}{},
\newblock
\unskip
\newblock
\APACjournalVolNumPages{\aaps}{29}{}{407-409}.
\PrintBackRefs{\CurrentBib}

\bibitem [\protect \citeauthoryear {%
{Irrgang}%
, {Desphande}%
, {Moehler}%
, {Mugrauer}%
\BCBL {}\ \BBA {} {Janousch}%
}{%
{Irrgang}%
\ \protect \BOthers {.}}{%
{\protect \APACyear {2016}}%
}]{%
irrgang}
\APACinsertmetastar {%
irrgang}%
\begin{APACrefauthors}%
{Irrgang}, A.%
, {Desphande}, A.%
, {Moehler}, S.%
, {Mugrauer}, M.%
\BCBL {}\ \BBA {} {Janousch}, D.%
\end{APACrefauthors}%
\unskip\
\newblock
\APACrefYearMonthDay{2016}{Jun}{},
\newblock
\unskip
\newblock
\APACjournalVolNumPages{\aap}{591}{}{L6}.
\PrintBackRefs{\CurrentBib}

\bibitem [\protect \citeauthoryear {%
{Jordi}%
\ \protect \BOthers {.}}{%
{Jordi}%
\ \protect \BOthers {.}}{%
{\protect \APACyear {2010}}%
}]{%
jordi}
\APACinsertmetastar {%
jordi}%
\begin{APACrefauthors}%
{Jordi}, C.%
, {Gebran}, M.%
, {Carrasco}, J\BPBI M.%
\ et al.\end{APACrefauthors}%
\unskip\
\newblock
\APACrefYearMonthDay{2010}{Nov}{},
\newblock
\unskip
\newblock
\APACjournalVolNumPages{\aap}{523}{}{A48}.
\PrintBackRefs{\CurrentBib}

\bibitem [\protect \citeauthoryear {%
{Karata{\textcommabelow s}}%
, {Bilir}%
\BCBL {}\ \BBA {} {Schuster}%
}{%
{Karata{\textcommabelow s}}%
\ \protect \BOthers {.}}{%
{\protect \APACyear {2005}}%
}]{%
karatas}
\APACinsertmetastar {%
karatas}%
\begin{APACrefauthors}%
{Karata{\textcommabelow s}}, Y.%
, {Bilir}, S.%
\BCBL {}\ \BBA {} {Schuster}, W\BPBI J.%
\end{APACrefauthors}%
\unskip\
\newblock
\APACrefYearMonthDay{2005}{{\APACmonth{07}}}{},
\newblock
\unskip
\newblock
\APACjournalVolNumPages{\mnras}{360}{4}{1345-1354}.
\PrintBackRefs{\CurrentBib}

\bibitem [\protect \citeauthoryear {%
{Kunder}%
\ \protect \BOthers {.}}{%
{Kunder}%
\ \protect \BOthers {.}}{%
{\protect \APACyear {2017}}%
}]{%
kunder}
\APACinsertmetastar {%
kunder}%
\begin{APACrefauthors}%
{Kunder}, A.%
, {Kordopatis}, G.%
, {Steinmetz}, M.%
\ et al.\end{APACrefauthors}%
\unskip\
\newblock
\APACrefYearMonthDay{2017}{{\APACmonth{02}}}{},
\newblock
\unskip
\newblock
\APACjournalVolNumPages{\aj}{153}{2}{75}.
\PrintBackRefs{\CurrentBib}

\bibitem [\protect \citeauthoryear {%
{Marsden}%
\ \protect \BOthers {.}}{%
{Marsden}%
\ \protect \BOthers {.}}{%
{\protect \APACyear {2014}}%
}]{%
marsden}
\APACinsertmetastar {%
marsden}%
\begin{APACrefauthors}%
{Marsden}, S\BPBI C.%
, {Petit}, P.%
, {Jeffers}, S\BPBI V.%
\ et al.\end{APACrefauthors}%
\unskip\
\newblock
\APACrefYearMonthDay{2014}{{\APACmonth{11}}}{},
\newblock
\unskip
\newblock
\APACjournalVolNumPages{\mnras}{444}{4}{3517-3536}.
\PrintBackRefs{\CurrentBib}

\bibitem [\protect \citeauthoryear {%
{Mugrauer}%
}{%
{Mugrauer}%
}{%
{\protect \APACyear {2019}}%
}]{%
mugrauer2019}
\APACinsertmetastar {%
mugrauer2019}%
\begin{APACrefauthors}%
{Mugrauer}, M.%
\end{APACrefauthors}%
\unskip\
\newblock
\APACrefYearMonthDay{2019}{{\APACmonth{12}}}{},
\newblock
\unskip
\newblock
\APACjournalVolNumPages{\mnras}{490}{4}{5088-5102}.
\PrintBackRefs{\CurrentBib}

\bibitem [\protect \citeauthoryear {%
{Mugrauer}%
, {Avila}%
\BCBL {}\ \BBA {} {Guirao}%
}{%
{Mugrauer}%
\ \protect \BOthers {.}}{%
{\protect \APACyear {2014}}%
}]{%
mugrauer2014}
\APACinsertmetastar {%
mugrauer2014}%
\begin{APACrefauthors}%
{Mugrauer}, M.%
, {Avila}, G.%
\BCBL {}\ \BBA {} {Guirao}, C.%
\end{APACrefauthors}%
\unskip\
\newblock
\APACrefYearMonthDay{2014}{Jan}{},
\newblock
\unskip
\newblock
\APACjournalVolNumPages{Astronomische Nachrichten}{335}{4}{417}.
\PrintBackRefs{\CurrentBib}

\bibitem [\protect \citeauthoryear {%
{Neuh{\"a}user}%
}{%
{Neuh{\"a}user}%
}{%
{\protect \APACyear {1997}}%
}]{%
neuhaeuser}
\APACinsertmetastar {%
neuhaeuser}%
\begin{APACrefauthors}%
{Neuh{\"a}user}, R.%
\end{APACrefauthors}%
\unskip\
\newblock
\APACrefYearMonthDay{1997}{Jan}{},
\newblock
\unskip
\newblock
\APACjournalVolNumPages{Science}{276}{}{1363-1370}.
\PrintBackRefs{\CurrentBib}

\bibitem [\protect \citeauthoryear {%
{Ochsenbein}%
, {Bauer}%
\BCBL {}\ \BBA {} {Marcout}%
}{%
{Ochsenbein}%
\ \protect \BOthers {.}}{%
{\protect \APACyear {2000}}%
}]{%
ochsenbein}
\APACinsertmetastar {%
ochsenbein}%
\begin{APACrefauthors}%
{Ochsenbein}, F.%
, {Bauer}, P.%
\BCBL {}\ \BBA {} {Marcout}, J.%
\end{APACrefauthors}%
\unskip\
\newblock
\APACrefYearMonthDay{2000}{Apr}{},
\newblock
\unskip
\newblock
\APACjournalVolNumPages{\aaps}{143}{}{23-32}.
\PrintBackRefs{\CurrentBib}

\bibitem [\protect \citeauthoryear {%
{Passegger}%
\ \protect \BOthers {.}}{%
{Passegger}%
\ \protect \BOthers {.}}{%
{\protect \APACyear {2019}}%
}]{%
passeger}
\APACinsertmetastar {%
passeger}%
\begin{APACrefauthors}%
{Passegger}, V\BPBI M.%
, {Schweitzer}, A.%
, {Shulyak}, D.%
\ et al.\end{APACrefauthors}%
\unskip\
\newblock
\APACrefYearMonthDay{2019}{Jul}{},
\newblock
\unskip
\newblock
\APACjournalVolNumPages{\aap}{627}{}{A161}.
\PrintBackRefs{\CurrentBib}

\bibitem [\protect \citeauthoryear {%
{Perryman}%
\ \protect \BOthers {.}}{%
{Perryman}%
\ \protect \BOthers {.}}{%
{\protect \APACyear {1997}}%
}]{%
perryman}
\APACinsertmetastar {%
perryman}%
\begin{APACrefauthors}%
{Perryman}, M\BPBI A\BPBI C.%
, {Lindegren}, L.%
, {Kovalevsky}, J.%
\ et al.\end{APACrefauthors}%
\unskip\
\newblock
\APACrefYearMonthDay{1997}{Jul}{},
\newblock
\unskip
\newblock
\APACjournalVolNumPages{\aap}{500}{}{501-504}.
\PrintBackRefs{\CurrentBib}

\bibitem [\protect \citeauthoryear {%
{Petigura}%
\ \BBA {} {Marcy}%
}{%
{Petigura}%
\ \BBA {} {Marcy}%
}{%
{\protect \APACyear {2011}}%
}]{%
petigura}
\APACinsertmetastar {%
petigura}%
\begin{APACrefauthors}%
{Petigura}, E\BPBI A.%
\BCBT {}\ \BBA {} {Marcy}, G\BPBI W.%
\end{APACrefauthors}%
\unskip\
\newblock
\APACrefYearMonthDay{2011}{{\APACmonth{07}}}{},
\newblock
\unskip
\newblock
\APACjournalVolNumPages{\apj}{735}{1}{41}.
\PrintBackRefs{\CurrentBib}

\bibitem [\protect \citeauthoryear {%
{Pfau}%
}{%
{Pfau}%
}{%
{\protect \APACyear {1984}}%
}]{%
pfau}
\APACinsertmetastar {%
pfau}%
\begin{APACrefauthors}%
{Pfau}, W.%
\end{APACrefauthors}%
\unskip\
\newblock
\APACrefYearMonthDay{1984}{Jan}{},
\newblock
\unskip
\newblock
\APACjournalVolNumPages{Jenaer Rundschau}{29}{3}{121-122}.
\PrintBackRefs{\CurrentBib}

\bibitem [\protect \citeauthoryear {%
{Poveda}%
, {Ruiz}%
\BCBL {}\ \BBA {} {Allen}%
}{%
{Poveda}%
\ \protect \BOthers {.}}{%
{\protect \APACyear {1967}}%
}]{%
poveda}
\APACinsertmetastar {%
poveda}%
\begin{APACrefauthors}%
{Poveda}, A.%
, {Ruiz}, J.%
\BCBL {}\ \BBA {} {Allen}, C.%
\end{APACrefauthors}%
\unskip\
\newblock
\APACrefYearMonthDay{1967}{Apr}{},
\newblock
\unskip
\newblock
\APACjournalVolNumPages{Boletin de los Observatorios Tonantzintla y
  Tacubaya}{4}{}{86-90}.
\PrintBackRefs{\CurrentBib}

\bibitem [\protect \citeauthoryear {%
{Rajpurohit}%
\ \protect \BOthers {.}}{%
{Rajpurohit}%
\ \protect \BOthers {.}}{%
{\protect \APACyear {2018}}%
}]{%
raj}
\APACinsertmetastar {%
raj}%
\begin{APACrefauthors}%
{Rajpurohit}, A\BPBI S.%
, {Allard}, F.%
, {Rajpurohit}, S.%
, {Sharma}, R.%
, {Teixeira}, G\BPBI D\BPBI C.%
, {Mousis}, O.%
\BCBL {}\ \BBA {} {Rajpurohit}, K.%
\end{APACrefauthors}%
\unskip\
\newblock
\APACrefYearMonthDay{2018}{{\APACmonth{12}}}{},
\newblock
\unskip
\newblock
\APACjournalVolNumPages{\aap}{620}{}{A180}.
\PrintBackRefs{\CurrentBib}

\bibitem [\protect \citeauthoryear {%
{Ram{\'\i}rez}%
, {Fish}%
, {Lambert}%
\BCBL {}\ \BBA {} {Allende Prieto}%
}{%
{Ram{\'\i}rez}%
\ \protect \BOthers {.}}{%
{\protect \APACyear {2012}}%
}]{%
ramirez}
\APACinsertmetastar {%
ramirez}%
\begin{APACrefauthors}%
{Ram{\'\i}rez}, I.%
, {Fish}, J\BPBI R.%
, {Lambert}, D\BPBI L.%
\BCBL {}\ \BBA {} {Allende Prieto}, C.%
\end{APACrefauthors}%
\unskip\
\newblock
\APACrefYearMonthDay{2012}{{\APACmonth{09}}}{},
\newblock
\unskip
\newblock
\APACjournalVolNumPages{\apj}{756}{1}{46}.
\PrintBackRefs{\CurrentBib}

\bibitem [\protect \citeauthoryear {%
{Rojas-Ayala}%
, {Covey}%
, {Muirhead}%
\BCBL {}\ \BBA {} {Lloyd}%
}{%
{Rojas-Ayala}%
\ \protect \BOthers {.}}{%
{\protect \APACyear {2012}}%
}]{%
rojas}
\APACinsertmetastar {%
rojas}%
\begin{APACrefauthors}%
{Rojas-Ayala}, B.%
, {Covey}, K\BPBI R.%
, {Muirhead}, P\BPBI S.%
\BCBL {}\ \BBA {} {Lloyd}, J\BPBI P.%
\end{APACrefauthors}%
\unskip\
\newblock
\APACrefYearMonthDay{2012}{{\APACmonth{04}}}{},
\newblock
\unskip
\newblock
\APACjournalVolNumPages{\apj}{748}{2}{93}.
\PrintBackRefs{\CurrentBib}

\bibitem [\protect \citeauthoryear {%
{Schweitzer}%
\ \protect \BOthers {.}}{%
{Schweitzer}%
\ \protect \BOthers {.}}{%
{\protect \APACyear {2019}}%
}]{%
schweitzer}
\APACinsertmetastar {%
schweitzer}%
\begin{APACrefauthors}%
{Schweitzer}, A.%
, {Passegger}, V\BPBI M.%
, {Cifuentes}, C.%
\ et al.\end{APACrefauthors}%
\unskip\
\newblock
\APACrefYearMonthDay{2019}{May}{},
\newblock
\unskip
\newblock
\APACjournalVolNumPages{\aap}{625}{}{A68}.
\PrintBackRefs{\CurrentBib}

\bibitem [\protect \citeauthoryear {%
{Soderblom}%
, {Hillenbrand}%
, {Jeffries}%
, {Mamajek}%
\BCBL {}\ \BBA {} {Naylor}%
}{%
{Soderblom}%
\ \protect \BOthers {.}}{%
{\protect \APACyear {2014}}%
}]{%
soderblom2014}
\APACinsertmetastar {%
soderblom2014}%
\begin{APACrefauthors}%
{Soderblom}, D\BPBI R.%
, {Hillenbrand}, L\BPBI A.%
, {Jeffries}, R\BPBI D.%
, {Mamajek}, E\BPBI E.%
\BCBL {}\ \BBA {} {Naylor}, T.%
\end{APACrefauthors}%
\unskip\
\newblock
\APACrefYearMonthDay{2014}{{\APACmonth{01}}}{},
\newblock
{\BBOQ}\APACrefatitle {{Ages of Young Stars}} {{Ages of Young Stars}}.{\BBCQ}
\newblock
\BIn{} H.~{Beuther}, R\BPBI S.~{Klessen}, C\BPBI P.~{Dullemond}\BCBL {}\
  \BOthers {.}\ (\BEDS), \APACrefbtitle {Protostars and Planets VI} {Protostars
  and Planets VI}\ \BPG~219.
\PrintBackRefs{\CurrentBib}

\bibitem [\protect \citeauthoryear {%
{Soderblom}%
\ \protect \BOthers {.}}{%
{Soderblom}%
\ \protect \BOthers {.}}{%
{\protect \APACyear {1993}}%
}]{%
soderblom}
\APACinsertmetastar {%
soderblom}%
\begin{APACrefauthors}%
{Soderblom}, D\BPBI R.%
, {Jones}, B\BPBI F.%
, {Balachand ran}, S.%
, {Stauffer}, J\BPBI R.%
, {Duncan}, D\BPBI K.%
, {Fedele}, S\BPBI B.%
\BCBL {}\ \BBA {} {Hudon}, J\BPBI D.%
\end{APACrefauthors}%
\unskip\
\newblock
\APACrefYearMonthDay{1993}{{\APACmonth{09}}}{},
\newblock
\unskip
\newblock
\APACjournalVolNumPages{\aj}{106}{}{1059}.
\PrintBackRefs{\CurrentBib}

\bibitem [\protect \citeauthoryear {%
{Stassun}%
\ \protect \BOthers {.}}{%
{Stassun}%
\ \protect \BOthers {.}}{%
{\protect \APACyear {2019}}%
}]{%
stassun}
\APACinsertmetastar {%
stassun}%
\begin{APACrefauthors}%
{Stassun}, K\BPBI G.%
, {Oelkers}, R\BPBI J.%
, {Paegert}, M.%
\ et al.\end{APACrefauthors}%
\unskip\
\newblock
\APACrefYearMonthDay{2019}{{\APACmonth{10}}}{},
\newblock
\unskip
\newblock
\APACjournalVolNumPages{\aj}{158}{4}{138}.
\PrintBackRefs{\CurrentBib}

\bibitem [\protect \citeauthoryear {%
{Szentgyorgyi}%
\ \BBA {} {Fur{\'e}sz}%
}{%
{Szentgyorgyi}%
\ \BBA {} {Fur{\'e}sz}%
}{%
{\protect \APACyear {2007}}%
}]{%
Szentgyorgyi2007}
\APACinsertmetastar {%
Szentgyorgyi2007}%
\begin{APACrefauthors}%
{Szentgyorgyi}, A\BPBI H.%
\BCBT {}\ \BBA {} {Fur{\'e}sz}, G.%
\end{APACrefauthors}%
\unskip\
\newblock
\APACrefYearMonthDay{2007}{Jun}{},
\newblock
{\BBOQ}\APACrefatitle {{Precision Radial Velocities for the Kepler Era}}
  {{Precision Radial Velocities for the Kepler Era}}.{\BBCQ}
\newblock
\BIn{} S.~{Kurtz}\ (\BED), \APACrefbtitle {Revista Mexicana de Astronomia y
  Astrofisica Conference Series} {Revista Mexicana de Astronomia y Astrofisica
  Conference Series}\ \BVOL~28, \BPG~129-133.
\PrintBackRefs{\CurrentBib}

\bibitem [\protect \citeauthoryear {%
{Tetzlaff}%
, {Neuh{\"a}user}%
\BCBL {}\ \BBA {} {Hohle}%
}{%
{Tetzlaff}%
\ \protect \BOthers {.}}{%
{\protect \APACyear {2011}}%
}]{%
tetzlaff}
\APACinsertmetastar {%
tetzlaff}%
\begin{APACrefauthors}%
{Tetzlaff}, N.%
, {Neuh{\"a}user}, R.%
\BCBL {}\ \BBA {} {Hohle}, M\BPBI M.%
\end{APACrefauthors}%
\unskip\
\newblock
\APACrefYearMonthDay{2011}{Jan}{},
\newblock
\unskip
\newblock
\APACjournalVolNumPages{\mnras}{410}{1}{190-200}.
\PrintBackRefs{\CurrentBib}

\bibitem [\protect \citeauthoryear {%
C\BPBI A\BPBI O.~{Torres}%
\ \protect \BOthers {.}}{%
C\BPBI A\BPBI O.~{Torres}%
\ \protect \BOthers {.}}{%
{\protect \APACyear {2006}}%
}]{%
torres2006}
\APACinsertmetastar {%
torres2006}%
\begin{APACrefauthors}%
{Torres}, C\BPBI A\BPBI O.%
, {Quast}, G\BPBI R.%
, {da Silva}, L.%
, {de La Reza}, R.%
, {Melo}, C\BPBI H\BPBI F.%
\BCBL {}\ \BBA {} {Sterzik}, M.%
\end{APACrefauthors}%
\unskip\
\newblock
\APACrefYearMonthDay{2006}{Dec}{},
\newblock
\unskip
\newblock
\APACjournalVolNumPages{\aap}{460}{3}{695-708}.
\PrintBackRefs{\CurrentBib}

\bibitem [\protect \citeauthoryear {%
G.~{Torres}%
, {Andersen}%
\BCBL {}\ \BBA {} {Gim{\'e}nez}%
}{%
G.~{Torres}%
\ \protect \BOthers {.}}{%
{\protect \APACyear {2010}}%
}]{%
torres2010}
\APACinsertmetastar {%
torres2010}%
\begin{APACrefauthors}%
{Torres}, G.%
, {Andersen}, J.%
\BCBL {}\ \BBA {} {Gim{\'e}nez}, A.%
\end{APACrefauthors}%
\unskip\
\newblock
\APACrefYearMonthDay{2010}{{\APACmonth{02}}}{},
\newblock
\unskip
\newblock
\APACjournalVolNumPages{\aapr}{18}{1-2}{67-126}.
\PrintBackRefs{\CurrentBib}

\bibitem [\protect \citeauthoryear {%
{Valenti}%
\ \BBA {} {Fischer}%
}{%
{Valenti}%
\ \BBA {} {Fischer}%
}{%
{\protect \APACyear {2005}}%
}]{%
valenti}
\APACinsertmetastar {%
valenti}%
\begin{APACrefauthors}%
{Valenti}, J\BPBI A.%
\BCBT {}\ \BBA {} {Fischer}, D\BPBI A.%
\end{APACrefauthors}%
\unskip\
\newblock
\APACrefYearMonthDay{2005}{{\APACmonth{07}}}{},
\newblock
\unskip
\newblock
\APACjournalVolNumPages{\apjs}{159}{1}{141-166}.
\PrintBackRefs{\CurrentBib}

\bibitem [\protect \citeauthoryear {%
{van Leeuwen}%
}{%
{van Leeuwen}%
}{%
{\protect \APACyear {2007}}%
}]{%
vanleeuwen}
\APACinsertmetastar {%
vanleeuwen}%
\begin{APACrefauthors}%
{van Leeuwen}, F.%
\end{APACrefauthors}%
\unskip\
\newblock
\APACrefYearMonthDay{2007}{Nov}{},
\newblock
\unskip
\newblock
\APACjournalVolNumPages{\aap}{474}{2}{653-664}.
\PrintBackRefs{\CurrentBib}

\bibitem [\protect \citeauthoryear {%
{Worley}%
\ \protect \BOthers {.}}{%
{Worley}%
\ \protect \BOthers {.}}{%
{\protect \APACyear {2012}}%
}]{%
worley}
\APACinsertmetastar {%
worley}%
\begin{APACrefauthors}%
{Worley}, C\BPBI C.%
, {de Laverny}, P.%
, {Recio-Blanco}, A.%
, {Hill}, V.%
, {Bijaoui}, A.%
\BCBL {}\ \BBA {} {Ordenovic}, C.%
\end{APACrefauthors}%
\unskip\
\newblock
\APACrefYearMonthDay{2012}{{\APACmonth{06}}}{},
\newblock
\unskip
\newblock
\APACjournalVolNumPages{\aap}{542}{}{A48}.
\PrintBackRefs{\CurrentBib}

\bibitem [\protect \citeauthoryear {%
{Zuckerman}%
, {Song}%
, {Bessell}%
\BCBL {}\ \BBA {} {Webb}%
}{%
{Zuckerman}%
\ \protect \BOthers {.}}{%
{\protect \APACyear {2001}}%
}]{%
zuckerman}
\APACinsertmetastar {%
zuckerman}%
\begin{APACrefauthors}%
{Zuckerman}, B.%
, {Song}, I.%
, {Bessell}, M\BPBI S.%
\BCBL {}\ \BBA {} {Webb}, R\BPBI A.%
\end{APACrefauthors}%
\unskip\
\newblock
\APACrefYearMonthDay{2001}{Nov}{},
\newblock
\unskip
\newblock
\APACjournalVolNumPages{\apjl}{562}{1}{L87-L90}.
\PrintBackRefs{\CurrentBib}

\end{thebibliography}

\section*{Author Biography}

Richard Bischoff is a PhD student at the Astrophysical Institute and University Observatory Jena. His main field of research are photometry and spectroscopy of exoplanet candidate host stars.

\end{document}